\documentclass[oldversion]{aa}
\usepackage{amsmath}
\usepackage{graphicx}
\usepackage{txfonts}
\usepackage{longtable}

\begin{document}

\title{Cyclotron lines in highly magnetized neutron stars}

\author{R.~Staubert\inst{1}, J.~Tr\"umper\inst{2}, E.~Kendziorra\inst{1}\thanks{In memoriam},
D.~Klochkov\inst{1}, K.~Postnov\inst{3}, P.~Kretschmar\inst{4}, K.~Pottschmidt\inst{5}, F.~Haberl\inst{2}, R.E.~Rothschild\inst{6}, 
A.~Santangelo\inst{1}, J.~Wilms\inst{7}, I. Kreykenbohm\inst{7}, F.~F\"urst\inst{4} }

\offprints{staubert@astro.uni-tuebingen.de}

\institute{
        Institut f\"ur Astronomie und Astrophysik, Universit\"at T\"ubingen,
        Sand 1, 72076 T\"ubingen, Germany
\and
        Max-Planck-Institut f\"ur extraterrestrische Physik, 
        Giessenbachstr. 1, 85748 Garching, Germany
\and
        Sternberg Astronomical Institute, Lomonossov University, 119992, Moscow, Russia
\and
        European Space Agency -- European Space Astronomy Center (ESA-ESAC),
        Camino Bajo del Castillo, s/n., Urb. Villafranca del Castillo, 28692 Villanueva de la Canada, 
        Madrid, Spain 
\and
        NASA Goddard Spaceflight Center, 8800 Greenbelt Rd, Greenbelt, Maryland, MD 20771, 
        USA, and CRESST, Department of Physics, and Center for Space Science and Technology, 
        UMBC, Baltimore, MD 21250, USA
\and
        Center for Astrophysics and Space Sciences, University of California at San Diego, 
        La Jolla, CA 92093-0424, USA
\and       
        Dr.\ Remeis Sternwarte, Astronomisches Institut der
        Universit\"at Erlangen-N\"urnberg, Sternwartstr. 7, 96049 Bamberg, Germany
}

\date{received: 2018 Oct 22, accepted: 2018 Nov 17}
\authorrunning{Staubert et al.}
\titlerunning{Cyclotron line sources}

\abstract{
Cyclotron lines, also called cyclotron resonant scattering features (CRSF) are spectral features, 
generally appearing in absorption, in the X-ray spectra of objects containing highly magnetized neutron stars,
allowing the direct measurement of the magnetic field strength in these objects. Cyclotron features are thought 
to be due to resonant scattering of photons by electrons in the strong magnetic fields. 
The main content of this contribution focusses on electron cyclotron lines as found in accreting X-ray 
binary pulsars (XRBP) with magnetic fields on the order of several $10^{12}$\,Gauss. Also, possible 
proton cyclotron lines from single neutron stars with even stronger magnetic fields are briefly discussed. 
With regard to electron cyclotron lines, we present an updated list of XRBPs that show evidence of such 
absorption lines. The first such line was discovered in a 1976 balloon observation of the accreting binary 
pulsar Hercules X-1,
it is considered to be the first direct measurement of the 
magnetic field of a neutron star. As of today (mid 2018), we list 36
XRBPs showing evidence of one ore more electron cyclotron absorption line(s).  A few have been measured 
only once and must be confirmed (several more objects are listed as candidates). In addition to the Tables 
of objects, we summarize the evidence of variability of the cyclotron line as a function of  various 
parameters (especially pulse phase, luminosity and time), and add a discussion of the different observed 
phenomena and associated attempts of theoretical modeling. We also discuss our understanding of the 
underlying physics of accretion onto highly magnetized neutron stars. For proton cyclotron lines, 
we present tables with seven neutron stars and discuss their nature and the physics in these objects. 
}

   
\maketitle
%

\section{Introduction}  
\label{sec:introduction}

In this contribution we review the status of cyclotron line research in the X-ray range, preferentially from
an observational point of view, together with the related theoretical background, and some
recent progress in modeling the relevant physics. Highly magnetized accreting neutron stars (NSs) in 
binary systems generally reveal themselves as X-ray pulsars. A substantial fraction of those objects
show line-like features in their high energy X-ray spectra, mostly in absorption at energies from $\sim$10\,keV
to $\sim$100\,keV, called cyclotron lines or cyclotron resonant scattering features (CRSFs). 
The following introductory statements are for objects with electron cyclotron lines (we will come to proton cyclotron 
line objects below). Such lines are generated close to the magnetic poles of accreting 
neutron stars in the hot, highly magnetized plasma 
where the kinetic energy of the in-falling material is converted to heat and radiation. Electrons assume 
discrete energies with respect to their movement perpendicular to the magnetic field, so called 
Landau levels. Resonant scattering of photons on these electrons then leads to scattering of
photons at the resonance  energy and to the generation of 
resonance features (in absorption) in the X-ray spectrum. The fundamental energy quantum corresponds 
to the energy difference between adjacent Landau levels, given by $\hbar \omega$, where $\omega = eB/m_{e}c$ is 
the gyro frequency, $e$ is the electron charge, B is the magnetic field in the scattering region, $m_e$ is the mass 
of the electron, and $c$ is the speed of light. The Landau levels (to first order equidistant) are linearly related to 
the strength of the magnetic field. The observation of such features then allows for the measurement of the magnetic 
field strength. The following centroid line energies $E_\mathrm{cyc}$ are expected:
\begin{equation}
\label{equ:Bcyc}
E_{cyc} = \frac{n}{(1+z)}  \frac{\hbar eB}{m_e c} \approx \frac{n}{(1+z)} 11.6~[keV]  \times B_{12},
\end{equation}
 where $B_\mathrm{12}$ is the magnetic field strength in units of $10^{12}$\,Gauss, z is the gravitational 
redshift due to the NS mass and n is the number of Landau levels involved: e.g., n = 1 
is the case of a scattering from the ground level to the first excited Landau level and
the resulting line is called the fundamental line. In the case of n = 2 (or higher) the lines are called harmonics.

The same physics is valid for other charged particles, for example protons, for which the difference between 
Landau levels is smaller by a factor of 1836 (the ratio of the proton to electron mass). Such lines are
thought to be observed from isolated, thermally radiating neutron stars in the range 0.1\,keV to a few 
keV. For the proton cyclotron lines to appear in the X-ray range, the magnetic field 
must typically be two orders of magnitude stronger than for electron cyclotron lines. 

Cyclotron line research with accreting X-ray binaries (or isolated  neutron stars) has become a field 
of its own within X-ray astronomy. It is a vibrant  field, in both theoretical and in observational work, dealing 
not only with the detection of new lines, but also with various recently discovered properties of these lines, 
such as the dependence of the line energy, width and depth on X-ray luminosity and its long-term
evolution with time. Cyclotron lines are an important diagnostic tool to investigate
the physics of accretion onto the magnetic poles of highly magnetized neutron stars.

\begin{figure}
   \includegraphics[width=0.4\textwidth]{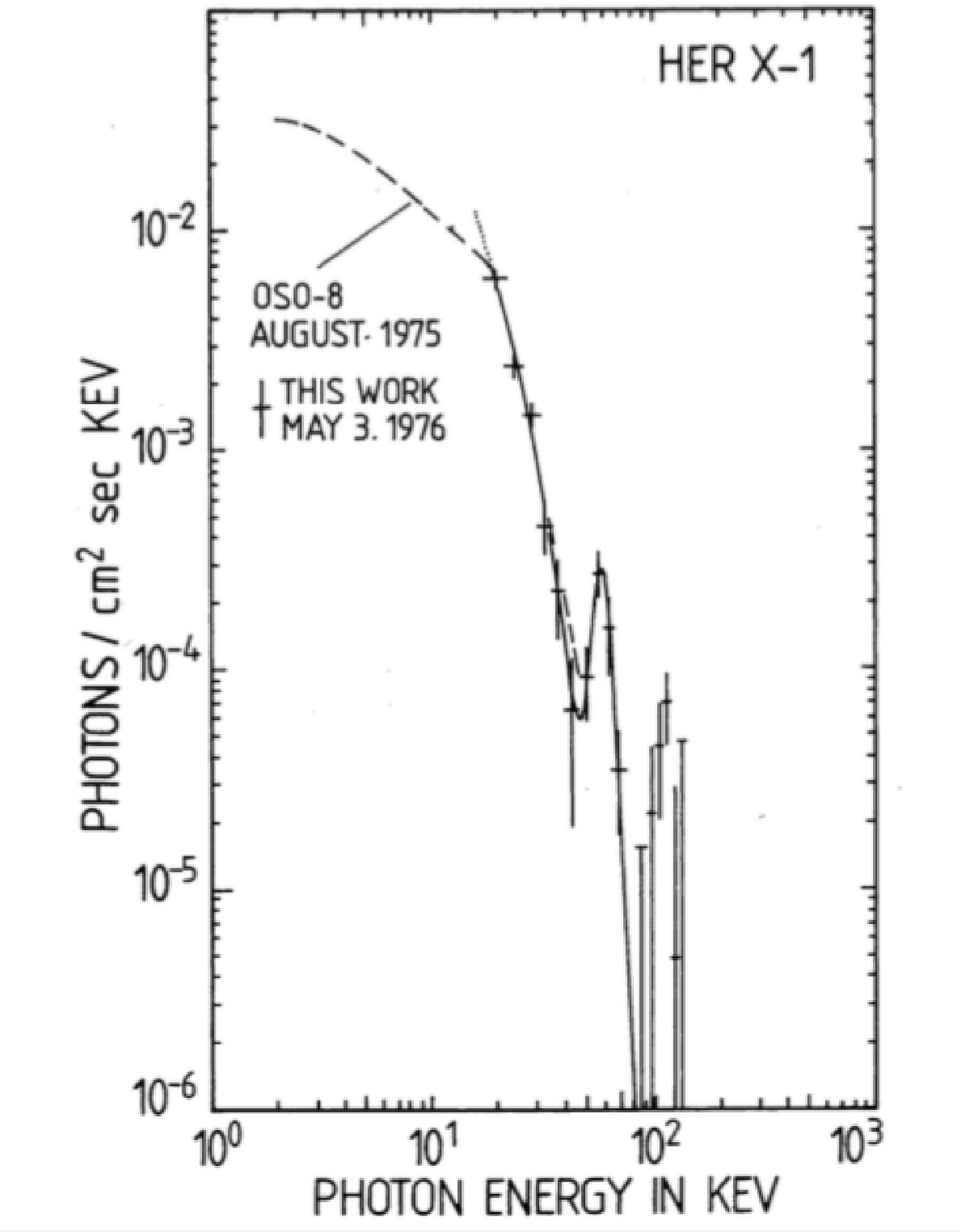}
   \caption{X-ray spectrum of Her X-1 as obtained in a balloon
     observation in 1976, constituting the first detection of a
     cyclotron line (from Tr\"{u}mper et al. 1978).}
   \label{fig:HerX1_CRSF}
\end{figure}

Here we give a brief history of the beginning of cyclotron line research and its 
evolution up to the present. The discovery of cyclotron lines in 1976 in the X-ray spectrum of Her X-1 
\citep{Truemper_etal77,Truemper_etal78} (see Fig.~\ref{fig:HerX1_CRSF})
was not the result of a targeted search, but came as a surprise. A joint group of the Astronomical Institute T\"ubingen 
(AIT) and the Max Planck Institute for Extraterrestrial Physics (MPE) had flown a hard X-ray balloon payload 
(``Balloon HEXE'') on 2-3 May 1976, observing Her X-1 and Cygnus X-1 at energies from 20\,keV to 200\,keV. 
While the AIT group concentrated on the analysis of the energy dependent pulse profiles \citep{Kendziorra_etal77},
their colleagues at MPE dealt with the energy spectrum. They found an unexpected shoulder-like deviation 
from the steep total energy spectrum around 50 keV which looked like a resonance feature. Detailed 
investigations showed that the feature was present in the pulsed spectrum and absent in the Cygnus X-1 
spectrum taken on the same balloon flight, excluding the possibility of a spurious origin. Because of its high 
energy and intensity it could be attributed only to the electron cyclotron resonance. 

This discovery was first reported in a very late talk -- proposed and accepted during the meeting -- at the Eighth 
Texas Symposium on Relativistic Astrophysics on 16 December 1976 \citep{Truemper_etal77}, and became 
a highlight of that meeting. An extended version that appeared later \citep{Truemper_etal78} has become the 
standard reference for the discovery which revealed a new phenomenon in X-ray astronomy and provided the 
first direct measurement of a neutron star magnetic field, opening a new avenue of research in the field.
The limited spectral resolution of the Balloon-HEXE did not allow us to distinguish between an emission or 
absorption line \citep{Truemper_etal78}. For the answer to this question we had to await radiative transfer calculations 
in strong magnetic fields. For that the pioneering work on the magnetic photon-electron cross sections was 
important \citep{Lodenquai_etal74,GnedinSunyaev_74}. The latter work -- based on the assumption of a 
(Thomson) optically thin radiation source -- also predicted that at the cyclotron a resonance emission line 
would occur (see also \citealt{BaskoSunyaev_75}). 

The discovery of the Her X-1 cyclotron lines spurred a burst of early theoretical papers on the radiative transfer 
in strongly magnetized plasmas. Some of them provided support for an emission line interpretation of the feature 
(e.g., \citealt{DaughertyVentura_77,Meszaros_78,Yahel_79a,Yahel_79b,WassermanSalpeter_80,MelroseZheleznyakov_81}).
The first support for an absorption line interpretation came from a talk by Andy Fabian at the first international 
workshop on cyclotron lines, organized by the
Max-Planck-Institut f\"ur extraterrestrische Physik (MPE) in fall 1978. His argument was: the radiating high 
temperature plasma would be optically thick. In the cyclotron resonance region the reflectivity would be high 
and according to the Kirchhoff law the emissivity would be low (Fabian 1978, unpublished). Other important 
early theoretical papers followed  (e.g., \citealt{Bonazzola_etal79, Ventura_etal79,Bussard_80,KirkMeszaros_80,
Langer_etal80,Nagel_80,Yahel_80a,Yahel_80b,PravdoBussard_81,Nagel_81a,Nagel_81b,Meszaros_etal83}). 
All these and many following works led to the current view that these lines are seen ``in absorption". This was 
also observationally verified for Her~X-1 in 1996 through data from \textsl{Mir}-HEXE \citep{Staubert_03}. 
The first significant confirmation of the 
cyclotron line in Her~X-1 came in 1980 on the basis of \textsl{HEAO-1}/A4 observations \citep{Gruber_etal80}.
The same experiment discovered a cyclotron line at 20 keV in the source 4U~0115+63 \citep{Wheaton_etal79}. 
A re-examination of the data uncovered the existence of two lines at 11.5 and 23 keV  which were interpreted 
as the fundamental and harmonic electron cyclotron resonances seen in absorption \citep{White_etal83}. Later 
observations of the source found it to show three \citep{Heindl_etal99a}, four \citep{Santangelo_etal99}, 
and finally up to five lines in total \citep{Heindl_etal04}. 

In many accreting X-ray binary pulsars (XRBPs) the centroid energy $E_\mathrm{cyc}$ of cyclotron 
lines is seen to vary. The first variations seen were variations of $E_\mathrm{cyc}$ with pulse phase 
(e.g., \citealt{Voges_etal82, Soong_etal90b,Vasco_etal13}), and are attributed to a varying viewing
angle to the scattering region. A negative correlation between $E_\mathrm{cyc}$ and the X-ray 
luminosity was reported during high luminosity outbursts of two X-ray transients: V~0332+53
and 4U~0115+63 \citep{Makishima_etal90,Mihara_95}. Today we consider this 
to be real only in V~0332+53 (see Sect.~\ref{sec:luminosity}). It was again in Her~X-1 that two 
more variability phenomena were first detected: firstly, a positive correlation of $E_\mathrm{cyc}$ 
with the X-ray luminosity \citep{Staubert_etal07} and secondly, a long-term decay of 
$E_\mathrm{cyc}$ with time \citep{Staubert_etal14} (revealing a $\sim$5\,keV reduction over the last 
20 years). Today we know a total of seven objects showing the same positive
correlation (see Sect.~\ref{sec:luminosity}), and one more example for a long-term decay 
(Vela~X-1, \citealt{LaParola_etal16}) (Sect.~\ref{sec:long_term}).

In the last 40 years the number of known electron cyclotron sources has increased 
to $\sim$36 (see Table~\ref{tab:collection}) due to investigations with many X-ray observatories like
\textsl{Ginga, MIR-HEXE, RXTE, BeppoSAX, INTEGRAL, Suzaku, and NuSTAR}. 
For previous reviews about cyclotron line sources, see for example, 
\citet{Coburn_etal02,Staubert_03,Heindl_etal04,Terada_etal07,Wilms_12,CaballeroWilms_12,
RevnivtsevMereghetti_16,Maitra_17}.
Lists of CRSF sources can also be found at the webpages of Dr. Remeis-Sternwarte, 
Bamberg\footnote{http://www.sternwarte.uni-erlangen.de/wiki/doku.php?id=xrp:start} and 
the Istituto Astrrofisica Spaziale, Bologna\footnote{http://www.iasfbo.inaf.it/\textasciitilde mauro/pulsar\_list.html}. 
Recent theoretical work has followed two lines: analytical calculations
\citep{Nishimura_08,Nishimura_11,Nishimura_13,Nishimura_14} or making use of Monte Carlo techniques 
\citep{ArayaHarding_99,ArayaGochezHarding_00,Schoenherr_etal07,Schwarm_etal17a,Schwarm_etal17b}. 
In addition, evidence for the detection of proton cyclotron lines in the thermal spectra of isolated neutron stars
was provided by \textsl{Chandra} and \textsl{XMM-Newton} \citep{2007Ap&SS.308..181H}
(see Tables~\ref{tab_M7} and \ref{tab_M7lines}).
In this paper we review the status of cyclotron line research by summarizing the current knowledge about cyclotron 
line sources (both electron- and proton-cyclotron line sources) and the state of our understanding of the
the details of the underlying physics.

This paper is structured as follows: in Sect.~\ref{sec:source_list} we present a series of tables, which contain
detailed information about all objects known to us which show conclusive evidence for the existence of  
electron cyclotron lines (together with appropriate references), plus several uncertain candidate objects. 
In Sect.~\ref{sec:spectral_fitting} we discuss issues of spectral fitting, including a list of popular functions 
for the modeling of the continuum and the cyclotron lines, as well as systematic differences that must be 
taken into consideration when different results from the literature are to be compared. 
Sect.~\ref{sec:variation} discusses observed variations of measured cyclotron line energies 
(as summarized in Table \ref{tab:variation}): we find that the line energy can vary with pulse phase, 
with luminosity (both positive and negative), with time and (in Her~X-1) with phase of the super-orbital 
period. The width and depth of the cyclotron line(s) can also vary systematically with luminosity. 
It is also found that the spectral hardness of the continuum can vary with X-ray luminosity, in close correlation 
with variations of the cyclotron line energy. 
Sect.~\ref{sec:long_term} we review the evidence for variations of the cyclotron line energy on medium to 
long timescales.
In Sect.~\ref{sec:param_cor} the current knowledge about correlations between the various spectral 
parameters is discussed, and in 
Sect.~\ref{sec:individual} we discuss a few individual sources, mostly those that show systematic 
variations with X-ray luminosity.
In Sect.~\ref{sec:physics} we discuss the relevant physics at work in accreting binary X-ray pulsars: 
the formation of spectral continua and cyclotron lines and the physics behind their systematic variations. 
In Sect.~\ref{sec:modeling} we briefly refer to theoretical modeling of cyclotron features, by both analytical
and Monte Carlo methods.
In Sect.~\ref{sec:strength} three methods of how to estimate the magnetic field strength of neutron stars
are discussed and compared.
In Sect.~\ref{sec:statistics} we present 
a statistical analysis, describing the overall mean properties of the discussed objects. 
Finally, in Sect.~\ref{sec:proton_lines} we discuss objects thought to show
proton cyclotron lines and conclude in 
Sect.~\ref{sec:summary} with a short general summary. 

\section{Collection of cyclotron line sources}
\label{sec:source_list}

This review attempts to summarize the current knowledge of cyclotron line sources -- both electron- 
and proton-cyclotron line sources. For both types of cyclotron line sources we provide a series of tables 
with information about a total of more than $\sim$40 objects that we consider to show one (or more) cyclotron 
lines in their X-ray spectra. Tables~\ref{tab_M7} and \ref{tab_M7lines} list seven objects with proton-cyclotron lines.
For electron-cyclotron line objects we present a total of five tables with the following content: \\
- Table \ref{tab:collection}: 36 objects with confirmed or reasonably secure CRSFs (14 of them still need further confirmation).
   The columns are: source name, type of object, pulse period, orbital period, eclipse (yes/no), 
   cyclotron line energy (or energies), instrument of first detection, confirmations yes/no, references. \\
- Table \ref{tab:candidates}: 11 objects which we call ``candidates", for which electron-CRSFs 
   have been claimed, but where sufficient doubts about the reality of the cyclotron line(s) exist or where additional
   observations are needed for confirmation (or not). \\
- Table \ref{tab:companion}: HEASARC type, position, optical counterpart, its spectral type, masses, distance, references. \\
- Table \ref{tab:variation}: Variation of E$_{\rm cyc}$ with pulse phase and with L$_{\rm x}$, variation of spectral hardness 
   with L$_{\rm x}$, references. \\
- Table \ref{tab:width}: E$_{\rm cyc}$, width, ``strength',' and optical depth of cyclotron lines at certain L$_{\rm x}$, 
   references, notes. \\

\section{Spectral fitting}
\label{sec:spectral_fitting}

X-ray spectra of XRBPs are of thermal nature. They are formed in a hot plasma ($T\sim10^{8}$\,K) 
over the NS magnetic poles where infalling matter arrives with half the speed of light at the stellar 
surface. The emission process is believed to be governed by Comptonization of thermal photons 
which gain energy by scattering off hot plasma electrons (thermal Comptonization). In addition,
bulk motion Comptonization in the fast moving plasma above the deceleration region
and cyclotron emission plays an important role \citep[e.g.,][]{BeckerWolff_07,Ferrigno_etal13}. 
The shape of the spectral continuum formed by Comptonized photons emitted in the
breaking plasma at the polar caps of an accreting magnetized NS was first computed in
the seminal work of \citet{LyubarskiiSunyaev_82}. It has been shown that
non-saturated Comptonization leads to a power law-like continuum which cuts off 
at energies $\gtrsim$$kT_e$, where $T_e$ denotes the
electron temperature in the plasma. 

The calculations of  \citet{LyubarskiiSunyaev_82} and the numerical computations performed later provided 
a physical motivation for the usage of analytic power law functions with exponential cutoff to model the spectral 
continua of XRBPs. Historically, the following realizations of such functions became most popular and are included 
in some of the standard spectral fitting packages such as \textsl{XSPEC}\footnote{http://heasarc.nasa.gov/xanadu/xspec},
\emph{Sherpa}\footnote{http://cxc.harvard.edu/sherpa} and ISIS\footnote{http://space.mit.edu/asc/isis}. 
The most simple one, with just three free fit parameters, is the so-called \texttt{cutoffpl} model (these names are
the same in all mentioned fitting packages):
\begin{equation}
I_E = K\cdot E^{-\Gamma}\exp{(-E/E_\text{fold})},
\end{equation}
where $E$ is the photon energy and the free fit parameters $K$, $\Gamma$ and $E_\text{fold}$ 
determine the normalization coefficient, the photon index, and the exponential folding 
energy,\footnote{In \textsl{XSPEC} the folding energy for this function is called $E_\text{cut}$ }
respectively. We note that this represents a continuously steepening continuum.

The next function has an additional free parameter, the cutoff energy
$E_\text{cutoff}$:
\begin{equation}
I_E =
\begin{cases}
K\cdot E^{-\Gamma}, & \text{if\,} E \leq E_\text{cutoff} \\
K\cdot E^{-\Gamma} \exp{\left(-\frac{E-E_\text{cutoff}}{E_\text{fold}}\right)}, & \text{if\,}
E > E_\text{cutoff}.
\end{cases}
\end{equation}
In the spectral fitting packages mentioned here, this function is realized as a product of a power 
law and a multiplicative exponential factor: \texttt{power law}$\times$\texttt{highecut}. Although 
this function is generally more flexible due to one more free parameters, it contains a 
discontinuity of its first derivative (a ``break'') at $E=E_\text{cutoff}$. Since the observed X-ray
spectra are generally smooth, an absorption-line like feature appears in the fit residuals when 
using this model with high quality data. To eliminate this feature, one either includes an artificial narrow absorption line 
in the model \citep[e.g.,][]{Coburn_etal02} or substitutes the part of the function around the 
``break'' with a third order polynomial such that no discontinuity in the derivative appears 
\citep[e.g.,][]{Klochkov_etal08c}. Here the power law is unaffected until the cutoff energy is reached.

Another form of the power law-cutoff function is a power law with a Fermi-Dirac form of the cutoff 
\citep{Tanaka_86}:
\begin{equation}
I_E = K\cdot E^{-\Gamma}\left[1+
  \exp{\left(\frac{E-E_\text{cutoff}}{E_\text{fold}}\right)}\right]^{-1} 
\end{equation}
which has the same number of free parameters as the previous function. This function is not included 
any more in the current versions of the fitting packages.

\begin{figure}
   \includegraphics[width=0.4\textwidth]{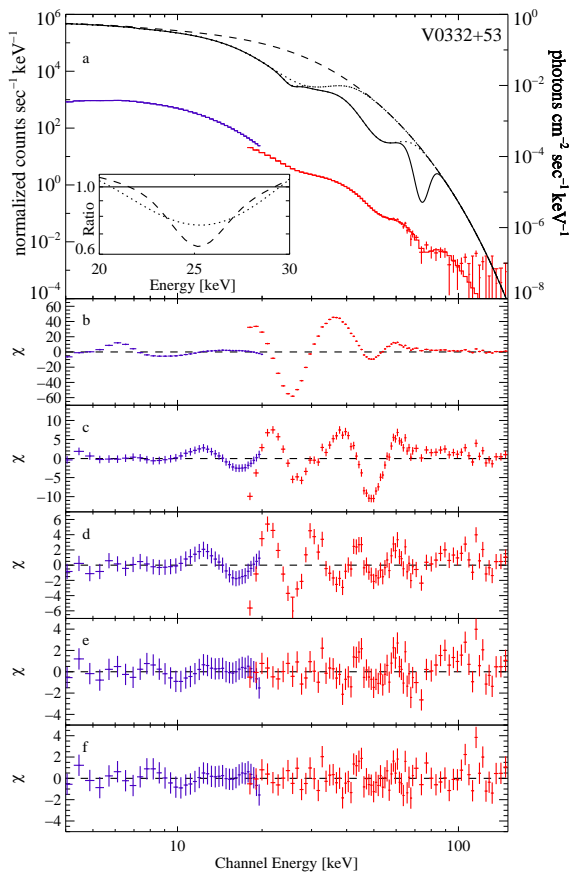}
   \caption{Example of a spectral modeling for continuum plus cyclotron lines. Reproduction of Fig.~2 of 
   \citet{Pottschmidt_etal05}, showing the phase averaged spectrum of V~0332+53 as observed with RXTE 
   for a Fermi-Dirac continuum and up to three cyclotron lines (modeled using the Gaussian optical depth 
   profile \texttt{gabs}). (a) Combined PCA/HEXTE spectrum (\textsl{croses}), 
   best-fit model (\textsl{histogram}), and unfolded spectrum (\textsl{dotted and solid lines, right y-axis label}),
   illustrating the cumulative effect of the CRSFs on the continuum; (b-f) Residuals for models taking increasing 
   numbers of CRSFs into account: (b) no line; (c) one line at 25.5\,keV; (d) two lines, at 25.6\,keV and at 50\,keV; 
   (e) two lines: the fundamental line modeled by two Gaussian components at 25.2\,keV and 26.6\,keV, and its 
   harmonic at 49.9\,keV; (f) three lines, the fundamental modeled by two components at $\sim$25\,keV and 
   $\sim$27\,keV and two harmonics at $\sim$51\,keV and at $\sim$74\,keV. For further explanation see 
   \citet{Pottschmidt_etal05}.
   }
   \label{fig:V0332_CRSFs}
\end{figure}

Finally, a sum of two power laws with a positive and a negative photon index multiplied by an 
exponential cutoff is used in the literature \citep[the NPEX model,][]{Mihara_95}:
\begin{equation}
I_E = K_1(E^{-\Gamma_1}+K_2E^{+\Gamma_2})\exp{(-E/E_\text{fold})}.
\end{equation}
\noindent In many applications using this model, $\Gamma_2$ is set to a value of two in order to 
represent the Wien portion of the thermal distribution.

In some spectra of XRBPs with high photon statistics, deviations from these simple phenomenological 
models are seen. In many cases, a wave-like feature in the fit residuals is present between a few and
$\sim$10--20\,keV. It is often referred to as the ``10-keV-feature''
\citep[e.g.,][]{Coburn_etal02}. 
The residuals can be flattened by including a broad emission or absorption line component. Although
sometimes interpreted as a separate physical component, the 10-keV-feature most probably reflects the 
limitations of the simple phenomenological models described above. 

To model the cyclotron line, one modifies the continuum functions described above by the inclusion of a 
corresponding multiplicative component. The following three functions are used in most cases to model 
CRSFs. The most popular one is a multiplicative factor of the form $e^{-\tau(E)}$, where the optical depth 
$\tau(E)$ has a Gaussian profile:
\begin{equation}
\tau(E) = \tau_0\exp\left[-\frac{(E-E_\text{cyc})^2}{2\sigma_\text{cyc}^2}\right],
\end{equation}
with $\tau_0$, $E_\text{cyc}$, and $\sigma_\text{cyc}$ 
being the central optical depth, the centroid energy, and the width of the line. 
We note that in the popular \textsl{XSPEC} realization of this function
\texttt{gabs}, $\tau_0$ is not explicitely used as a fit parameter. Instead, a product 
$\tau_0~\sigma_\text{cyc}\sqrt{2\pi}$ 
is defined that is often called the \textsl{``strength''} of the line. The physical motivation for the 
described line models stems from the formal solution of the transfer equation in the case of 
true absorption lines, for example, due to transitions between the atomic levels in stellar atmospheres. The Gaussian 
profile forms as a result of the thermal motion of atoms and reflects their Maxwellian distribution. Such a 
physical picture cannot be applied to the cyclotron features whose nature is completely different. 
The \texttt{gabs} function should thus be considered as a pure
phenomenological model in the case of CRSFs. \\ 

The second widely used phenomenological model has been specifically created to model cyclotron 
features. It is implemented in the  \textsl{XSPEC} \texttt{cyclabs} function which, similarly to \texttt{gabs}, 
provides a multiplicative factor of the form  $e^{-\tau(E)}$  for the continuum \citep{Mihara_90,Makishima_etal90a}.
In this model, the line profile is determined by a Lorentzian function multiplied by a factor $(E/E_\text{cyc})^2$ 
(not a pure Lorentzian, as it is often claimed in the literature). Eq.\ref{equ:cyclabs} shows the
formula used for $\tau(E)$ for the case of two lines, the fundamental at $E_\text{cyc}$ and 
the first harmonic at $2E_\text{cyc}$.
\begin{equation}
\label{equ:cyclabs}
  \tau(E) = \tau_1\frac{(W_1E/E_\text{cyc})^2}{(E-E_\text{cyc})^2+W_1^2}
  + \tau_2\frac{(W_2E/[2E_\text{cyc}])^2}{(E-2E_\text{cyc})^2+W_2^2},
\end{equation}
where $\tau_\text{1,2}$ are the central optical depths of the two lines while $W_\text{1,2}$ 
characterize their width.  One can fix $\tau_1$ or $\tau_2$ to zero, if necessary. When using 
\texttt{cyclabs}, however, one needs to be careful, since  the ratio of the true energy centroids
may not always be exactly equal to two.
Figure~\ref{fig:V0332_CRSFs} gives an example for a spectral fitting of the spectrum of V~0332+53 as 
observed with \textsl{RXTE} PCA and HEXTE \citep{Pottschmidt_etal05} for a Fermi-Dirac continuum
model and up to three cyclotron lines, modeled by Gaussian optical depth profiles \texttt{gabs}.

For a substantial fraction of confirmed CRSF sources listed in Table~\ref{tab:collection}, 
multiple cyclotron lines are reported. The harmonic lines correspond to resonant
scattering between the ground Landau level and higher excited levels (see Eq.\ref{equ:Bcyc}). 
The energies of the harmonic lines are thus expected to be multiples of the fundamental line 
energy. However, the broad and complex profiles of CRSFs, the influence of the spectral 
continuum shape on the measured line parameters, and various physical effects lead to deviations 
of the line centroid energies from the harmonic spacing \citep[see e.g.,][]{Nishimura_08,Schoenherr_etal07}. 
To model multiple cyclotron lines, it is best to introduce several independent absorption line models 
(as described above) into the spectral function, leaving the line centroid energies uncoupled. 

Soon after the discovery of the cyclotron line in Her X-1, a third possibility was occasionally used. 
The X-ray continuum was multiplied by a factor $[1-G(E)]$, where $G(E)$ is a Gaussian 
function with an amplitude between zero and one \citep[e.g.,][]{Voges_etal82}.

\begin{figure}
 \includegraphics[width=0.45\textwidth]{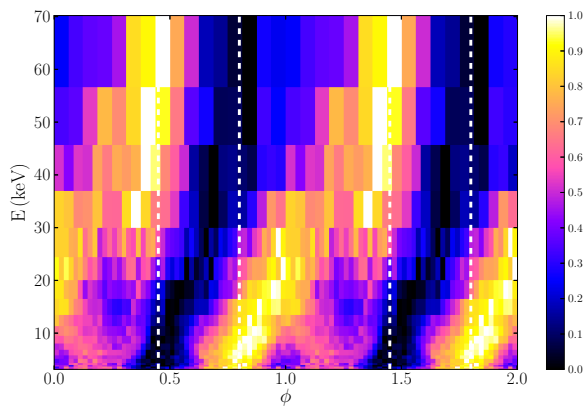}    
\hfill
\vspace{-2mm}
\caption{Example of a two-dimensional plot, showing color-coded flux for photon energy versus pulse phase, 
\citep[for GS~0834-430, Fig.~3 of][]{Miyasaka_etal13}. Horizontal cuts are pulse profiles for the selected energy 
range, vertical cuts are spectra for the selected phase interval.}
   \label{fig:2D_color}
\end{figure}

The usage of different spectral functions both for the continuum and for the cyclotron lines poses a natural 
problem when observational results are to be compared to those that were obtained using different fit functions. 
It is obvious that the width and the strength/depth of the line are defined in different ways in the models 
above, such that they cannot be directly compared with each other. Even more challenging is the determination 
of the centroid energy of the line. In case of a symmetric and sufficiently narrow absorption line, 
for which variations of the continuum over the energy range affected by the line can be neglected, one naturally 
expects that the models above would return very similar centroid energies. But, the cyclotron lines are (i) mostly 
quite broad and (ii) are overlaid on a highly inclined continuum which changes exponentially with energy. 
Furthermore, real lines can be noticeably asymmetric \citep[e.g.,][]{Fuerst_etal15}. The line centroids measured 
with different spectral functions are therefore systematically different. Specifically, our systematic analysis 
\footnote{Staubert et al. 2007, Poster at San Diego Conf. Dec. 2007, ``The Suzaku X-ray universe"} had shown that 
a fit with the \texttt{cyclabs} model or using the aforementioned multiplicative $[1-G(E)]$ factor result in a 
systematically lower (typically, by a few keV) centroid energies $E_\text{cyc}$ compared to the \texttt{gabs} model
(see also \citealt{Lutovinov_etal17b}). For the cyclotron feature in Her X-1, \citet{Staubert_etal14} had therefore
added 2.8\,keV to the \textsl{Suzaku} values published by \citet{Enoto_etal08}. 

Starting with the work by \citet{WangFrank_81} and \citet{LyubarskiiSunyaev_82}, attempts have been 
made to fit the observed spectra with physical models of the polar emitting region in accreting pulsars by computing 
numerically the properties of the emerging radiation \citep[e.g., also,][]{BeckerWolff_07,Farinelli_etal12,
Farinelli_etal16,Postnov_etal15}. 
Even though the use of heuristic mathematical functions has been quite successful in describing the observed spectral 
shapes, the resulting fit parameters generally do not have a unique physical meaning. Achieving exactly this is the goal 
of physical models. A few such physical spectral models are publicly available for
fitting observational data through implementations in \textsl{XSPEC}, for example by \citet{BeckerWolff_07}
(\texttt{BW}\footnote{http://www.isdc.unige.ch/\~ferrigno/images/Documents/\\BW\_distribution/BW\_cookbook.html}),
\citet{Wolff_etal16} (\texttt{BWsim})
or by \citet{Farinelli_etal16} (\texttt{compmag}\footnote{https://heasarc.gsfc.nasa.gov/xanadu/xspec/manual/
\\XSmodelCompmag.html})
The number of free parameters in these models is, however, relatively large such that some of them 
need to be fixed or constrained a priori to obtain a meaningful fit \citep[see, e.g.,][for details]{Ferrigno_etal09,Wolff_etal16}. 
A number of Comptonization spectral models are available in \textsl{XSPEC} which are used in the 
literature to fit spectra of accreting pulsars and of other astrophysical objects whose spectrum is shaped by
Comptonization: \texttt{compbb,compls,compps,compst,comptb,comptt}, and including CRSFs,
\texttt{cyclo} (see the \textsl{XSPEC} manual webpage for 
details\footnote{https://heasarc.gsfc.nasa.gov/xanadu/xspec/manual/Additive.html}). 
These models are calculated for a set of relatively simple geometries and do not (except \texttt{cyclo})
take into account magnetic field and other features characteristic for accreting pulsars. The best-fit parameters 
obtained from spectral fitting should therefore be interpreted with caution. Otherwise, with their relatively low 
number of input parameters and (mostly) short computing time, these models provide a viable  alternative to the 
phenomenological models described earlier.

In Sect.~\ref{sec:param_cor} we discuss physical correlations between various spectral parameters.
Since mathematical inter-dependencies between fit parameters are unavoidable in multiparameter fits,
it is worth noting that fitted values of the centroid cyclotron line energy  $E_\mathrm{cyc}$, the focus of this
contribution, appear to be largely insensitive to the choice of different continuum functions, as was shown in the 
two systematic studies of ten (respective nine) X-ray binary pulsars showing CRSFs using observational data of 
\textsl{RXTE} \citep{Coburn_etal02} and \textsl{Beppo}/SAX \citep{DoroR_17}.

\begin{figure}
 \includegraphics[angle=90,width=0.6\textwidth]{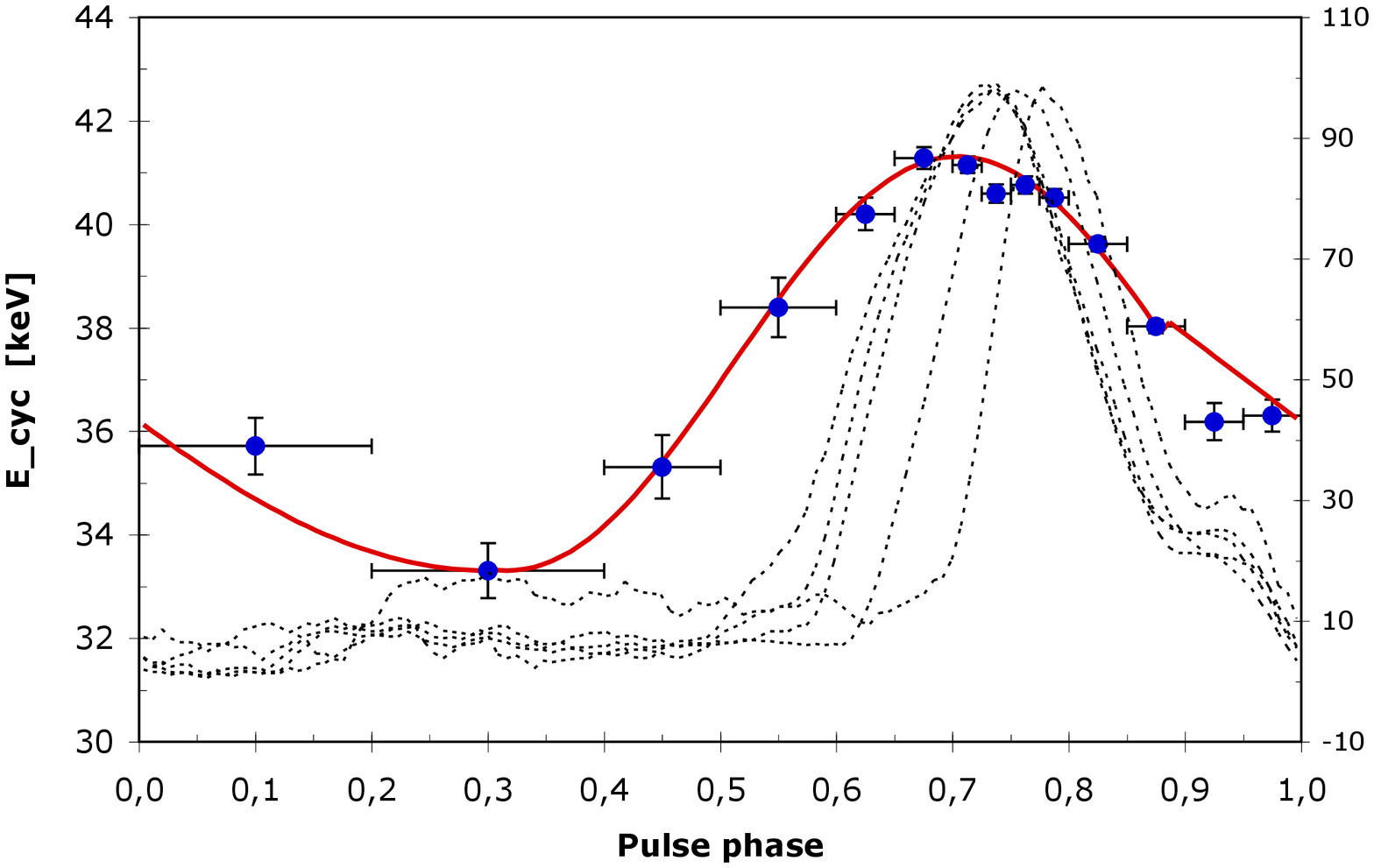}    
\hfill
\vspace{-11mm}
\caption{Dependence of cyclotron line energy on pulse phase for Her~X-1 together with pulse
profiles for different phases of the 35d modulation (see Fig. 2 of \citealt{Staubert_etal14}). 
The right-hand scale is normalized flux (0-100) for the pulse profiles.}
   \label{fig:pulse_phase}
\end{figure}

\section{Observed variations in cyclotron line energy}
\label{sec:variation}

The cyclotron line energy has been found to vary with pulse phase (in almost all objects), with luminosity 
(both positive and negative, in nine objects so far), with time (so far in two objects) 
and with phase of the super-orbital period (so far only in Her~X-1). In addition, the width and depth of the 
cyclotron line(s) can also systematically vary with luminosity. The spectral hardness of the continuum can 
vary with X-ray luminosity, in close correlation with variations of the cyclotron line energy.

\subsection{Cyclotron line parameters as function of pulse phase} 
\label{sec:pulse_phase}

Throughout a full rotation of the neutron star we see the accretion mound or column under different angles.
We therefore expect to observe significant changes in flux, in the shape of the continuum and 
in the CRSF parameters as function of pulse phase. The variation of flux as function of pulse phase is called
the pulse profile. Each accreting pulsar has its own characteristic pulse profile which is often highly energy dependent.
With increasing energy the pulse profiles tend to become smoother (less structured) and the pulsed fraction
(the fraction of photons contributing to the modulated part of the flux) increases \citep{Nagase_89,Bildsten_etal97}.
Furthermore, those profiles can vary from observation to observation, due to changes in the physical conditions
of the accretion process, for example varying accretion rates. One way of presenting energy dependent pulse profiles
is a two-dimensional plot: energy bins versus phase bins, with the flux coded by colors (see Fig.~\ref{fig:2D_color}).
In this plot horizontal lines represent pulse profiles (for selected energies) and columns
represent spectra (for selected pulse phases). The centroid energy of cyclotron lines are generally also pulse phase 
dependent. The range of variability is from a few percent to 40\% (see Table~\ref{tab:variation}).
As an example, Fig.~\ref{fig:pulse_phase} shows the variation of E$_{\rm cyc}$ for Her~X-1 which is on the order 
of 25\% (see Fig.~2 of \citealt{Staubert_etal14}).

\begin{figure}[h!]
    \includegraphics[width=0.45\textwidth]{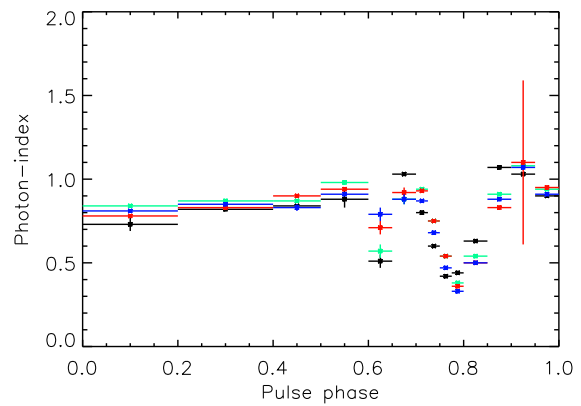}
   \caption{Her~X-1: the photon index $\Gamma$ as function of pulse phase for the four different 35\,d phase
   intervals:  interval 1 in black  (0.007-0.056), interval 2 in green (0.09-0.112), 
   interval 3 in red (0.137-0.161), and interval 4 in blue (0.186-0.217) 
   (see Fig.~7 and Table~1 of \citet{Vasco_etal13}).
   }
   \label{compar_g}
\end{figure}

In a simple picture, the changes of the CRSF energy are a result of sampling different heights of the line forming region 
as a function of pulse phase. Modeling these variations can help to constrain the geometry (under the assumption of a 
dipole magnetic field) of the accretion column with respect to the rotational axis as well as the inclination under which 
the system is viewed \citep[see, e.g.,][]{Suchy_etal12}. However, for a more physical constraint on those variations, the 
general relativistic effects such as light-bending around the neutron star have to be taken into account. These effects 
result in a large number of degrees of freedom, making it difficult to find a unique solution of the accretion geometry
\citep{Falkner_etal18}.

In most sources, the strength of the CRSF depends strongly on pulse phase. In particular, the fundamental line is 
sometimes only seen at certain pulse phases, for example, Vela~X-1 \citep{Kreykenbohm_etal02} or KS~1947+300 
\citep{Fuerst_etal14a}. This behavior could indicate that the contributions of the two accretion columns vary
and/or that the emission pattern during large parts of the pulse is such
that the CRSF is very shallow or filled by spawned photons \citep{Schwarm_etal17a}. The latter 
model agrees with the fact that the harmonic line typically shows less depth variability with phase.

Continuum parameters are also known to change as function of pulse phase, but not necessarily in step with the 
observed CRSF variation (see Fig.~\ref{compar_g}). These changes can occur on all timescales, often varying 
smoothly as function of pulse phase 
\citep[e.g.,][for a summary see Table~\ref{tab:variation}]{Suchy_etal08, MaitraPaul_13b,JaisawalNaik_16}. 
However, also sharp features are sometimes observed, where the spectrum is changing dramatically over only a 
few percent of the rotational period of the NS. The most extreme example is probably EXO\,2030+375, where the 
photon-index changes by $\Delta\Gamma > 1.2$ and the absorption column also varies by over an order of magnitude 
\citep{Ferrigno_etal16b,Fuerst_etal17}. This effect is interpreted as an accretion curtain moving through our line of 
sight, indicating a unique accretion geometry in EXO\,2030+375.

Another peculiar effect has been observed in the pulse profiles of some X-ray pulsars: the pulse profile shifts in 
phase at the energy of the CRSFs \citep[e.g., in 4U~0115+63;][]{Ferrigno_etal11}. This can be explained by the 
different cross sections at the CRSF energies, leading to changes in emission pattern, as calculated by 
\citet{Schoenherr_etal14}. \\

\begin{figure}[h!]
    \includegraphics[width=0.47\textwidth]{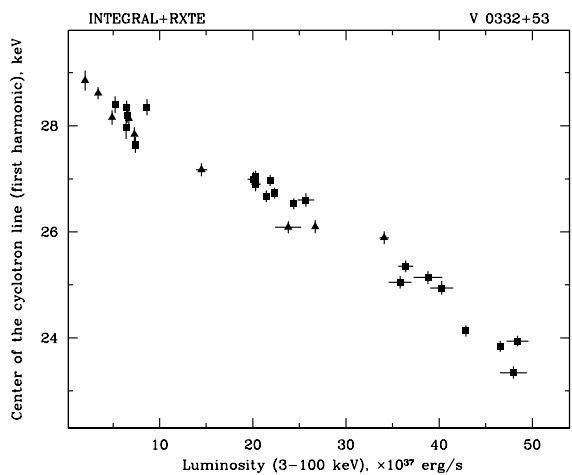}
   \caption{Negative $E_\mathrm{cyc}$/L$_{\rm x}$ correlation as observed in V~0332+57
   by \textsl{ INTEGRAL} and \textsl{RXTE} during an outburst in 2004/2005 
   (Fig.~4 of \citealt{Tsygankov_etal06}).
   }
   \label{V0332_neg}
\end{figure}

\subsection{Cyclotron line parameters as function of luminosity}
\label{sec:luminosity}

The spectral properties of several cyclotron line sources show a dependence on X-ray luminosity ($L_\mathrm{x}$).
In particular, the cyclotron line energy $E_\mathrm{cyc}$ was found to vary with $L_\mathrm{x}$ in a systematic 
way, in correlation with the spectral hardness  of the underlying continuum.  Also the line width and depth of the 
cyclotron line(s) can vary. The first detection of a negative dependence of $E_\mathrm{cyc}$ on 
L$_{\rm x}$ was claimed by \citet{Makishima_etal90} and \citet{Mihara_95} in high luminosity outbursts of 
three transient sources: 4U~0115+63, V~0332+53 and Cep~X-4, observed by \textsl{Ginga}. 
``Negative'' dependence means that  $E_\mathrm{cyc}$ decreases with increasing L$_{\rm x}$. 
Fig.~\ref{V0332_neg} shows the clear and strong negative correlation in V~0332+53 
from observations by \textsl{ INTEGRAL} and \textsl{RXTE} (Fig.~4 of \citealt{Tsygankov_etal06}).
However, of the three sources' dependencies originally quoted, only one, V~0332+53, can today be 
considered a secure result \citep{Staubert_etal16}: in Cep~X-4 the effect was not confirmed (the source 
is instead now considered to belong to the group of objects with a ``positive'' dependence \citealt{Vybornov_etal17}).
In 4U~0115+63, the reported negative (or anticorrelated) dependencies  
\citep{Tsygankov_etal06,Nakajima_etal06,Tsygankov_etal07,Klochkov_etal11},  have been shown to be
most likely an artifact introduced by the way the continuum was modeled (\citealt{SMueller_etal13},
see also \citealt{Iyer_etal15}). More recently, a second source with a negative E$_{\rm cyc}$/L$_{\rm x}$ correlation 
was found: SMC~X-2 (see fig~\ref{fig:Ecyc-Lx}).

A simple idea about the physical reason for a negative correlation was advanced by \citet{Burnard_etal91}.
Based on \citet{BaskoSunyaev_76}, who had shown that the height of the radiative shock above the neutron 
star surface (and with it the line forming region) should grow linearly with increasing accretion rate. They noted 
that this means a reduction in field strength and therefore a reduction of $E_\mathrm{cyc}$. 
Thus, with changing mass accretion rate, the strength of the (assumed dipole) magnetic field and 
$E_\mathrm{cyc}$ should vary according to the law 
\begin{equation}
\label{eqn:dipole}
E_{cyc}(\dot M) = E_0\left(\frac{R_{NS}}{H_l(\dot M) + R_{NS}}\right)^3
\end{equation}
where $E_0$ corresponds to the line emitted from the neutron star surface magnetic field 
$B_\mathrm{B_s}$, $R_{NS}$ is the radius of the NS, and $H_ l$ is the actual height of the line 
forming region above the NS surface.
Regarding the physics of the scaling of the height of the line forming region see Sect.~\ref{sub:shocks}.

\begin{figure}
\includegraphics[width=0.42\textwidth]{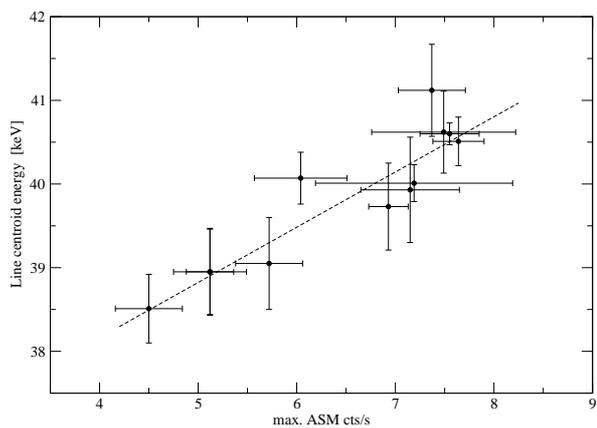}
\caption{The originally discovered E$_{\rm cyc}$ / L$_{\rm x}$ correlation in Her~X-1 (reproduced from \citet{Staubert_etal07}).}
\label{fig:pos_cor_ori}
\end{figure}

\begin{figure}
\includegraphics[width=0.4\textwidth]{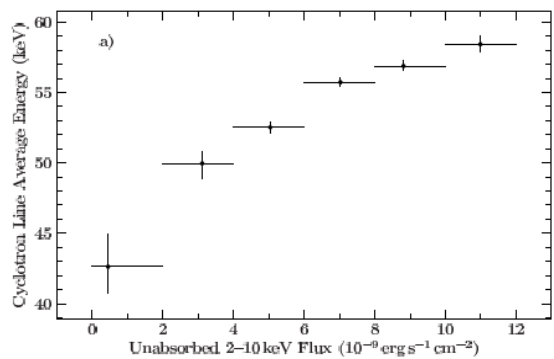}
\caption{E$_{\rm cyc}$ / L$_{\rm x}$ correlation observed in GX~304$-$1
(reproduced from \citealt{Rothschild_etal17}).}
\label{fig:GX304_cor}
\end{figure}

\begin{figure*}
\includegraphics[width=0.9\textwidth]{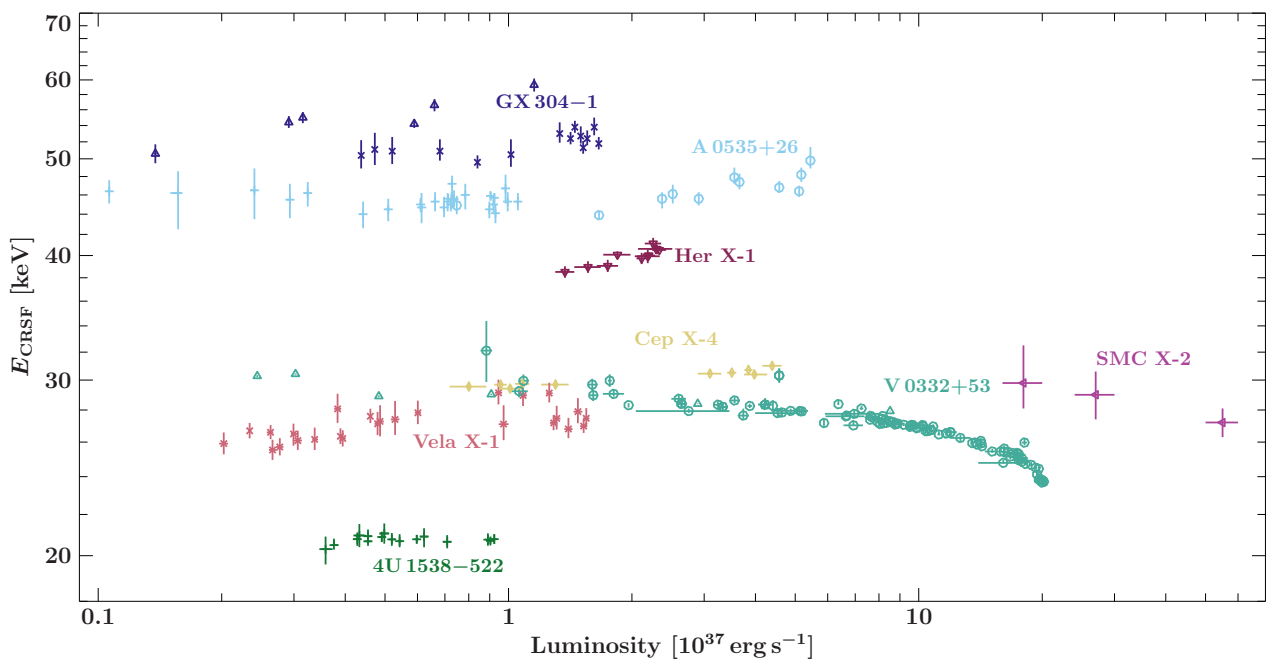}       
\hfill
\caption{Compilation of correlations between E$_{\rm cyc}$ and L$_{\rm x}$. Data are taken from the following 
sources (the numbers in parentheses are the numbers of the references as given in Table~\ref{tab:references}: 
SMC~X-2 (91); 4U~1538$-$522 (81); GX\,304$-$1 (20, 24, 54); Cep~X-4 (48, 130); Vela~X-1 (32);  A\,0535+26 
(73, 74); V\,0332+53 (99, 123); Her~X-1 (25). The luminosities are calculated using the distances measured by 
Gaia, as given in parentheses in Table~\ref{tab:companion}. }
   \label{fig:Ecyc-Lx}
\end{figure*}

The first positive correlation (E$_{\rm cyc}$ increases with increasing L$_{\rm x}$) was finally discovered in 
2007 by \citet{Staubert_etal07} in Her~X-1, a persistent medium luminosity source (Fig.~\ref{fig:pos_cor_ori}). 
Since then six (possibly seven) more sources (all at moderate to low luminosities) have been found with 
E$_{\rm cyc}$ increasing with increasing luminosity (see Table~\ref{tab:variation}): Vela~X-1 
\citep{Fuerst_etal14b, LaParola_etal16}, A~0535+26 \citep{Klochkov_etal11,Sartore_etal15}, 
GX~304-1 \citep{Yamamoto_etal11,Malacaria_etal15,Rothschild_etal17}, Cep~X-4 \citep{Vybornov_etal17},
4U~1626.6-5156 \citep{DeCesar_etal13}, and
V~0332+53 (the only source with a strong negative E$_{\rm cyc}$ / L$_{\rm x}$ correlation at high luminosities, 
see above) has been confirmed to switch to a positive correlation at very low flux levels at the end of an outburst 
\citep{Doroshenko_etal17,Vybornov_etal18}.
A possible positive correlation in 4U~1907+09 \citep{Hemphill_etal13} needs to be confirmed.

An interesting deviation from a purely linear correlation (only detectable when the dynamical range in 
L$_{\rm x}$ is sufficiently large) has recently been noticed in GX~304-1 and Cep~X-4, namely a 
flattening toward higher luminosity \citep{Rothschild_etal17,Vybornov_etal17} 
(see e.g., Fig.~\ref{fig:GX304_cor}). This behavior is  theoretically well explained by the model of
a collisionless shock (see Sect.~\ref{sec:physics}).

There are indications that in some objects the line width (and possibly the depth) also correlate with the luminosity 
(see Table~\ref{tab:width}) positively (at least up to a few times $10^{37}$\,erg/s). This is not surprising
since the line energy itself correlates with luminosity and the line width and depth generally correlate with 
the line energy (see Sect.~\ref{sec:param_cor}). 
Measured values for the width and the depth of cyclotron lines are compiled in 
Table~\ref{tab:width}. However, the information on line width and depth are overall incomplete and 
rather scattered. 

In the discovery paper of the positive E$_{\rm cyc}$/L$_{\rm x}$ correlation in Her~X-1, \citet{Staubert_etal07}
had proposed, that for low to medium luminosity sources the physics of deceleration of the accreted matter
close to the surface of the NS is very different from the case of the (transient) high luminosity sources: there is no
radiation dominated shock that breaks the in-falling material (moving to larger heights when the accretion rate 
increases). Rather, the deceleration of the material is achieved through Coulomb interactions that produce the
opposite behavior: for increasing accretion rate, the line forming region is pushed further down toward the NS 
surface (that is to regions with a stronger B-field), leading to an increase in E$_{\rm cyc}$. This picture is 
supported by the theoretical work by \citet{Becker_etal12}, that demonstrated that there are two regimes of
high and low/medium accretion rates (luminosities), separated by a ``critical luminosity'' L$_{\rm crit}$ on the 
order of $\sim10^{37}$\,erg/s. Fig.~\ref{fig:Ecyc-Lx} is an updated version of Fig.~2 in \citet{Becker_etal12},
showing our current knowledge of those sources that appear to show E$_{\rm cyc}$/L$_{\rm x}$ correlations. 
For further details of the physics, which is clearly more complicated than the simple picture above, see
Sect.~\ref{sec:physics} and the remarks about individual sources in Sect.~\ref{sec:individual}.

For completeness, we mention alternative ideas regarding the luminosity dependence of E$_{\rm cyc}$:
for example, \citet{Poutanen_etal13}, who -- for high luminosity sources -- follow the idea of an increasing height of the radiation 
dominated shock, but produce the CRSFs in the radiation reflected from larger and variable areas of the NS surface.
We note, however, that it still needs to be shown that cyclotron absorption features can indeed be produced by reflection,
and the model does not work for the positive correlation, which is -- by far -- the more frequently observed correlation. 
\citet{Mushtukov_etal15b}, on the other hand, suggest that the varying cyclotron line energy is produced by 
Doppler-shift due to the radiating plasma, the movement of which depends on the accretion rate.
A further extensive effort to model the luminosity dependence of E$_{\rm cyc}$ on luminosity by analytical calculations 
is from \citet{Nishimura_08,Nishimura_11,Nishimura_13,Nishimura_14}. By combining changes of the height and the area 
of the accretion mound with changes in the emission profile, he claims to be able to explain both -- negative and the positive --
E$_{\rm cyc}$/L$_{\rm x}$ correlations.

\subsection{E$_{\rm cyc}$/L$_{\rm x}$ correlations on different timescales and line-continuum correlation}
\label{sec:pulse_amplitude}

The original discoveries of the E$_{\rm cyc}$/L$_{\rm x}$ correlations were based on observations with variation
of L$_{\rm x}$ on long timescales: days to months for the negative correlation (e.g., outburst of V~0332+53,
\citealt{Tsygankov_etal06}) and months to years for the first positive correlation in Her~X-1 \citep{Staubert_etal07}.
It was first shown by \citet{Klochkov_etal11} that the same correlations are also observed on short timescales, that is on 
timescales of individual pulses: a new technique of analysis was introduced the so-called pulse-to-pulse
or \textsl{pulse-amplitude-resolved} analysis. Most accreting pulsars exhibit strong variations in pulse flux (or amplitude),
due to variations in the accretion rate (or luminosity) on timescales at or below the duration of single pulses.
Any variations of the rate of capture of material at the magnetospheric boundary to the accretion disk are instantaneously
mirrored in the release of gravitational energy close to the surface of the neutron star (the free fall timescale is
on the order of milliseconds). So, in selecting pulses of similar amplitude and generating spectra of all photons in
those different groups, one can study the luminosity dependence of spectral parameters.  \citet{Klochkov_etal11} analyse
observational data from \textsl{INTEGRAL} and \textsl{RXTE} of the two high-luminosity transient sources V~0332+53
and 4U~0115+63 as well as the medium to low luminosity sources Her~X-1 and A~0535+26. The significance of this 
work is twofold. First, the luminosity dependence of the cyclotron line energy as known from the earlier work (dealing with
long-term variations) is reproduced: the correlation is negative for the high luminosity sources and positive for the
medium to low luminosity sources. Second, the spectral index of the power law continuum varies with luminosity in the 
same way as the cyclotron line energy: --$\Gamma$ correlates negatively (the spectra become softer) for the high 
luminosity sources and positively (the spectra become harder, --$\Gamma$ get less negative) for the medium to low 
luminosity sources.\footnote{In contrast to earlier wording (e.g., \citealt{Klochkov_etal11}), we prefer to say: 
E$_{\rm cyc}$ and --$\Gamma$ vary in the ``same way''.} 
A supporting result with respect to the continuum was found by \citet{Postnov_etal15} who used 
hardness ratios F(5-12\,keV)/F(1.3-3\,keV): in all sources analyzed the continuum becomes harder for increasing
luminosity up to L$_{\rm x}$ of $\sim$(5-6)~$10^{37}$\,erg/s. For three objects where data beyond this luminosity were
available (EXO~2030+375, 4U~0115+63, and V~0332+53), the correlation changed sign (here we may see the
transition from subcritical to supercritical accretion -- see Sect.~\ref{sub:shocks}). \\

In Table \ref{tab:variation} we collate information about variations of E$_{\rm cyc}$ with pulse phase
and with luminosity L$_{\rm x}$ and about changes of $\Gamma$ (or spectral hardness) with L$_{\rm x}$.
Those sources that show a systematic E$_{\rm cyc}$/L$_{\rm x}$ correlation are discussed individually below. 
It is interesting to note that the sources with a positive correlation greatly outnumber those with the (first detected) 
negative correlation. We suggest that the positive correlation is a common property of low to medium luminosity 
accreting binary pulsars. It is also worth noting that the luminosity dependence of E$_{\rm cyc}$ is found on 
different timescales, ranging from years over days to seconds (the typical time frame of one pulse). 
Further details about the width and the depth of the cyclotron lines are compiled in Table~\ref{tab:width}.
These parameters are also variable, but the information on this is, unfortunately, still rather scattered.

\subsection{Long-term variations of the cyclotron line energy}
\label{sec:long_term}

So far, only two sources show a clear variability of the CRSF centroid energy over long
timescales (tens of years): Her~X-1 and Vela~X-1. We discuss both sources below.
For completeness, we mention 4U~1538-522 as a possible candidate for a long-term increase \citep{Hemphill_etal16}.
Only further monitoring of this source over several years will tell whether the suspicion is justified.
A peculiar variation on medium timescales (100\,d) was observed in V~0332+53 during an
outburst \citep{Cusumano_etal16,Doroshenko_etal17,Vybornov_etal18} and from one outburst to the next 
($\sim$400\,d) \citep{Vybornov_etal18}. During the outburst of June-September 2015 the source showed the well 
known anticorrelation of E$_{\rm cyc}$ with luminosity during the rise and the decay of the burst, but at the end 
of the burst the CRSF energy did not come back to its initial value (as was observed several times before). 
The data are consistent with a linear decay of E$_{\rm cyc}$ over the 100\,d outburst by $\sim$5\%
\citep{Cusumano_etal16}. But at the next outburst, 400\,d later, E$_{\rm cyc}$ had in fact resumed its original 
value of $\sim$30\,keV. The physics of this phenomenon is unclear. 
 
\begin{figure}
\includegraphics[angle=90,width=0.55\textwidth]{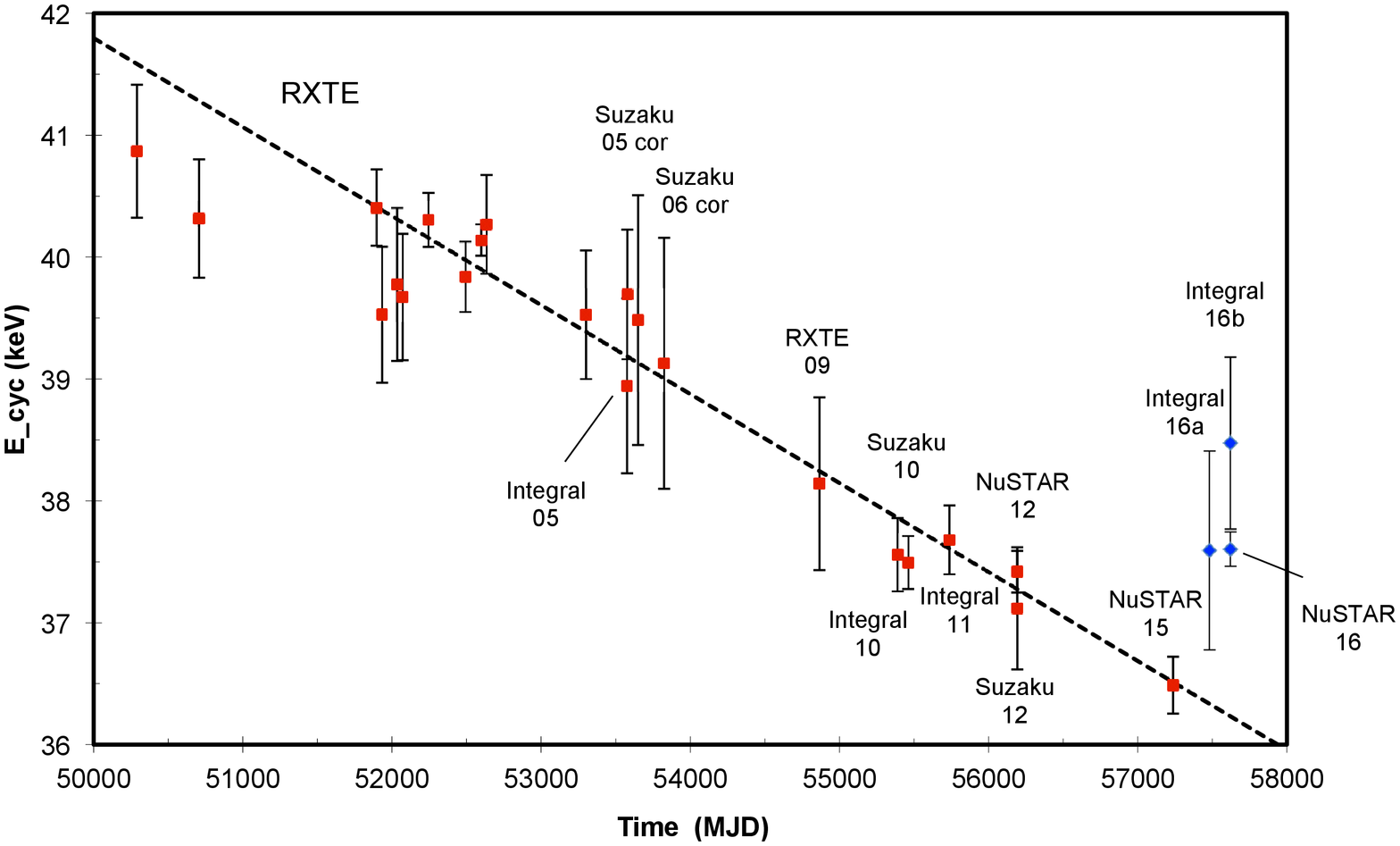}
\caption{Cyclotron line energy E$_{\rm cyc}$ in Her~X-1. The long-term decay stopped in 2016
\citep[see Fig.\,1 in][]{Staubert_etal17}.}
\label{fig:decay_stop}
\end{figure}

The first and best documented case for a long-term variation of the CRSF energy is Her~X-1 
\citep{Staubert_etal07,Staubert_etal14,Staubert_etal16}. At the time of 
discovery in 1976 the cyclotron line energy was around 37\,keV (interpreted as absorption line).
During the following $\sim$14 years several instruments (including the enlarged \textsl{Balloon-HEXE}, 
\textsl{HEAO}-A4, \textsl{Mir-HEXE} and \textsl{Ginga}) measured values between 33--37\,keV 
(with uncertainties which allowed this energy to be considered well established and constant) 
\citep{Gruber_etal01,Staubert_etal14}. After a few years of no coverage, a surprisingly high value ($\sim$44\,keV)
from observations with the \textsl{GRO}/BATSE instrument in 1993--1995 was announced \citep{Freeman_96}.
Greeted with strong doubts at the time, values around 41\,keV were subsequently measured by \textsl{RXTE} 
\citep{Gruber_etal99} and \textsl{BeppoSAX} \citep{DalFiume_etal98}, confirming that the cyclotron line energy
had indeed increased substantially between 1990 to 1993. Further observations then yielded hints for a possible slow 
decay. While trying to consolidate the evidence for such a decay, using a uniform set of \textsl{RXTE} observations, 
\citet{Staubert_etal07} instead found a new effect: a positive correlation between the pulse phase averaged E$_{\rm cyc}$ 
and the X-ray flux (or luminosity) of the source (Sect.~\ref{sec:luminosity}), letting the apparent decrease largely appear as 
an artifact. However, with the inclusion of further data and a refined method of analysis, namely fitting E$_{\rm cyc}$ values 
with two variables (flux and time) simultaneously, it was possible to establish statistically significant evidence of a 
true long-term decay of the phase averaged cyclotron line energy by $\sim$5\,keV over 20 years 
\citep{Staubert_etal14,Staubert_14,Staubert_etal16}. Both dependencies -- on flux and on time -- seem to be always 
present. The decay of E$_{\rm cyc}$ was independently confirmed by the analysis of \textsl{Swift}/BAT data
\citep{Klochkov_etal15}. Further measurements, however, yield evidence that the decay of E$_{\rm cyc}$
may have stopped with a hint to an inversion \citep{Staubert_etal17} (see Fig.~\ref{fig:decay_stop}). Since then, 
an intensified effort has been underway to monitor the cyclotron line energy with all instruments currently available.
The analysis of the very latest observations between August 2017 and September 2018 seems to confirm this trend.

The physics of the long-term variation of E$_{\rm cyc}$ is not understood. We do, however, believe that this is not
a sign of a change in the global (dipole) field of the NS, but rather a phenomenon associated with the field configuration
localized to the region where the CRSF is produced. Apparently, the magnetic field strength at the place of the resonant 
scattering of photons trying to escape from the accretion mound surface must have changed with time. A few thoughts 
about how such changes could occur are advanced by \citet{Staubert_etal14, Staubert_etal16} and \citet{Staubert_14}. 
Putting internal NS physics aside, changes could be connected to either a geometric displacement of the cyclotron 
resonant scattering region in the dipole field or to a true physical change in the magnetic field configuration at the polar cap. 
The latter could be introduced by continued accretion with a slight imbalance between the rate of accretion and the rate of 
``losing'' material at the bottom of the accretion mound (either by incorporation into the neutron star crust or leaking of 
material to larger areas of the NS surface), leading to a change of the mass loading and consequently of the 
structure of the accretion mound: height or B-field configuration \citep[e.g., ``ballooning",][]{Mukherjee_etal13b,Mukherjee_etal14}.
It is also interesting to note that the measured E$_{\rm cyc}$ corresponds to a magnetic field strength 
($\sim3.8\times10^{12}$\,Gauss) 
which is a factor of two larger than the polar surface field strength estimated from applying accretion torque models 
(see Sect.~\ref{sec:strength}). This discrepancy could mean, that the B-field measured by E$_{\rm cyc}$ is for a local
quadrupole field that is stronger and might be vulnerable to changes on short timescales (but see Sect.~\ref{sec:strength}).
Around 2015 the cyclotron line energy in Her~X-1 seems to have reached a bottom value, comparable
to the value of its original discovery ($\sim$37\,keV), with a possible slight increase \citep{Staubert_etal17},
leading to the question of whether we could -- at some time -- expect another jump upwards (as seen between 1990 and 1993). 

Besides Her~X-1, we know one more source with a long-term (over 11\,yr) decay of the CRSF energy: Vela~X-1
\citep{LaParola_etal16}.  Vela~X-1 is also one of seven sources showing a positive 
E$_{\rm cyc}$/L$_{\rm x}$ dependence, originally discovered by \citet{Fuerst_etal14b} and confirmed by 
\citet{LaParola_etal16}. The same physics is probably at work in both sources.
Most likely, more sources with this phenomenon will be found as monitoring, for example, with \textsl{Swift}/BAT, continues.

\subsection{Correlations between E$_{\rm cyc}$ and other spectral parameters}
\label{sec:param_cor}

Before entering discussion about the physics associated with the generation of X-ray
spectra in general and cyclotron lines in particular, we briefly  mention an interesting observational 
phenomenon: there are several correlations between spectral parameters of the continuum and those
of the CRSFs. Historically, the first such correlation is that between the cutoff energy E$_{\rm cutoff}$
(see Eqs. (3) and (4)) and the CRSF centroid energy E$_{\rm cyc}$, first noted by \citet{MakishimaMihara_92}
(also \citet{Makishima_etal99}, who realized that the relationship is probably not linear, but closer to 
E$_{\rm cutoff}$ $\propto$ E$_{\rm cyc}^{0.7}$. Then a clear (linear) relationship was found between the width
of the CRSF $\sigma_{\rm cyc}$ and the centroid energy E$_{\rm cyc}$ \citep{Heindl_etal00, DalFiume_etal00}.
It followed a systematic study in 2002 of all accreting pulsars showing CRSFs that were observed by 
\textsl{RXTE} \citep{Coburn_etal02}. Here, the above mentioned correlations were confirmed and
one additional identified: the relative line width $\sigma_{\rm cyc}$/E$_{\rm cyc}$ versus the optical depth 
$\tau$ of the line. As a consequence, almost every spectral parameter is -- to some degree --
correlated with all the others \citep[see the correlation matrices, Tables 8 and 9 in][]{Coburn_etal02}.
Through Monte Carlo simulations, \citet{Coburn_etal02} have shown that the correlations between the
parameters are not an artifact (e.g., due to a mathematical coupling in the fitting process), but are actually
of physical nature. Even though some ideas about the physical meaning of the observed correlations had been 
put forward, a rigorous check of the viability of those ideas is still needed.

\begin{figure}
\includegraphics[angle=90,width=0.58\textwidth]{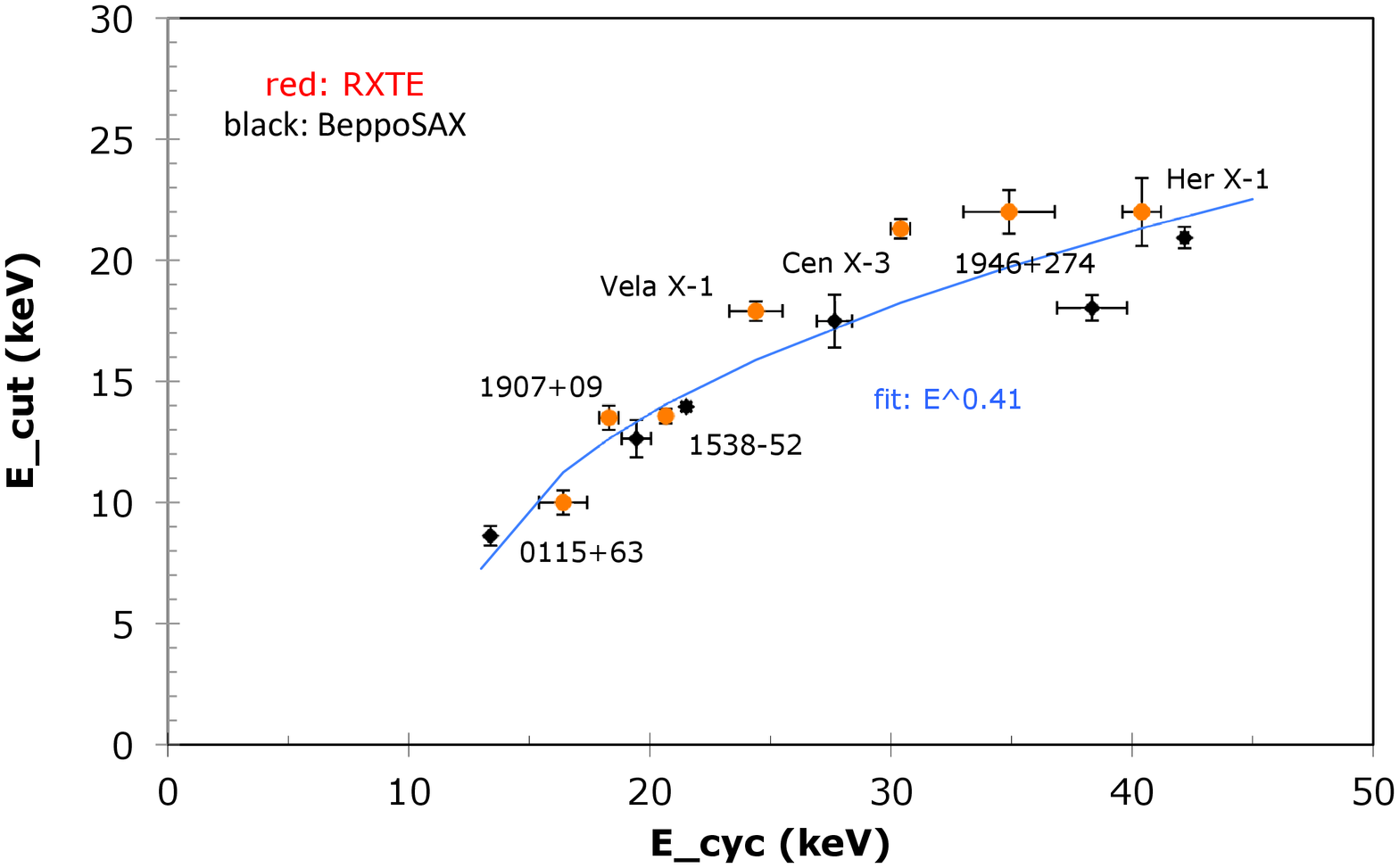}
\caption{Cutoff energy E$_{\rm cut}$ versus cyclotron line energy
E$_{\rm cyc}$ for objects studied by \textsl{RXTE} \citep{Coburn_etal02}
and \textsl{Beppo}/SAX \citep{DoroR_17}. The cutoff energy is roughly
proportional to the square root of E$_{\rm cyc}$. The two objects 4U~1744-28 and  4U~1626-67 are extreme outliers to this relationship, and not included in the fit.}
\label{fig:Ecut_E}
\end{figure}

\begin{figure}
\includegraphics[angle=90,width=0.56\textwidth]{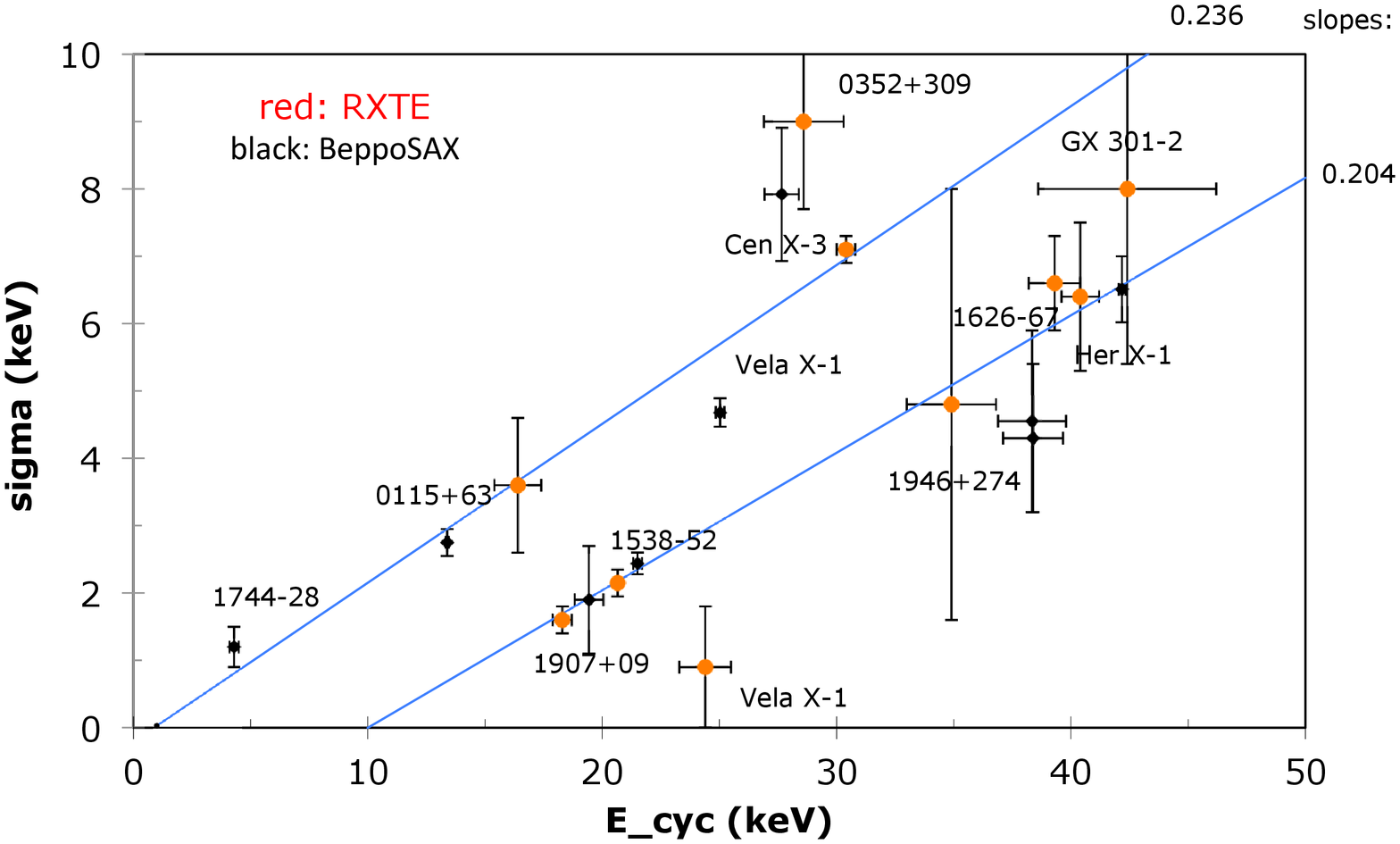}
\caption{Width $\sigma$ of the CRSFs versus cyclotron line energy
E$_{\rm cyc}$ for objects studied by \textsl{RXTE} \citep{Coburn_etal02}
and \textsl{Beppo}/SAX \citep{DoroR_17}. There seem to be two groups
of objects each with a linear relationship and similar slopes, but separated
by an offset in sigma by $\sim$2\,keV.}
\label{fig:sig_E}
\end{figure}

Recently, a similar systematic study using observational data from \textsl{Beppo}/SAX (on nearly the same 
set of objects, observed around the same time as by \textsl{RXTE}), was completed (PhD thesis Univ. of
T\"ubingen by \citealt{DoroshenkoA_17}).\footnote{http://astro.uni-tuebingen.de/diss.shtml} 
This study largely confirms the earlier results. Combining the results of both studies, two interesting 
features appear. First, the dependence of E$_{\rm cutoff}$ on E$_{\rm cyc}$ is even weaker, more like 
E$_{\rm cutoff}$ $\propto$ E$_{\rm cyc}^{0.4}$ (see Fig.~\ref{fig:Ecut_E}). And second, the linear correlation  
$\sigma_{\rm cyc}$ $\propto$ E$_{\rm cyc}$ is well demonstrated (the data points from \textsl{Beppo}/SAX 
and \textsl{RXTE} are in good agreement). However, there is considerable scatter when all eleven objects 
are to be fit by one relationship. The scatter can be drastically reduced by assuming that there are two groups, 
both following the same slope, but with a constant offset of $\sim$2\,keV in $\sigma$ to one another 
(Fig.~\ref{fig:sig_E}). The objects in the group with the smaller $\sigma$ values appear to be the more 
``regular'' objects, that tend to obey most other correlations fairly well. The objects in the other group are more 
often ``outliers'' in other correlations and have other or extreme properties, such as very high luminosity (Cen~X-3, 
4U~0115+63), a different continuum (4U~0352+309), or are otherwise peculiar, as is the bursting pulsar 
(4U~1744-28).
Generally we find that as the width $\sigma$ and depth $\tau$ of the CRSF increase with increasing centroid 
energy E$_{\rm cyc}$, two continuum parameters also increase: the cut-off energy E$_{\rm cutoff}$ and the 
power law index $-\Gamma$ (the continuum hardens).

\subsection{Individual sources}
\label{sec:individual}

Here we summarize the main characteristics of all sources showing a correlation between E$_{\rm cyc}$ and L$_{\rm x}$.
We start with sources of low to medium luminosity showing a positive correlation (seven sources), 
followed by the two sources at very high luminosities showing a negative correlation. \\

{\bf\textsl{Her X-1}}: This source is in many ways unique: it belongs to the class of low mass accreting binary pulsars 
with the highly magnetized neutron star accreting through an accretion disk. 
The optical companion, HZ Hercules, is a low mass main sequence star with spectral type A to F. It was known well before
the discovery of the X-ray source as an interesting variable star. The optical light is modulated at the binary period of 1.70\,d
due to heating by the X-rays from the neutron star. A long-term optical history, showing periods of extended lows, is documented 
on several photographic plates from different observatories, dating back to 1890 \citep{Jones_etal73}. 
The binary system shows a large number of observational 
features (also due to the low inclination of its binary orbit), that are characteristic for the whole class of objects.  
Apart from being one (together with Cen~X-3) of the two first detected binary X-ray pulsars \citep{Tananbaum_etal72}, 
Her~X-1 is associated with several other  ``first detections'' \citep[see e.g.,][]{Staubert_14}, in particular with respect to cyclotron 
line research: 1) the detection of the first cyclotron line ever (Fig.~\ref{fig:HerX1_CRSF}) \citep{Truemper_etal77,Truemper_etal78}, 
constituting the first ``direct'' measurement of the magnetic field strength in a neutron star, 2) the detection of the first positive 
correlation of the cyclotron line energy E$_{\rm cyc}$ with X-ray luminosity \citep{Staubert_etal07} (see Sect.~\ref{sec:pulse_amplitude}), 
and 3) the first detection of a long-term decay of E$_{\rm cyc}$ (by $\sim5$\,keV over 20\,yrs) \citep{Staubert_etal14,Staubert_etal16} 
(see Sect.~\ref{sec:long_term}).

The positive correlation is also found on a timescale of seconds using the ``pulse-to-pulse'' (or ``amplitude-resolved''
analysis) \citep{Klochkov_etal11} (see Sect.~\ref{sec:pulse_amplitude}). In this analysis it is demonstrated that, together with
the cyclotron line energy, the continuum also varies: for Her~X-1 the continuum gets harder (the power law index $\Gamma$
decreases, $-\Gamma$ increases) with increasing L$_{\rm x}$. This appears to hold for all objects with a positive 
E$_{\rm cyc}$/L$_{\rm x}$ correlation and the opposite is true for that with a negative E$_{\rm cyc}$/L$_{\rm x}$ correlation.
In addition, Her~X-1 is one of the few binary X-ray pulsars exhibiting a super-orbital period of $\sim$35\,days, strongly modulating
the overall X-ray flux as well as the shape of the pulse profile \citep[see e.g.,][]{Staubert_14}. This modulation is thought to be
due to an obscuration of the areas on the neutron star surface where X-rays are produced by the accretion disk. The super-orbital 
modulation is not a good clock and has its own quite intriguing timing properties \citep{Staubert_etal10c,Staubert_etal13,Staubert_etal14}.
It is an open question whether the suggestion of free precession of the neutron star 
\citep{Truemper_etal86,Staubert_etal09,Staubert_etal10a,Staubert_etal10b,Postnov_etal13} is indeed viable. 
It is interesting to note that E$_{\rm cyc}$ appears to change slightly with 35d-phase \citep{Staubert_etal14}. \\

{\bf\textsl{GX~304$-$1}}: Originally discovered at hard X-ray energies (above 20 keV) during a set of MIT balloon scans 
of the Galactic plane in the late 1960s and early 1970s \citep{McClintock_etal71}, and seen as an \textit{UHURU} 
source at 2$-$10 keV \citep{Giacconi_etal72}, the source went into quiescence for nearly 30 years \citep{Pietsch_etal86}
until its re-emergence in 2008 \citep{Manousakis_etal08}. The binary orbit is 132.2\,d (from the separation between outbursts,
\citep{Sugizaki_etal15,Yamamoto_etal11,Malacaria_etal15}), and the pulse period is 275\,s \citep{McClintock_etal77}. 
The star V850~Cen was identified as the optical companion \citep{Mason_etal78}. The first detection of a cyclotron line at $\sim$54\,keV 
and its possible luminosity dependence is based on \textsl{RXTE} observations of an outburst in August 2010 \citep{Yamamoto_etal11}. 
The positive luminosity correlation was confirmed through \textsl{INTEGRAL} results \citep{Klochkov_etal12,Malacaria_etal15}.
Recent analysis of all of the \textsl{RXTE} observations of GX~304$-$1 by \citet{Rothschild_etal16,Rothschild_etal17} covered four 
outbursts in 2010 and 2011. This analysis not only confirmed the positive correlation of the CRSF energy with luminosity, 
but also showed a positive correlation of the line width and depth with luminosity. In addition, a positive correlation was seen 
for the power law index (the spectrum getting harder with increasing luminosity) and for the iron line flux.  For the first time, 
all the CRSF correlations were seen to flatten with increasing luminosity (see Fig.\ref{fig:GX304_cor}).
As will be discussed in Sect.~\ref{sec:physics}, this behavior can be successfully modeled assuming a slow down of the
accretion flow by a collisionless shock, in which the radiation pressure is of minor importance. \\

{\bf\textsl{Vela~X-1}}: An eclipsing high-mass X-ray binary discovered in the early years of X-ray astronomy by 
\citet{Chodil_etal67}, this source is considered as an archetypal wind accretor. It is a persistent source with essentially
random flux variations explained by accretion from a dense, structured wind surrounding the optical companion 
HD~77581\citep[e.g.,][]{Fuerst_etal14b}. HD~77581 is a 24 $M_{\odot}$ main sequence giant in a binary orbit 
of 8.96\,d and moderate eccentricity (e = 0.09).

The first evidence for cyclotron lines in this source -- a fundamental around 
$\sim$25\,keV and a first harmonic near 50\,keV -- was found by \citet{Kendziorra_etal92}
using \textsl{Mir-HEXE} and further detailed by \citet{Kretschmar_etal96}. 
Early observations with \textsl{RXTE} confirmed these detections
\citep{Kretschmar_etal97}. Based on \textsl{NuSTAR} observations, 
\citet{Fuerst_etal14b} reported a clear increase of the energy of the 
harmonic line feature with luminosity, the first clear example for such
behavior in a persistent HMXB. In contrast, the central energy of the 
fundamental line shows a complex evolution with luminosity, which might be 
caused by limitations of the phenomenological models used.
From a study of long-term data of \textsl{Swift}/BAT, \citet{LaParola_etal16}
confirmed the positive correlation of the harmonic line energy with luminosity
and noted, that the fundamental line is not always present.
In addition, they found a secular variation of the line energy, decreasing by 
$\sim$0.36~keV~yr$^{-1}$ between 2004 and 2010. This establishes Vela~X-1
as the second source (after Her~X-1) for which a long-term decay is observed. \\

{\bf\textsl{A~0535+26}}: 
The transient X-ray binary A\,0535$+$26 was discovered in outburst
during observations of the Crab with the Rotation Modulation
Collimator on \textsl{Ariel~V}, showing X-ray pulsations of
$\sim$104\,s \citep{Rosenberg_etal75}. The system was found to consist of
an O9.7IIIe donor star and a magnetized neutron star
\citep{Liller_75,Steele_etal98}. It has an eccentricity of $e=0.47\pm0.02$
and an orbital period of $P_\mathrm{orb}=110.3\pm0.3$\,d
\citep{Finger_etal96}. The distance to A\,0535$+$26 is $\sim$2\,kpc 
\citep{Giangrande_etal80,Steele_etal98}, which has been recently confirmed by
Gaia \citep{BailerJones_etal18}. 

Absorption line-like features at $\sim$50 and $\sim$100\,keV, where
the former was of low significance, were first detected with
\textsl{MIR} TTM/HEXE during a 1989 outburst of the source
\citep{Kendziorra_etal94}. They were interpreted as the fundamental and
first harmonic cyclotron resonances. An $\sim$110\,keV feature was
confirmed with \textsl{CGRO} OSSE during a 1994 outburst \citep[][with
no definite statement on the presence of the low energy line due to
OSSE's 45\,keV threshold]{Grove_etal95}. In the following years the fundamental
line was independently confirmed near $\sim$46\,keV with different
missions during a 2005 August/September outburst
\citep{Kretschmar_etal05,Wilson_etal05,Inoue_etal05} and it has been studied for
several outbursts since then \citep[e.g.,][]{Ballhausen_etal17}.

The outbursts of A\,0535$+$26 show varying peak brightnesses, reaching
15--50\,keV fluxes from a few 100\,mCrab to $\sim$5.5\,Crab
\citep{Camero_etal12,Caballero_etal13b}. Bright, so-called giant
outbursts are known to have occured in 1975, 1980, 1983, 1989, 1994,
2005, 2009, and 2011 \citep[][and references
therein]{Caballero_etal07,Caballero_etal13,Caballero_etal13b}. The source has
been observed over a large range of luminosities. High quality data
could be obtained down to comparatively low outburst luminosities
\citep{Terada_etal06,Ballhausen_etal17} and even in quiescence
\citep{Rothschild_etal13}. The fundamental cyclotron line energy generally
does not change significantly with luminosity (over a wide range) \citep[see
Fig.~\ref{fig:Ecyc-Lx} and, e.g.,][]{Caballero_etal07}. There are,
however, tantalizing indications for a more complex behavior: (i) 
clear increase of the line energy up to $52^{+1.6}_{-1.4}$\,keV was
observed for a flare during the rising phase of the 2005 giant
outburst \citep{Caballero_etal08}, (ii) similar to Her~X-1 a positive
$E_\mathrm{cyc}$/$L_\mathrm{x}$ correlation was found using the
``pulse-to-pulse'' analysis \citep{Klochkov_etal11,DMueller_etal13}, and (iii)
the positive correlation might also be visible at the highest
luminosities \citep[see Fig.~\ref{fig:Ecyc-Lx} and, e.g.,][]{Sartore_etal15}.
The continuum emission of A\,0535$+$26 has been observed to harden
with increasing luminosity \citep{Klochkov_etal11,Caballero_etal13}, with
possible saturation at the highest luminosities \citep{Postnov_etal15} and
a more strongly peaked rollover at the lowest outburst luminosities
\citep{Ballhausen_etal17}. \\

{\bf\textsl{Cep X-4}}:  This source was discovered by OSO-7 in 1972 \citep{Ulmer_etal73}. During an outburst in 
1988 observed by \textsl{Ginga}, regular pulsations with a pulse period around 66\,s were discovered,
and evidence for a CRSF around 30\,keV was found \citep{Mihara_etal91}. The optical counterpart was 
identified by \citet{Bonnet-BidaudMouchet_98}, who measured a distance of $3.8\pm0.6$\,kpc.
\citet{Mihara_95}, on the basis of  \textsl{Ginga} observations, had claimed that Cep~X-4 (together with the high
luminosity sources 4U~0115+63 and V~0332+53) showed a negative E$_{\rm cyc}$/L$_{\rm x}$ correlation.
This has never been confirmed. Instead a positive correlation was discovered in \textsl{NuSTAR} data
\citep{Fuerst_etal15} -- although with two data points only. \citet{Jaisawal_etal15} detected the first harmonic 
in observations by \textsl{Suzaku}. \citet{Vybornov_etal17}, analyzing an outburst observed in 2014 by 
\textsl{NuSTAR} using the pulse-amplitude-resolving technique \citep{Klochkov_etal11} confirmed the existence 
of the two cyclotron lines at $\sim$30\,keV and $\sim$55\,keV and found a strong positive E$_{\rm cyc}$ / 
L$_{\rm x}$ correlation, that is well modeled by assuming a collisionless shock. \\

{\bf\textsl{Swift~1626.6$-$5156}}: The source was discovered by \textsl{Swift}/BAT during an outburst in
2005 as a transient pulsar with $\sim$15\,s pulse period \citep{Palmer_etal05}. The optical companion was identified 
as a Be-star \citep{NegueruelaMarco_06}. After different suggestions, a convincing orbital solution was finally found 
by \citet{Baykal_etal10} with a period of 132.9\,d and a very small eccentricity (0.08).
Several observations over three years (2006--2008) by \textsl{RXTE}/PCA led to the discovery of a CRSF at
about 10\,keV \citep{DeCesar_etal13}. Even though the discovery of this CRSF needs confirmation by
further observations (and other instruments), we list this source not under the category ``candidates'' because the
evidence from the different observations is quite strong and there are clear signatures of the usual behavior
of CRSFs, including a strong correlation of the phase resolved CRSF energy with pulse phase, an indication for a positive 
correlation with luminosity  and a hint to a first harmonic at $\sim$18\,keV. \\

{\bf\textsl{V~0332+53}}: The transient X-ray binary V\,0332$+$53 was discovered in outburst in
1983 by \textsl{Tenma} \citep{Tanaka_83,Makishima_etal90a} and in parallel an earlier outburst was 
revealed in \textsl{Vela~5B} data from 1973 \citep{TerrellPriedhorsky_83,TerrellPriedhorsky_84}. 
In follow-up observations by \textsl{EXOSAT} during the 1983 activity, 4.4\,s pulsations were
discovered, an accurate position was measured, and the orbital period and eccentricity were determined 
to be 34 days and 0.37, respectively \citep[][see also \citealt{Zhang_etal05}]{Stella_etal85}. 
The O8--9Ve star BQ~Cam was identified as the optical counterpart \citep{Argyle_etal83,Negueruela_etal99}. 
The distance to the system was estimated to be 6--9\,kpc \citep[][but see also 
\citealt{Corbet_etal86}]{Negueruela_etal99}.

V\,0332$+$53 displays normal as well as giant outbursts. Occurrences of the latter have been observed 
in 1973, 2004/2005, and 2015 \citep{Ferrigno_etal16a}. During giant outbursts the source can become one
of the most luminous X-ray sources in the Galaxy, reaching a few times $10^{38}\,\mathrm{erg\,s}^{-1}$. 
Quasi-periodic oscillations with frequencies of $\sim$0.05\,Hz and $\sim$0.22\,Hz have been observed
\citep{Takeshima_etal94,Qu_etal05,CaballeroGarcia_etal16}.
 
The \textsl{Tenma} observation of V\,0332$+$53 also provided evidence for a fundamental cyclotron line 
feature at $\sim$28\, keV \citep{Makishima_etal90a}. Its presence was confirmed with high
significance by \textsl{Ginga} observations of the source during a 1989 outburst, which also showed 
indications for a harmonic feature at $\sim$53\,keV \citep{Makishima_etal90}. The giant outburst in 
2004/2005 allowed for the confirmation of this harmonic as well as the detection of a rare second harmonic 
at $\sim$74\,keV with \textsl{INTEGRAL} and \textsl{RXTE} \citep{Kreykenbohm_etal05,Pottschmidt_etal05}.

These observations of the giant outburst in 2004/2005 further revealed that the energy of the 
fundamental cyclotron line decreased with increasing luminosity 
\citep{Tsygankov_etal06,Tsygankov_etal10,Mowlavi_etal06}.
Additional studies showed that the correlation was also present for the first harmonic line
\citep[although characterized by a weaker fractional change in energy,][]{Nakajima_etal10} 
as well as for the pulse-to-pulse analysis of the fundamental line \citep{Klochkov_etal11}. 
A qualitative discussion of results from pulse phase-resolved spectroscopy of this outburst in
terms of the reflection model for cyclotron line formation has been presented 
\citep{Poutanen_etal13,Lutovinov_etal15}. \textsl{Swift}, \textsl{INTEGRAL}, and \textsl{NuSTAR} 
observations of the most recent giant outburst in 2015 also showed the negative correlation and
provided evidence that the overall line energy, and thus the associated $B$-field, was lower just 
after the outburst, indicating some decay over the time of the outburst
\citep{Cusumano_etal16,Ferrigno_etal16a,Doroshenko_etal17,Vybornov_etal18}
(see also the discussion in Sect.~\ref{sec:long_term}).

V\,0332$+$53 is singled out by the fact that it is the only source to date in which we
find both E$_{\rm cyc}$ / L$_{\rm x}$ correlations: the negative at high luminosities and the 
positive at low luminosity \citep{Doroshenko_etal17,Vybornov_etal18}.
It is also the only one with a very strong negative correlation,
accompanied by a second source -- SMC~X-2 -- which shows a much weaker dependence 
(see Fig.~\ref{fig:Ecyc-Lx}). This two-fold behavior is in line with the correlation between
the spectral index (or spectral hardening) as found in several other accreting pulsars:
a hardening at low luminosities and a softening at very high luminosities
\citep{Klochkov_etal11,Postnov_etal15}. \\

{\bf\textsl{SMC~X-2}}: This transient source in the Small Magellanic Cloud was detected
by \textsl{SAS-3} during an outburst in 1977 \citep{Clark_etal78,Clark_etal79}. Later outbursts were observed by
several satellites. In 2000 the source was identified as an X-ray pulsar by \textsl{RXTE} and \textsl{ASCA} with a 
period of 2.37\,s \citep{Torii_etal00,Corbet_etal01,Yokogawa_etal01}. The optical companion suggested by 
\citet{Crampton_etal78} was later resolved into two objects and the northern one identified as the true companion
based on an optical periodicity of $\sim$18.6\,d \citep{Schurch_etal11}, that appeared to coincide with an
X-ray period of $\sim$18.4\,d found in \textsl{RXTE} data \citep{Townsend_etal11}. The optical classification
of the companion was determined to be O9.5~III-V \citep{McBride_etal08}. During an outburst in 2015 
\citet{JaisawalNaik_16} found a cyclotron line at $\sim$27\,keV in \textsl{NuSTAR} data, that showed a weak
negative correlation with luminosity (see Fig.~\ref{fig:Ecyc-Lx}). If this is confirmed, SMX~X-2 is the second 
high luminosity source showing this negative correlation with luminosity. As with Swift~1626.6-5156, the detection 
of the cyclotron line needs confirmation. \citet{Lutovinov_etal17a}, using observations by \textsl{Swift}/XRT of the same
outburst in 2015, detected a sudden drop in luminosity to below a few times $10^{34}$\, erg/s, which, 
interpreted as the signature of the propeller effect, allows us to estimate the strength of the magnetic field to be around 
3~$\times 10^{12}$\,~Gauss. This is quite close to the B-field value found from the cyclotron line energy
(see also Sect.~\ref{sec:strength}). \\

{\bf\textsl{4U~0115+63}}: This source is included here since, historically, it was thought to be a high luminosity
source showing a negative E$_{\rm cyc}$ / $L_x$ correlation. As we discuss below, we now believe, however, that 
this is probably not correct. 4U~0115+63 is a high mass X-ray binary system, first discovered in the sky survey by 
\textsl{UHURU} \citep{Giacconi_etal72}, 
with repeated outbursts reaching high luminosities \citep{Boldin_etal13}. The system consists of a pulsating neutron 
star with a spin period of 3.61\,s \citep{Cominsky_etal78} and a B0.2Ve main sequence star \citep{Johns_etal78}, orbiting 
each other in 24.3\,d \citep{Rappaport_etal78}. The distance to this system has been estimated at $\sim$7\,kpc 
\citep{NegueruelaOkazaki_01}. 
As early as 1980 a cyclotron line at 20\,keV was discovered in 4U~0115+63 in observational data of \textsl{HEAO-1}/A4
\citep{Wheaton_etal79}. A re-examination of the data uncovered the existence of two lines at 11.5 and 23 keV  which 
were immediately interpreted as the fundamental and harmonic electron cyclotron resonances  \citep{White_etal83}. 
Later observations of the source found three \citep{Heindl_etal99a}, then four  \citep{Santangelo_etal99}, 
and finally up to five lines in total \citep{Heindl_etal04}. 4U~0115+63 is still the record holder in the numbers of harmonics.

A negative correlation between the pulse phase averaged cyclotron line energy and the X-ray luminosity 
(a decrease in E$_{\rm cyc}$ with increasing $L_x$) was claimed for the first time in this source by \citet{Mihara_95} 
on the basis of observations with \textsl{Ginga} (together with two other high luminosity transient sources: Cep~X-4, 
and V~0332+53). This negative correlation was associated with the high accretion rate during the X-ray outbursts, 
and as due to a change in height of the shock (and emission) region above the surface of the neutron star with changing 
mass accretion rate, $\dot{M}$. In the model of \citet{Burnard_etal91}, the height of the polar accretion structure is 
tied to $\dot{M}$ (see above). A similar behavior was observed in further outbursts of 4U~0115+63 in March-April 1999 
and Sep-Oct 2004: both \citet{Nakajima_etal06} and \citet{Tsygankov_etal07} had found a general anticorrelation 
between E$_{\rm cyc}$ and luminosity. The negative correlation was also confirmed by \citet{Klochkov_etal11} 
using the pulse-amplitude-resolved analysis technique, together with the softening of the continuum when
reaching very high luminosities (see also \citealt{Postnov_etal15}).

However, \citet{SMueller_etal13}, analyzing data of a different outburst of this source 
in March-April 2008, observed by \textsl{RXTE} and \textsl{INTEGRAL}, have found that the negative correlation 
for the fundamental cyclotron line is likely an artifact due to correlations between continuum and line parameters 
when using the NPEX continuum model. Further, no anticorrelation is seen in any of the harmonics. 
\citet{Iyer_etal15} have suggested an alternative explanation: there may be two systems of cyclotron lines with 
fundamentals at $\sim$11\,keV and $\sim$15\,keV, produced in two different emission regions (possibly at different 
poles). In this model, the second harmonic of the first system coincides roughly with the 1st harmonic in the second system.
In summary, we conclude that 4U~0115+63 does not show an established dependence of a CRSF energy on luminosity.

\section{Physics of the accretion column}
\label{sec:physics}

In this Section we discuss the basic physics with relevance to the accretion onto highly 
magnetized neutron stars. The generation of the X-ray continuum and the cyclotron lines, as well 
as their short- and long-term variability, depends on the structure of the accretion column, the
physical state of the hot magnetized plasma, the magnetic field configuration and many details 
with regard to fundamental interaction processes between particles and photons that govern
the energy and radiation transport within the accretion column.

Our basic understanding is that material transferred from a binary companion couples to the
magnetosphere of the neutron star (assuming a simple dipole field initially). The accreted plasma
falls along the magnetic field lines down to the surface of the NS onto a small area 
at the polar caps, where it is stopped and its kinetic energy is converted to heat and radiation. From
the resulting ``accretion mound'' the radiation escapes in directions that depend on 
the structure of the mound, the B-field and the gravitational bending by the NS mass.
If the magnetic and spin axes of the neutron star are not aligned, a terrestrial observer sees a 
flux modulated at the rotation frequency of the star.

We refer to the following fundamental contributions to the vast literature on this topic: 
\citet{GnedinSunyaev_74,Lodenquai_etal74,BaskoSunyaev_75,ShapiroSalpeter_75,WangFrank_81,
LangerRappaport_82,BraunYahel_84,Arons_etal87,Miller_etal87,Meszaros_92,Nelson_etal93,Becker_etal12}. 
Further references are given in the detailed discussion below.

\subsection{Accretion regimes}
\label{sub:shocks}

In the simplest case of a dipole magnetic field, the plasma filled polar cap area is 
$A=\pi r_p^2$ with polar cap radius $r_p\simeq R_{NS}\sqrt{R_{NS}/R_m}$, where $R_m$ is 
the magnetospheric radius. The latter is determined by the balance of the magnetic field and matter 
pressure, and is a function of the mass accretion rate $\dot M$ and the NS magnetic field, 
which can be characterized by the surface NS value, $B_s$, or, equivalently, by the dipole 
magnetic moment $\mu=(B_sR_{NS}^3)/2$ - see footnote.\footnote{We note that the 
factor of a half, required by electrodynamics describing the B-field at the magnetic poles (see e.g., Landau-Lifschits, 
Theory of fields), is often missing in the literature (see also the comments in Table~\ref{tab:estimates}).}
The accreting matter moves toward the NS with 
free-fall velocity, which is about $\mathrm{v_0}=10^{10}$~cm s$^{-1}$ close to the NS. The matter 
stops at (or near) the NS surface, with its kinetic energy being ultimately released in the form of radiation 
with a total luminosity of $L_a=\dot M\mathrm{v_0}^{2}/2$.
The continuum is believed to be due to thermal bremsstrahlung radiation from the $\sim$10$^{8}$\,K
hot plasma, blackbody radiation from the NS surface and cyclotron continuum radiation - all
modified by Comptonisation \citep{BaskoSunyaev_75,BeckerWolff_07, Becker_etal12}
and the cyclotron line by resonant scattering of photons on electrons with discrete energies of
Landau levels \citep{Ventura_etal79,Bonazzola_etal79,Langer_etal80,Nagel_80,Meszaros_etal83}.

Since the discovery of negative and positive $E_\mathrm{cyc}$/$L_\mathrm{x}$ correlations,
first found for V~0332+53 \citep{Makishima_etal90,Mihara_95} and Her~X-1 \citep{Staubert_etal07}, respectively,
it has become very clear that there are different accretion regimes, depending on the mass accretion
rate (X-ray luminosity). An important value separating low and high accretion rates is the 
``critical luminosity'', $L*\sim 10^{37}$~erg~s$^{-1}$ (see below). The different accretion regimes correspond to 
different breaking regimes, that is, different physics by which the in-falling material is decelerated.
So, the study of the $E_\mathrm{cyc}$/$L_\mathrm{x}$ dependence provides a new tool for probing physical 
processes involved in stopping the accretion flow above the surface of strongly magnetized neutron stars in 
accreting X-ray pulsars.

The different breaking regimes are as follows.
If the accretion rate is very small, the plasma blobs frozen in the NS magnetosphere arrive almost 
at the NS surface without breaking, and the final breaking occurs in the NS atmosphere via 
Coulomb interactions, various collective plasma processes and nuclear interactions. 
This regime was first considered in the case of spherical accretion onto NS without magnetic 
fields by \citet{ZeldovichShakura_69} and later in the case of 
magnetized neutron stars (e.g., \citealt{Miller_etal87,Nelson_etal93}). 
The self-consistent calculations of the layer in which energy of the accreting flow was released 
carried out in these papers showed that the stopping length of a proton due to Coulomb 
interactions amounts to $y=\int\rho dz \sim$50\,g~cm$^{-2}$, where $\rho$ is the plasma density 
and the height of this layer is comparable to the NS  atmosphere size. Clearly, in this case the 
CRSF feature (if measured) should reflect the magnetic field strength close to the NS surface and 
should not appreciably vary with changing accretion luminosity 

With increasing mass accretion rate $\dot M$, the accreting flow (treated gas-dynamically) 
starts decelerating with the formation of a collisionless shock (CS). The possibility of 
deceleration via a CS was first considered in the case of accretion of plasma onto a neutron star
with a magnetic field by \citet{BisnovatyiFridman_70}. 
Later several authors (e.g., \citealt{ShapiroSalpeter_75,LangerRappaport_82,BraunYahel_84}) 
postulated the existence of a stationary CS above the neutron star
surface if the accretion luminosity is much less than the Eddington luminosity. The 
formation and structure of the CS, in this case, accounting for detailed microphysics (cyclotron 
electron and proton cooling, bremsstrahlung losses, resonant interaction of photons in diffusion 
approximation, etc.), is calculated numerically by \citet{BykovKrasil_04} in 1D-approximation. 
These calculations confirmed the basic feature of CS: (1) the formation above the neutron star 
surface at the height $H_l\sim (\mathrm{v_0}/4)t_{ei}$,
where $\mathrm{v_0}\approx c/3$ is the upstream velocity, and $t_{ei}$ 
is the equilibration time between protons and electrons via Coulomb interactions, which follows 
from the requirement to transfer most of the kinetic energy carried by ions to radiating 
electrons, (2) the release of a substantial fraction (about half) of the in-falling kinetic energy 
of the accretion flow downstream of the CS in a thin interaction region of the flow. 

The CS breaking regime persists until the role of the emitted photons becomes decisive, which 
can be quantified (to an order of magnitude) by equating
the photon diffusion time across the accretion column, 
$t_d\sim r_P^2/(cl_\gamma)$, where $l_\gamma$ is the photon mean free path in the strong 
magnetic field, and the free-fall time of plasma from the shock height, 
$t_{ff}\sim H_l/(\mathrm{v_0}/4)\sim r_p/\mathrm{v_0}$.
This relation yields the so-called 
``critical luminosity'', $L*\sim 10^{37}$~erg~s$^{-1}$, above which the in-falling accretion flow is 
decelerated by a \textsl{radiative shock} (RS) \citep{BaskoSunyaev_75,BaskoSunyaev_76,Arons_etal87}. 
In the literature, there are several attempts to calculate this critical luminosity (which should 
depend on the magnetic field, the geometry of the flow, etc.) more precisely 
\citep[see e.g.,][]{WangFrank_81,Becker_etal12,Mushtukov_etal15a}.
It should be kept in mind, however, that the transition to a radiation dominated (RD)-dominated
regime occurs rather smoothly, and this critical luminosity should be perceived as a guidance value 
(and not a strict boundary) for the separation between the pure CS or RS regimes in a particular source.  

\subsubsection{Scaling CRSF relations in collisionless shock regime}
\label{ss:CS}

In the CS regime that can be realized in accreting X-ray pulsars with low or moderate X-ray 
luminosity (e.g., Her X-1, GX 304-1, Cep X-4, etc.), there are simple and physically clear 
relations between the CRSF properties (energy, width, depth) and X-ray luminosity, which 
have been checked by X-ray observations. Indeed, the typical CS height is a few hundred meters above 
the neutron star surface and scales with the plasma number density as $H_l\sim 1/n_e$ and, 
through the mass conservation $\dot M\sim r_p^2 n_e \mathrm{v_0}$, as $H_l\sim 1/\dot M$
\citep{ShapiroSalpeter_75,BaskoSunyaev_76}, confirmed through detailed numerical simulations by 
\citealt{BykovKrasil_04} (see their Fig.~5). 
At these heights, the dipole structure of the neutron star magnetic field is already important. 
The theory of CRSF formation in an inhomogeneous magnetic field can be found in 
\citet{Zheleznyakov_96}. In the inhomogeneous magnetic field, the 
cyclotron line is formed in a resonant layer. The width of the resonant layer for the assumed 
dipole magnetic field is $\Delta r_\mathrm{res}\sim \beta_{T_\mathrm{e}} r_\mathrm{res}/3$, 
where $\beta_{T_\mathrm{e}}=\mathrm{v_{T_\mathrm{e}}}/c\sim 1/10$ is the thermal velocity of 
post-shock electrons and $r_{res}$ is the radial height of the resonant layer; for typical 
temperatures $T_\mathrm{e}\sim 10$~keV and cyclotron photon energies  
$\hbar\omega_\mathrm{cyc}\sim$30-50\,~keV, the size of the resonant layer 
$\Delta  r_\mathrm{res}\sim 6\times 10^4$~cm can be comparable with the shock height 
$H_\mathrm{s}$ and thus can substantially affect the CRSF formation. 
We note that the post-shock electron temperature $T_\mathrm{e}$ does not vary substantially.

The characteristic optical depth of the resonant layer in the inhomogeneous dipole magnetic 
field $B$ is \citep{Zheleznyakov_96} 
\begin{multline}
\label{taures}
\tau_\mathrm{res}=\frac{16}{3}\frac{\pi^2 e^2}{m_\mathrm{e}c}\frac{n_\mathrm{e}\Delta r_\mathrm{res}}{\omega_\mathrm{cyc}}\sim \\10^4\left(\frac{n_\mathrm{e}}{10^{20}\mathrm{cm}^{-3}}\right) \times \frac{E_{cyc}}{50 keV} \times \frac{R_{NS}}{10^6 cm} \times \frac{B_{s}}{10^{12}G}
\end{multline} 
This means that diffusion of resonant photons occurs. 
As the CS downstream sidewall  area $2\pi r_pH_l$ is typically smaller than the polar
cap area  $A = \pi r_p^2$ (at least if $H_l\ll r_p$), most of the diffusing photons escape
from the uppermost parts of the structure. \\

Clearly, the line dependence on the observed X-ray flux is entirely determined 
by how the collisionless shock height $H_\mathrm{s}$ responds to variable mass accretion rate. 
This scaling indeed was first observed in Her X-1 and explained in terms of the line formation in 
the varying magnetic field with height in \citet{Staubert_etal07}. As discussed above, we now know four additional 
X-ray pulsars presumably in the CS breaking regime, showing a positive $E_\mathrm{cyc}$ / $L_\mathrm{x}$
correlation: Vela~X-1, A~0535+26, GX~304-1, Cep~X-4 (see \ref{sec:luminosity}).
A factor five to six in the dynamic range in $L_\mathrm{x}$ has allowed for the last two objects to detect the
theoretically expected deviations from a purely linear correlation. 
We note that not only the energy $E_\mathrm{cyc}$,
but also its width $W$ and depth $\tau_\mathrm{cyc}$
show variations with the observed X-ray flux, consistent with this non-linear scaling (see  \citealt{Rothschild_etal17} 
for more detail), thus lending credence to the simple physical picture outlined above.

\subsubsection{Scaling CRSF relations in radiative shock regime}
\label{ss:RS}

In the case of RS deceleration of the accretion flow at high X-ray luminosities, the situation 
with CRSF formation is not so straightforward as in the CS-regime considered above. Indeed, 
in this case an optically thick extended intrinsically 2D structure is formed 
\citep{Davidson_73,BaskoSunyaev_76,WangFrank_81,Postnov_etal15}, 
and the line formation should be calculated by solving the 2D radiation transfer problem.
Still, the assumption that the shock height (and the line emitting region) should increase with
increasing mass accretion rate \citep{BaskoSunyaev_76} should hold, and $E_\mathrm{cyc}$
should vary according to Eq.~\ref{eqn:dipole}, as first suggested by \citet{Burnard_etal91}.
In addition, reflection of radiation from the neutron star surface could play a role. 
While physically feasible, the reflection model for CRSF formation and change with X-ray flux in 
high-luminosity accreting X-ray pulsars advanced by \citet{Poutanen_etal13} may not be universally 
applied in all cases, since the negative CRSF correlation with luminosity is now reliably confirmed 
by X-ray observations of only one bright transient X-ray pulsar V0332+53  \citep{Tsygankov_etal06} 
and tentatively diagnosed for SMC~X-2 \citep{JaisawalNaik_16}. Clearly, future observations (possibly, 
with X-ray polarization measurements) are needed here to study the intriguing CRSF behavior in the 
RS-case.    

\subsection{Cyclotron line modeling}
\label{sec:modeling}

This paper concentrates on the observational aspects of objects from which spectra with cyclotron lines are 
observed. A description of the state of theoretical modeling  of the CRSF is beyond the scope
of this contribution -- a comprehensive review as a counterpart would, however, be highly useful.

In the Introduction and in Sect.~\ref{sec:physics} above mention the history of 
the early theoretical work that attempted to calculate expected spectra analytically and that
has been ground breaking for our understanding of the basic physics in highly magnetized hot plasma -- 
the source of high energy X-ray spectra with cyclotron lines. \citet{Wang_etal88} give a good summary of
the early work, which is mostly from the 70s and 80s of the last century. More recent analytical calculations 
are from \citet{Nishimura_08,Nishimura_11,Nishimura_13,Nishimura_14,Nishimura_15}. Nishimura invokes 
several specific conditions, for example, changes of the height and the area of the accretion mound, changes in the 
emission profile, non-uniform illumination of the emission region, superposition of lines from different regions, 
changing relative contributions from the two accreting poles and other, to explain observational details. 
The success of this may, however, be due to a large number of free parameters and/or
assumptions that enter the calculations from the start.

A completely different technique to understand and model the physical conditions that lead to the observed
cyclotron line spectra are Monte Carlo calculations 
\citep{Isenberg_etal98,ArayaHarding_99,ArayaGochezHarding_00,Schoenherr_etal07,Nobili_etal08,Schwarm_etal17a,Schwarm_etal17b}. 
Here individual particles (electrons, photons, protons) are followed through some extended volume of
highly magnetized hot plasma, calculating their interactions with each other and with the magnetic field.
At the boundaries of the volume photons escape, some of them finding their way into the telescopes of
observers, where the energy distribution (spectra) and the timing distribution (pulse profiles and other
time variations) are measured. Usually, some input continuum spectrum is chosen that illuminates a certain volume 
under specific geometries. The method is rather flexible, allowing for the testing of different input continua and geometries
(e.g., illumination from the bottom, from the top, from a centrally extended source, or others),
and angle- and energy dependent interaction cross-sections can easily be adjusted.
The most recent and complete calculations are by Fritz Schwarm\footnote{Schwarm: PhD Thesis, University of
Erlangen-N\"urnberg, https://www.sternwarte.uni-erlangen.de/docs/theses/2017-05\_Schwarm}
\citep{Schwarm_etal17a,Schwarm_etal17b}. This work represents the current state of the art and provides, for the
first time, a framework (usable by the general scientific 
community)\footnote{https://www.sternwarte.uni-erlangen.de/~schwarm/public/cyclo} for a direct comparison and
fitting of observed spectra with Monte-Carlo-simulated spectra, such that physical conditions and parameters can be 
estimated.

\begin{table*}[]
 \caption[]{Magnetic field strength for a few selected sources: comparing measurements based on the observed cyclotron line energy 
 with estimates based on applying accretion torque models. }
  \label{tab:estimates}
  \begin{center}
  \begin{tabular}{llcccccc}
 \hline\noalign{\smallskip}
  System  & Refer.$^{a}$  & $\mu_{30}$(GL)$^{c}$         & B$_{12}$~=$^{e}$ & Cyclotron   & B$_{12}$        & B(Wang)  & B$_{torque}$ \\
                      & ``rotator'' $^{b}$  &     or                         & 2$\times\mu_{30}$ & line            & from                & / B(GL)    & / B$_{cyc}$    \\
                      & & $\mu_{30}$(Wang) =$^{d}$                 & = B$_{torque}$      & Energy       & E$_{cyc}$       &    or         &                       \\
                      & &  5$\times\epsilon_{30}$(Wang)            &                                & E$_{cyc}$  & (= B$_{cyc}$)  & B(Klus)   &                       \\
                      & & [$10^{30}\,G~cm^{3}$]                         & [$10^{12}$\,G]        & [keV]          & [$10^{12}$\,G] & / B(GL)   &                       \\
  \hline\noalign{\smallskip}                       
Her~X-1         & GL``fast''             & 0.47$^{g}$                & 0.94                        & 37               & 3.83                & 2.2          & 0.25 \\
   " " "              & Wang                  & 1.05                          & 2.10                        & 37               & 3.83                &                & 0.55 \\ 
GX~301-2      & GL ``slow''           & 0.3                            & 0.6                          & 37               & 3.83                &                & 0.16  \\ 
4U~0115+63  & GL ``fast''            & 1.4                            & 2.8                          & 12               & 1.24                & 2.1          & 2.3   \\ 
   " " "              & Wang                  & 2.95                          & 5.9                          & 12               & 1.24                &                & 4.8    \\ 
4U~1626-67   & Wang                  & 4.35                          & 8.7                          & 37               & 3.83                 &                & 2.3   \\
A~0535+26     & GL ``slow''          & 3.3                            & 6.6                          & 50               & 5.17                 & 2.1          & 1.3   \\ 
   " " "              & Klus                     &                                  & 14                           & 50               & 5.17                &                & 2.7    \\ 
Cen~X-3         & GL ``fast''             & 4.5                           & 9                             & 28               & 2.90                & 2.1          & 3.1    \\
   " " "              & Wang                   & 9.55                         & 19.1                       & 28                & 2.90                &                & 6.6    \\
X Per              & GL ``slow''            & 4.8                          & 9.6                         & 29                & 2.19                 &                & 4.4    \\ 
   " " "              & GL ``slow''            & 4.8                          & 9.6                         & 29                & 2.19                 &                & 4.4    \\ 
   " " "              & Klus                      &                                & 42                           & 29               & 2.19                 &                & 19.2  \\ 
Vela~X-1$^{h}$ & GL ``fast''          & 86                           & 172                         & 25               & 2.60                 &                & 66      \\
A~0535+26     & GL ``fast''             & 148                         & 296                         & 50               & 5.17                &                 & 57     \\ 
GX~301-2       & GL ``fast''             & 394                         & 788                         & 37               & 3.83                 &                & 206    \\ 
  \noalign{\smallskip}\hline
  \end{tabular}
  \end{center}
$^{a}$ References: ``GL''  : \citet{GhoshLamb_79b}, ``Wang''  : \citet{Wang_87}, ``Klus''  : \citet{Klus_etal14};
$^{b}$ "slow" or ``fast'' solution according to \citet{GhoshLamb_79b}, 
$^{c}$ original values from the literature;
$^{d}$ the original values of \citet{Wang_87} are $\epsilon_{30}$, $\mu_{30}$ = 5$\times\epsilon_{30}$; 
$^{e}$ we use the definition: B = 2~$\mu$~$R^{-3}$ (unlike \citealt{GhoshLamb_79a}, see text) and consider this to be the field strength 
B$_{0}$  at the surface of the neutron star at the magnetic poles;
$^{g}$ we note that $\mu_{30}$=0.47 is a valid solution for Her~X-1 within the GL-model, but it requires L$_{37}$=2.7 
(not 1.0, as in Table~1 of \citealt{GhoshLamb_79b}); 
$^{h}$ for Vela X-1 the  GL ``slow'' solution yields $\mu_{30}$ = $2.7~10^{-5}\,G~cm^{3}$, obviously not valid. \\
\end{table*}

\section{Empirical knowledge of the magnetic field strength of a neutron star}
\label{sec:strength}
Even though we assume that the observation of the cyclotron line energy gives a direct measure
of the strength of the magnetic field at the site where the resonant scattering occurs, it does not
provide an accurate value for the global magnetic field (thought to be a dipole). This is because
the cyclotron scattering region may be at some distance above the neutron star surface, and/or
there may be local field structures which could have a different (even stronger) field strength. 
A classical method to estimate the magnetic moment of a neutron star is to infer it through timing
observations of pulsars. This has been done for non-accreting pulsars by applying the theory of 
dipole radiation to observational data of the change of the pulse period with time 
\citep{Gold_68,GoldreichJulian_69,OstrikerGunn_69}:
B = 3.2$\times10^{19}({{\rm P}\dot{\rm P}})^{1/2}$, with B in Gauss, P in s and $\dot{\rm P}$ in s\,s$^{-1}$
(leading, e.g., to an estimate of the field strength of the Crab pulsar of $\sim3.8\times10^{12}$ Gauss).
In the case of accreting pulsars -- relevant for cyclotron line objects discussed here -- accretion torque models
are applied to describe the spin-up/spin-down behavior observed in these objects. The most widely
used model, applicable to the accretion through an accretion disk, is the one developed by \citet{GhoshLamb_79b}
(Paper~III) (with the preceding Paper~I: \citealt{GhoshLamb_77} and Paper~II: \citealt{GhoshLamb_79a}).
The model provides a set of 
equations for the relationship between the magnetic moment of the neutron star and 
the three observables: pulse period P, its time derivative dP/dt and the X-ray luminosity L$_{x}$ (estimated 
through the X-ray flux and the distance to the source), and two unknown quantities: the magnetic moment 
$\mu$ and a dimensionless torque parameter $n(\omega_{s})$, which itself is a complicated function of a 
``fastness parameter'' $\omega_{s}$ (the ratio of the magnetospheric radius to the co-rotation radius to the 
power three-half; in addition there are two dimensionless factors $S_{1}$ and $S_{1}$ (of order one) that depend 
on the neutron star mass and radius.
Similar models (some also for the case of wind accretion) have been presented by 
\citet{Wang_87,Lovelace_etal95,KluzniakRappaport_07,Postnov_etal11,Shakura_etal12,Shi_etal15,Parfrey_etal16}.

For a given magnetic moment $\mu$ of an accreting pulsar the history of the accretion torque over time
(largely determined by the adopted values of the mass accretion rate and the ratio of the magnetospheric radius
to the co-rotation radius) determines the final value of the pulse period, that is when the system reaches
a state at or near an equilibrium: the spin-up and spin-down torques are nearly equal leading to a net torque 
and a period derivative dP/dt near zero. Many of the cyclotron line sources discussed here are actually close to equilibrium. 

The question of how B-field estimates gained through accretion torque models compare with direct measurements through
cyclotron line energies, has been addressed in the literature. In Table~\ref{tab:estimates} we present an updated summary
of values from \citet{GhoshLamb_79b},  \citet{Wang_87} and  \citet{Klus_etal14} for a few objects, and perform a comparison 
with the information from cyclotron lines. The polar magnetic field strength at the surface of the neutron star is 
calculated\footnote{We note that we need the field strength at the poles which is two times that at the equator, apparently
used by \citet{GhoshLamb_79a}.} as B$_{12}$ = 2~$\mu_{30}$~$R^{-3}$, while
the cyclotron line measurements (values are taken from Table~\ref{tab:collection}) lead to B$_{12}$ = 1.2~E$_{cyc}$/11.6. 
It is important to note that the Ghosh \& Lamb-model generally provides two solutions for the magnetic moment $\mu$:
a ``slow rotator''  solution and a ``fast rotator''  solution (see \citealt{GhoshLamb_79b}, Table~2 and Figs.12,13). 
In the upper part of Table~\ref{tab:estimates} we have reproduced the solutions which best match the results from the cyclotron
line measurements and we find a discrepancy by factors of a few:
for Her~X-1 and GX~301-2 the ratio B$_{torque}$/B$_{cyc}$ (column 9) is $\sim$1/2 and $\sim$1/3, while for the other objects listed 
this ratio is above a factors of three and more. Toward the end of Table~\ref{tab:estimates} we list the ``fast rotator''  solutions
for the four long period objects A~0535+26 (P = 104\,s), Vela~X-1 (P = 283\,s), GX~301-2 (P = 681\,s) and X~Per (P = 837\,s),
for which the ratio B$_{torque}$/B$_{cyc}$ reaches values above 100.\footnote{We note that the ``slow rotator'' solution for Vela~X-1
yields $\mu_{30}$ = $2.7~10^{-5}\,G~cm^{3}$, obviously not valid, and for X~Per the two solutions are equal.}

\begin{figure}
\includegraphics[width=0.45\textwidth]{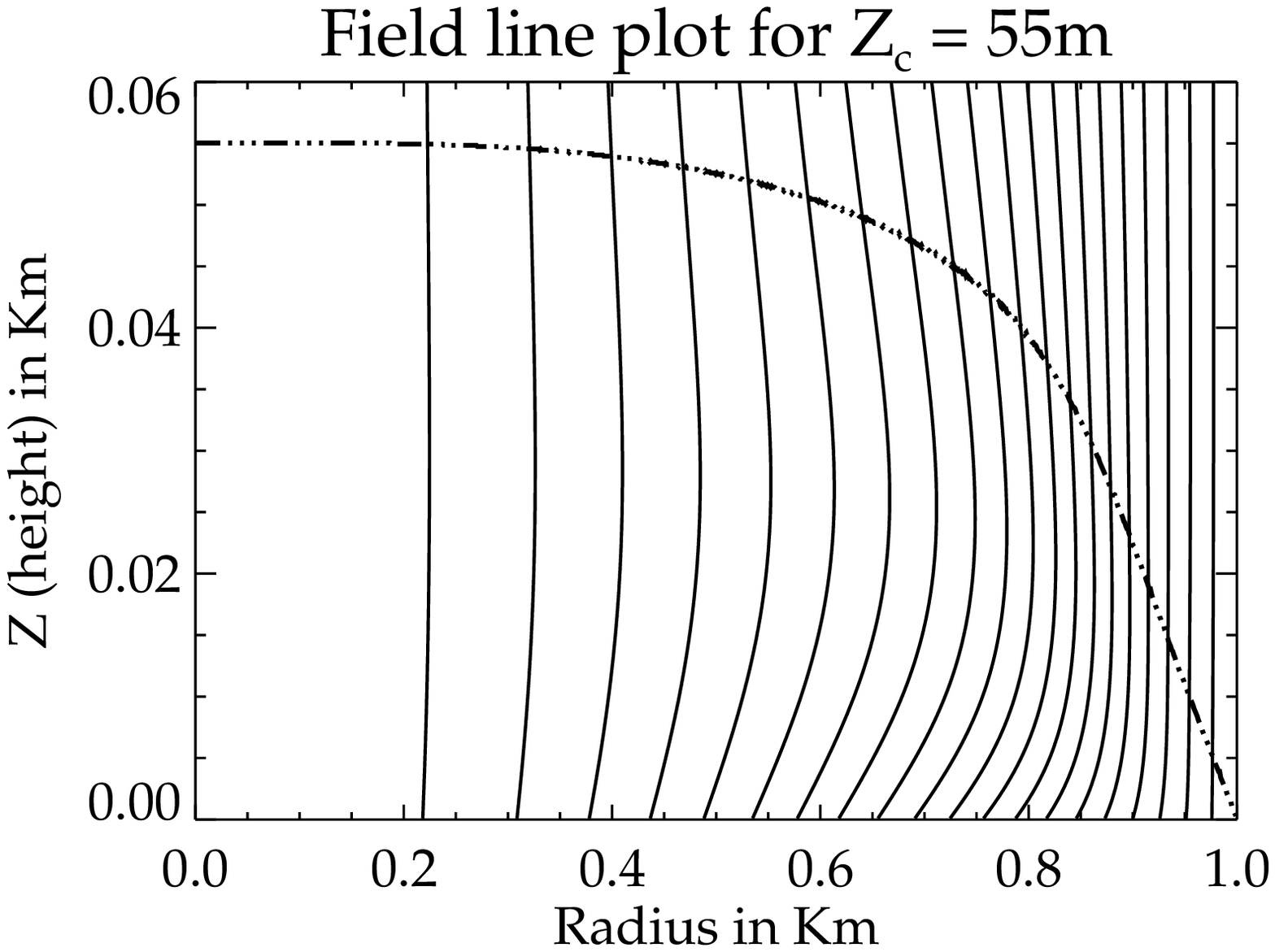}
\includegraphics[width=0.45\textwidth]{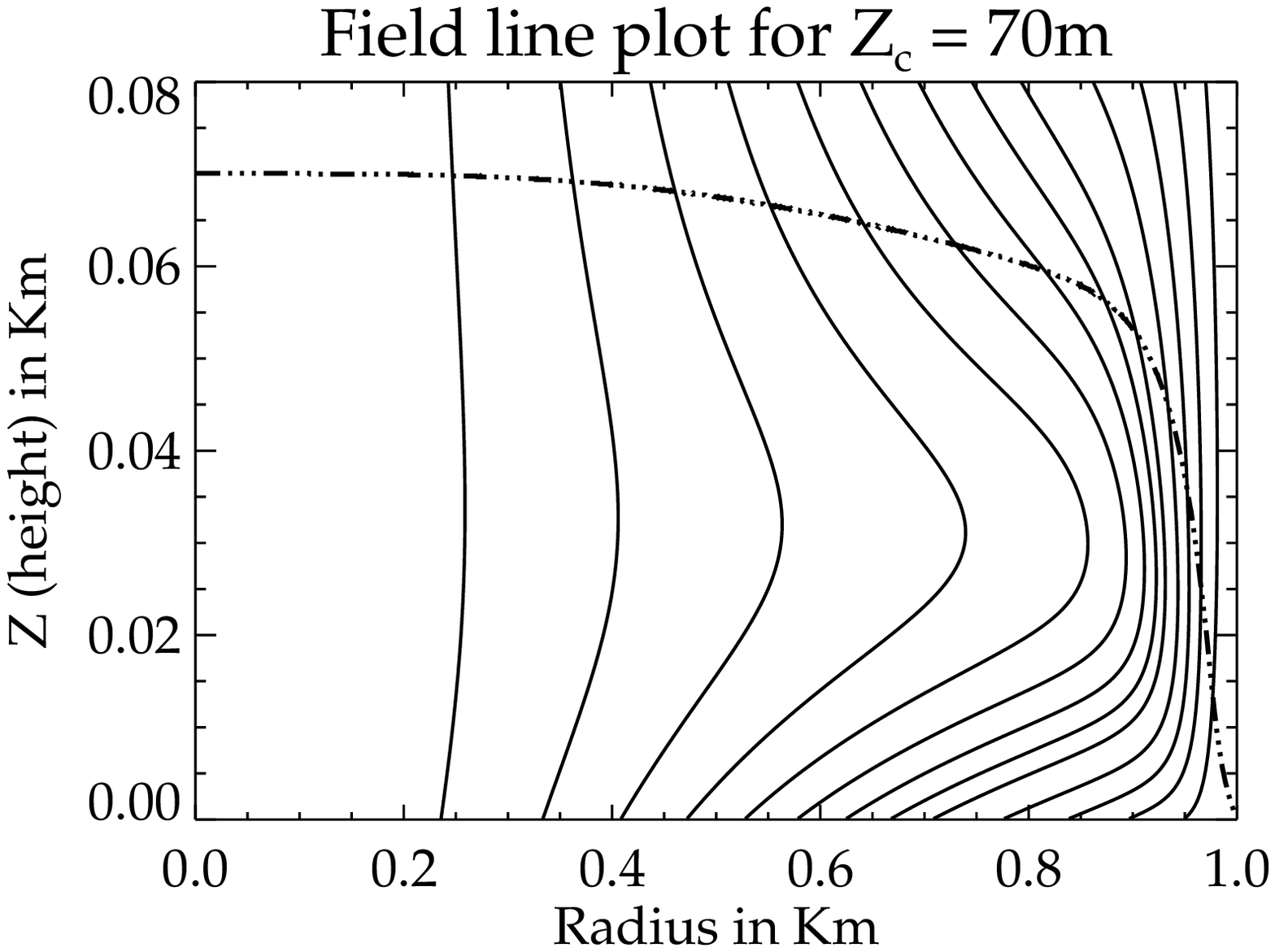}
\caption{Calculation of field distortion at the polar cap of an accreting neutron star: reproduction of Fig.~3 of \citet{MukhBhatt_12},
visulizing model calculations for two accretion mounds of different heights (55\,m and 70\,m) and mass loading (0.9 and 2.31~$10^{-12}$
$M_{\odot}$), respectively.}
\label{fig:MB12_fig3}
\end{figure}

It has been correctly argued that the magnetic field strength determined from the CRSF must not be identical to the polar dipole
field: E$_{cyc}$ measures the local field strength at the place where the resonant scattering occurs which produces the line feature, but
this place could be at a significant height above the neutron star surface, where the field is weaker. 
A possible explanation for ratios smaller than one in the final column of Table~\ref{tab:estimates} 
is a possible enhancement of the polar magnetic field by a change in the field structure due to the accumulated material, 
as suggested by \citealt{MukhBhatt_12,Mukherjee_etal13b,Mukherjee_etal14}: field lines are pushed out radially from the center of the accretion
mound, resulting in bunching of field lines and increasing the local field strength in the outer parts of the accretion column, which is the likely 
place where the cyclotron line is generated. See also calculations of field ``ballooning'' by \citet{PayneMelatos_07, Litwin_etal01}) and the
discussion of the long-term change of E$_{cyc}$ (Sect.~\ref{sec:long_term}). \citet{GhoshLamb_79b} had noted that the E$_{cyc}$ measured
in Her~X-1 indicates a stronger field than obtained from applying the accretion torque model and suggested that the field structure could 
contain ``higher multipole moments" with stronger magnetic fields than the global dipole field.\footnote{A deviation from a pure dipole structure 
is in the literature often referred to as contribution from ``multipole components''.} 
\citet{MukhBhatt_12} showed that for specific geometries the field strength could be increased by up to a factor 4.6. In Fig.~\ref{fig:MB12_fig3}
we reproduce Fig.~3 from \citet{MukhBhatt_12}, showing calculated field distortions for two accretion mounds with different height and
mass loading.

However, we point out that differences by a factor of a few are not necessarily due to physical effects. Torque model 
calculations make assumptions about the mass and radius of the neutron star (which also determine the moment of inertia and 
the gravitational redshift), changes of which, individually or in combination can change the determined magnetic field strength by
several tens of percent.\footnote{For example: assume M = 1.3~$M_{\odot}$, a change of $R_{NS}$ from 10\,km to 12\,km results in a 
reduction of the B estimate in Her~X-1 by $\sim$40\%, or increasing M to 1.5~$M_{\odot}$ combined with reducing R to 9\,km 
will produce an increase in B by a similar amount.} In addition, some of the observed values may have rather large uncertainties,
especially the luminosity which scales with the square of the assumed distance to the source. With these observational uncertainties an
accretion torque model will not yield a unique solution for the magnetic moment $\mu$ or magnetic field strength B (note for this conversion 
the radius $R_{NS}$ enters again with the third power), but rather an allowed range that may cover almost an order of magnitude.

\begin{figure}
   \includegraphics[width=0.47\textwidth]{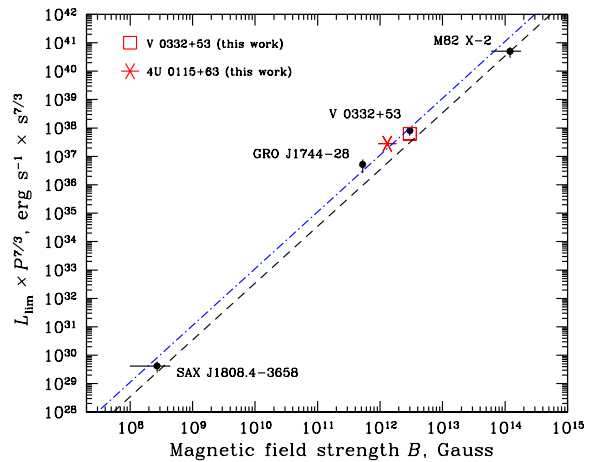}
\vspace{-14mm}
   \caption{``Propeller effect''  (reproduced Fig.~4 of \citealt{Tsygankov_etal16a}): observations of five sources 
   confirm the expected relationship L$_{lim}$~P$^{7/3}$~$\propto$ B,  with L$_{lim}$ being the limiting luminosity 
   at which this effect sets in. The dashed lines are for two slightly different values of $\xi$ in Eqn.~\ref{eqn:Llim}.}
   \label{fig:propeller}
\end{figure}

We also note, quite fundamentally, that accretion torque models generally assume an aligned rotator, while the spinning neutron stars 
in accreting binary pulsars are surely not aligned rotators. Unfortunately,  3D-calculations of magnetospheres, as done for T~Tauri stars 
or CVs \citep[see e.g.,][]{Romanova_etal08,KulkarniRomanova_13}, have so far not been done for the geometrically much larger systems 
of neutron stars with terra-Gauss fields.

The above considerations make apparent that one should not take the observed cyclotron line energy and the derived field strength as a 
measure of the global dipole field and to calculate other system parameters like the distance -- or even basic neutron star parameters like 
mass and radius -- by applying an accretion torque model (as has been done in the literature, for example \citealt{Takagi_etal16}). 

For completeness, we mention a third method to estimate the magnetic moment $\mu$, based on flux and timing observations: when the mass accretion 
rate becomes very low, the magnetospheric radius increases, reaching and exceeding the co-rotation radius. Then the so called propeller effect  
\citep{IllarionovSunyaev_75,Stella_etal86,Christodoulou_etal17} sets in, by which material at the inner edge of the accretion disk is thrown out 
by the rotating magnetic field (often referred to as ``centrifugal barrier''), efficiently inhibiting accretion and leading to a dramatic drop 
(approximately two orders of magnitude) in X-ray luminosity. The onset of the ``propeller effect'' starts at a limiting accretion rate which can be calculated 
by equating the co-rotation radius R$_{c}$ = $(GMP^{2}/4\pi^{2})^{1/3}$ (where the angular velocity of the magnetosphere equals that of the 
material in the Kepler orbit at the inner edge of the accretion disk) with the magnetospheric radius R$_{m}$ = k~$\dot M^{-2/7}~\mu^{4/7}~(2GM)^{-1/7}$.
G is the gravitational constant, P the spin period, M the mass of the NS, $\dot M$ the accretion rate, $\mu$ the magnetic
moment of the NS and k a numerical factor (usually 0.5). The limiting accretion rate $\dot M_{lim}$ determines a limiting luminosity L$_{lim}$
\citep{Campana_etal02,Fuerst_etal17}:
\begin{equation}
\label{eqn:Llim}
L_{lim} = \frac{GM\dot M_{lim}}{R}~\simeq~3.9 \times10^{37}~\xi^{7/2}~B_{12}^2~P^{7/3}~M^{-2/3}~R_{6}^5~[erg/s]
\end{equation}
with quantities as defined above, plus R$_{6}$ being the NS radius in units of $10^{6}$\,cm, B$_{12}$ being the polar
magnetic field strength in units of $10^{12}$\,Gauss  (related to the magnetic moment $\mu$ = 0.5~BR$^{3}$), $\xi$ being a 
numerical factor on the order of one. \citet{Tsygankov_etal16a,Tsygankov_etal16c} give an impressive example of five objects
showing the ``propeller effect'', which demonstrates that the relation L$_{lim}$$\times$P$^{7/3}$~$\propto$ B holds for 
$\sim$7\,orders of magnitude in B (see Fig.~\ref{fig:propeller}).
This method, however, does not have the power to resolve the above discussed differences between B-estimates through torque 
models and CRSF energies, since the systematic uncertainties are of similar magnitude.

\section{Statistical analysis}
\label{sec:statistics}

Here we summarize some statistics regarding electron CRSF sources, numbers of sources showing specific characteristics.
\begin{itemize}
\item $\sim$350 known X-ray binary pulsars (XRBP)
        (e.g., \citealt{Bildsten_etal97}, or see footnotes\footnote{http://www.iasf.inaf.it/\~mauro/pulsar\_list.html\#CRSF} and 
        \footnote{http://www.sternwarte.uni-erlangen.de/wiki/doku.php?id=xrp:start}). 
\item 36 cyclotron line sources ($\sim$10\% of all XRBP) (see Table~\ref{tab:collection}): 
         13 persistent High Mass X-ray Binaries (HMXB), 19 Be transients, 4 Low Mass X-ray Binaries (LMXB), 
         plus $\sim$16 candidates) (see Table~\ref{tab:candidates})
\item 11 sources with multiple cyclotron lines (see Table~\ref{tab:collection}): \\
         7 sources with 2 lines; 3 sources with 3 lines (V~0332+53, MAXI~J1409-619, GRO~J1744-28), 1 source with more than 
         3 lines (4U~0115+63). 
\item 7 sources with positive correlation of E$_{\rm cyc}$ with L$_{\rm x}$ (see Table~\ref{tab:variation}): 
         Her~X-1, GX~304-1, Vela~X-1, Cep~X-4, V~0332+53, A~0535+26, 4U~1626.6-5156 (plus  4U~1907+09, to be confirmed). 
\item 2 sources with negative correlation of E$_{\rm cyc}$ with L$_{\rm x}$ (see Table~\ref{tab:variation}): 
         V~0332+53 (at very high L$_{\rm x}$) and SMC ~X-2. 
\item 2 sources with long-term variation of E$_{\rm cyc}$ (see Sect.~\ref{sec:long_term}): 
         Her~X-1 \citep{Staubert_etal17}, Vela~X-1 \citep{LaParola_etal16}, (plus one source with intermediate-term variation: 
         V~0332+53 \citep{Cusumano_etal16}, plus possibly 4U~1538-522 \citep{Hemphill_etal14}).
\item 18 sources with correlation of photon index with L$_{\rm x}$ (see Table~\ref{tab:variation}).
\end{itemize}

\section{Proton cyclotron lines}
\label{sec:proton_lines}

\begin{figure}
\includegraphics[angle=-90,width=0.45\textwidth]{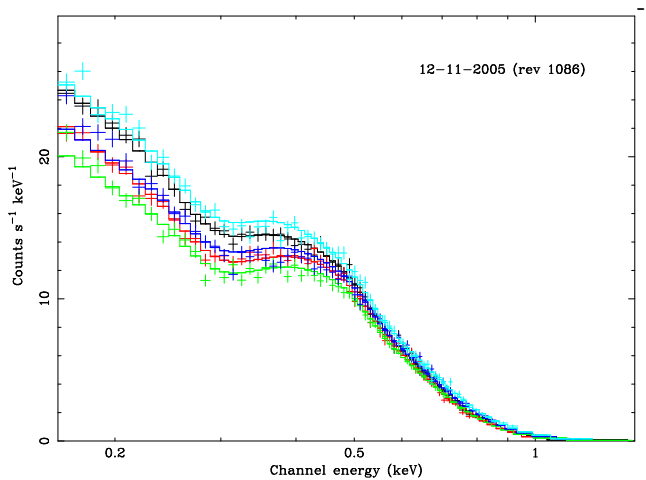}-
\vspace{4mm}
\caption{Pulse phase resolved EPIC-pn spectra of RX~J0720.4-3125 observed by \textsl{XMM-Newton}: a reproduction 
of Fig.~5 of \citet{2007Ap&SS.308..181H}, showing an absorption feature around 300\,eV, which is interpreted as a proton cyclotron 
line produced in a magnetic field of about $5.6 \times10^{13}$\,Gauss. The different colors give the spectra for different 
pulse phase intervals: 0.0-0.2: black, 0.2-0.4: red, 0.4-0.6: green, 0.6-0.8: blue, and 0.8-1.0: light blue.}
\label{fig:p-CRSF}
\end{figure}

The resonances produced by protons gyrating in a strong magnetic field are lower in energy
compared with those of electrons by a factor given by their mass ratio m$_{\rm e}$/m$_{\rm p}$. 
Absorption lines seen by Chandra, XMM-\textsl{Newton} in the 0.2 $-$ 10 keV band require very strong magnetic 
fields $\gtrapprox 10^{14}$ G. In contrast to electron cyclotron lines, those produced by protons (or nuclei) appear 
only at the fundamental frequency \citep{2010A&A...518A..24P}. The problem with proton line observations is that 
the absorption lines may be explained as well by other effects \citep{2015SSRv..191..171P}, 
for example by photo-ionisation in a dense cloud in the vicinity of the neutron star \citep{2009A&A...497L...9H}, 
or as the result of a complex inhomogeneous temperature distribution on the surface of the neutron star 
\citep{2014MNRAS.443...31V}. 
The discovery of possible proton cyclotron lines has been reported for a number of objects (including ULXs and magnetars,
which we do not discuss in detail here because of the problems of the just described confusion). We only mention the
magnetar SGR\,1806$-$20 \citep{Ibrahim_etal02,Ibrahim_etal03}
for which the magnetic field strength estimated from P and $\dot {\rm P}$ is consistent with that inferred from an
absorption line seen at $\sim$5\,keV, assuming that it is a proton cyclotron line.
The criterion also holds for a class of seven isolated neutron stars which were discovered in ROSAT data. 
They are bright X-ray sources but given their relatively small distances of typically a few hundred parsec they are 
intrinsically X-ray dim. Hence they are often called X-ray dim isolated neutron stars (XDINS) or simply the 
Magnificent Seven (M7). At least five of the seven objects exhibit absorption line features in the 
Chandra/XMM-\textsl{Newton} energy band (0.2-10\,keV). When interpreted as proton cyclotron lines their 
corresponding magnetic field strengths agree with those derived from the measured P, $\dot {\rm P}$ data within a 
factor of a few \citep{2007Ap&SS.308..181H}. 
In the following an update concerning these sources is given. 
In Sect.\,\ref{m7-intro} we introduce the M7 stars and summarize their timing properties which help to constrain 
the large scale structure of their magnetic fields. 
The observed absorption features in their X-ray spectra are described in Sects.\,\ref{m7-abs} 
and Sect.\,\ref{m7-discus} the evidence for them being proton cyclotron lines is discussed, 
followed by a short summary. 
Fig.~\ref{fig:p-CRSF} shows an example of pulse phase resolved spectra of RX~J0720.4-3125, with an
absorption feature around 300\,eV, as observed by \textsl{XMM-Newton} \citep{2007Ap&SS.308..181H}.

\begin{table*}
\centering
\caption{The Magnificent Seven}
\begin{tabular}{lcccccccccc}
\hline\hline\noalign{\smallskip}
\multicolumn{1}{l}{Object} &
\multicolumn{1}{c}{kT$_\infty$} &
\multicolumn{1}{c}{P} &
\multicolumn{1}{c}{p.f.\tablefootmark{a}} &
\multicolumn{1}{c}{$\dot{\rm P}$} &
\multicolumn{1}{c}{B$_{\rm dip}$} &
\multicolumn{1}{c}{$\tau$} &
\multicolumn{1}{c}{t$_{\rm kin}$} &
\multicolumn{1}{c}{m$_{\rm B}$\tablefootmark{b}} &
\multicolumn{1}{c}{d\tablefootmark{c}} &
\multicolumn{1}{c}{Ref.\tablefootmark{d}} \\
\multicolumn{1}{l}{ } &
\multicolumn{1}{c}{(eV)} &
\multicolumn{1}{c}{(s)}  & 
\multicolumn{1}{c}{(\%)} &
\multicolumn{1}{c}{(s\,s$^{-1}$)} &
\multicolumn{1}{c}{($10^{13}$\,G)} &
\multicolumn{1}{c}{(Myr)} &
\multicolumn{1}{c}{(Myr)} &
\multicolumn{1}{c}{(mag)} &
\multicolumn{1}{c}{(pc)} &
\multicolumn{1}{c}{} \\ 
\noalign{\smallskip}\hline\noalign{\smallskip}
 \object{RX\,J0420.0$-$5022} & 48      & 3.45  & 17     & $-$2.8$\times10^{-14}$  & 1.0   & 1.95 & ?              & 26.6     & $\sim$345          & (1) \\
 \object{RX\,J0720.4$-$3125} & 84$-$94 & 16.78 & 8$-$15 & $-$1.40$\times10^{-13}$ & 5.0   & 1.91 & 0.85           & 26.6     & 286$^{+27}_{-23}$  & (2) \\
 \object{RX\,J0806.4$-$4123} & 95      & 11.37 & 6      & $-$5.50$\times10^{-14}$ & 2.5   & 3.24 & ?              & $>$24    & $\sim$250          & (3) \\
 \object{RX\,J1308.6+2127}   & 100     & 10.31 & 18     & $-$1.12$\times10^{-13}$ & 3.5   & 1.45 & 0.55/0.90/1.38 & 28.4     &       ?            & (4) \\
 \object{RX\,J1605.3+3249}   & 100     & ?     & $<$2   & ?                 & ?     & ?    & 0.45           & 27.2     & $\sim$390          & (5) \\
 \object{RX\,J1856.5$-$3754} & 61      & 7.06  & 1      & $-$2.97$\times10^{-14}$ & 1.5   & 3.80 & 0.42$-$0.46    & 25.2     & 120$^{+11}_{-15}$  & (6) \\
 \object{RX\,J2143.0+0654}   & 104     & 9.43  & 4      & $-$4.00$\times10^{-14}$ & 1.9   & 3.72 & ?              & $>$26    & $\sim$430          & (7) \\
\noalign{\smallskip}\hline                                                                                        
\end{tabular}
\tablefoot{Values in the table were taken from \citet{2007Ap&SS.308..181H} and \citet{2014A&A...563A..50P} and updated 
when more recent results were available.\\
\tablefoottext{a}{Pulsed fraction.}\\ 
\tablefoottext{b}{Optical magnitudes are taken from \citet{2008AIPC..968..129K}.}\\
\tablefoottext{c}{Distance estimates indicated by $\sim$ are from \citet{2007Ap&SS.308..171P}.}\\
\tablefoottext{d}{References: 
(1) \citet{1999A&A...351L..53H,2004A&A...424..635H,2011ApJ...740L..30K}
(2) \citet{1997A&A...326..662H,2001A&A...365L.302C,2002MNRAS.334..345Z,2004A&A...415L..31D,2004A&A...419.1077H,2005ApJ...628L..45K,2011MNRAS.417..617T,2012MNRAS.423.1194H,2015ApJ...807L..20B,2017A&A...601A.108H}
(3) \citet{1998AN....319...97H,2004A&A...424..635H,2009ApJ...705..798K}
(4) \citet{1999A&A...341L..51S,2003A&A...403L..19H,2005ApJ...635L..65K,2007Ap&SS.308..619S,2010MNRAS.402.2369T,2011A&A...534A..74H,2017MNRAS.468.2975B}
(5) \citet{1999A&A...351..177M,2004ApJ...608..432V,2009A&A...497..423M,2012PASA...29...98T,2014A&A...563A..50P,2017xru.Pires}
(6) \citet{1996Natur.379..233W,2001ApJ...549..433W,2001A&A...379L..35B,2003A&A...399.1109B,2007ApJ...657L.101T,2010ApJ...724..669W,2011MNRAS.417..617T,2013MNRAS.429.3517M}
(7) \citet{2001A&A...378L...5Z,2005ApJ...627..397Z,2009ApJ...692L..62K}
}}
\label{tab_M7}
\end{table*}

\subsection{The Magnificent Seven}
\label{m7-intro}

RX\,J1856.5$-$3754 was the first isolated neutron star (INS) discovered in ROSAT data 
\citep{1996Natur.379..233W}. Despite extensive searches only six further objects with similar properties, 
as summarized in Table~\ref{tab_M7}, were found in the ROSAT data 
\citep{2003ApJ...598..458R,2005A&A...444...69C,2008A&A...482..617P,2011AJ....141..176A}.
The X-ray emission of these seven objects is characterized by
soft, blackbody-like emission attenuated by low photo-electric absorption by the interstellar medium, 
but in at least five cases with broad absorption features. The X-ray spectra show no indication of
hard, non-thermal X-ray emission components which could originate from magneto-spheric activity.
No flux variations on timescales of years are observed (but see RX\,J0720.4$-$3125 below), which suggests 
that we see thermal emission directly from the surface of cooling isolated neutron stars. 
Their distances range from $\sim$120\,pc for RX\,J1856.5$-$3754 determined from parallax measurements 
\citep{2010ApJ...724..669W} and several hundred pc constrained from absorption column densities 
\citep{2007Ap&SS.308..171P}.

At least six of the M7 stars are X-ray pulsars with spin periods between 3.45\,s (RX\,J0420.0$-$5022) and 16.78\,s 
(RX\,J0720.4$-$3125). The modulation in the X-ray flux varies strongly from star to star with 
semi-amplitudes between 1\% (RX\,J1856.5$-$3754) and 18\% (RX\,J1308.6+2127). The latter shows
a clear double-peaked pulse profile \citep{2003A&A...403L..19H}, indicating a spin period of the NS
twice as long as originally thought. The others are characterized by more gradual sinusoidal variations. 
\citet{2017A&A...601A.108H} recently reported a double-hump pulse profile also 
for RX\,J0720.4$-$3125 (similar to RX\,J1308.6+2127), suggesting a spin period of 16.78\,s.
Spin down is observed from the M7 stars with known spin periods 
using a series of Chandra and XMM-\textsl{Newton} observations, with accurate spin-down 
rates obtained from coherent timing solutions. Assuming the model of magnetic dipole breaking 
(see also Sect.~\ref{sec:strength}),
the knowledge of the spin period P and its derivative $\dot {\rm P}$ allows one to estimate the magnetic field strength 
(B = 3.2$\times10^{19}({{\rm P}\dot{\rm P}})^{1/2}$; 
with B in Gauss, P in s and $\dot{\rm P}$ in s\,s$^{-1}$), 
which yields values between 
1$\times10^{13}$\,Gauss (RX\,J0420.0$-$5022) and 
5.0$\times10^{13}$\,Gauss (RX\,J0720.4$-$3125). The characteristic ages estimated 
from the dipole model ($\tau$ = P/(2$\dot{\rm P}$)) of typically a few million years are a factor of a few 
longer than kinetic ages derived from back tracing the proper motion of the stars to likely birth places 
\citep{2009A&A...497..423M,2010MNRAS.402.2369T,2011MNRAS.417..617T,2012PASA...29...98T}.

{\bf RX\,J0420.0$-$5022} was discovered serendipitously in ROSAT data \citep{1999A&A...351L..53H} as X-ray 
source with a soft blackbody-like spectrum and no optical counterpart (m$_{\rm B}$ $>$ 25.25 mag). 
XMM-\textsl{Newton} observations with the EPIC-pn instrument revealed pulsations with a period of 3.45\,s, the shortest confirmed spin 
period among the M7 stars \citep{2004A&A...424..635H}. Using 14 observations with 
XMM-\textsl{Newton} \citet{2011ApJ...740L..30K} 
obtained a phase-coherent timing solution and measured a spin-down rate of 
$\dot{\rm P}$ = (2.8$\pm$0.3)$\times10^{-14}$ s\,s$^{-1}$, which yields the lowest dipolar magnetic field strength of 
1.0$\times10^{13}$ G.

{\bf RX\,J0720.4$-$3125} is the second brightest M7 star discovered in ROSAT all-sky survey data and 
pulsations with a period of 8.39\,s were indicated in the data of follow-up ROSAT PSPC and HRI pointed 
observations \citep{1997A&A...326..662H}. However, from a re-analysis of all available 
XMM-\textsl{Newton} high statistical quality data, \citet{2017A&A...601A.108H} concluded that the true spin period is 16.74\,s. The pulse 
profile folded at this period shows two humps with similar, but 
still clearly distinguishable shape. The object is unique as the only M7 star 
showing long-term variations in the X-ray spectrum on timescales of years \citep{2004A&A...415L..31D}.
A final analysis of monitoring observations with the EPIC-pn instrument of 
XMM-\textsl{Newton} spanning more than 11 years 
revealed a temperature increase from kT = 85\,eV to 94\,eV within about 1.5\,yr. After reaching the maximum 
temperature, the spectra indicate a very slow cooling by $\sim$2\,eV over 7\,yr \citep{2012MNRAS.423.1194H}.

{\bf RX\,J0806.4$-$4123} was discovered in a dedicated search for INS in the ROSAT all-sky survey 
\citep{1998AN....319...97H}. The spin period of 11.37\,s was detected in 
XMM-\textsl{Newton} data \citep{2002A&A...391..571H} 
and the period derivative could be constrained from a series of follow-up observations, although with large uncertainties 
\citep{2009ApJ...705..798K}.

{\bf RX\,J1308.6+2127 = RBS\,1223} was identified as source with empty error circle in the ROSAT Bright Survey (RBS) 
catalog \citep{1999A&A...341L..51S,2000AN....321....1S}. Follow-up Chandra observations first suggested a period of 
5.16\,s \citep{2002A&A...381...98H} before 
XMM-\textsl{Newton} observations provided data of high statistical quality which 
revealed a double-peak pulse profile with true period of 10.31\,s \citep{2003A&A...403L..19H}. 
A coherent timing solution was obtained from Chandra and 
XMM-\textsl{Newton} observations spanning a period of five years, yielding accurate 
values for spin period and derivative \citep{2005ApJ...635L..65K}.

{\bf RX\,J1605.3+3249 = RBS\,1556} was selected as INS from the ROSAT all-sky survey on the basis of its spectral 
softness and lack of bright optical counterpart \citep{1999A&A...351..177M}. An indication for a spin period (3.39\,s) was 
reported at a 4$\sigma$ confidence level together with uncertain constraints for the pulsar spin-down rate 
\citep{2014A&A...563A..50P}. However, using recent deep 
XMM-\textsl{Newton} observations \citet{2017xru.Pires} could not confirm the 
period with upper limits for the pulsed fraction of $\sim$2.6\%.

{\bf RX\,J1856.5$-$3754} is the X-ray brightest M7 star and was first found in a ROSAT pointed PSPC observation as 
brightest source in the field of view \citep{1996Natur.379..233W}. Together with the ROSAT all-sky survey detection and a 
ROSAT HRI observation no evidence for variability was found and the limit for F$\rm x$/F$_{opt}$ of $>$7000 strongly 
suggested an INS nature of the object. This first discovery triggered the search for other sources in the ROSAT all-sky 
survey with similar properties. The ROSAT PSPC spectrum could be modeled by pure blackbody emission and also the 
high-resolution Chandra LETG spectra did not reveal any significant deviations from a Planckian energy distribution 
\citep{2001A&A...379L..35B,2003A&A...399.1109B}. Although being the brightest M7 star, no absorption feature similar 
to those found in other M7 stars was significantly detected so far. A very shallow modulation ($\sim$1\%) with a period 
of 7.06\,s was discovered in 
XMM-\textsl{Newton} EPIC-pn data \citep{2007ApJ...657L.101T} and constraints on the period derivative 
indicate a magnetic field strength comparable to what was found for the other M7 \citep{2008ApJ...673L.163V}. 

{\bf RX\,J2143.0+0654 = 1RXS\,J214303.7+065419 = RBS\,1774} was the last of the M7 discovered in ROSAT data 
\citep{2001A&A...378L...5Z}. Again 
XMM-\textsl{Newton} observations led to the discovery of X-ray pulsations (period 9.437\,s) and an 
absorption feature at 700--750\,eV \citep{2005ApJ...627..397Z,2007Ap&SS.308..161C}. From an additional series of 
XMM-\textsl{Newton} observations \citet{2009ApJ...692L..62K} measured the spin-down rate, which indicates a magnetic dipole field strength 
of 
$\sim$2$\times10^{13}$\,G.

\begin{table*}[t]
\centering
\caption{The Magnificent Seven -- cyclotron lines }
\begin{tabular}{lcccccc}
\hline\hline\noalign{\smallskip}
\multicolumn{1}{l}{Object} &
\multicolumn{3}{c}{----------------------- Broad lines -----------------------} &
\multicolumn{3}{c}{--------- Narrow lines ---------} \\
\multicolumn{1}{l}{} &
\multicolumn{1}{c}{E$_{\rm cyc}$} &
\multicolumn{1}{c}{eqw$_{\rm cyc}$\tablefootmark{a}} &
\multicolumn{1}{c}{B$_{\rm cyc}$} &
\multicolumn{1}{c}{E$_{\rm cyc}$} &
\multicolumn{1}{c}{eqw$_{\rm cyc}$} &
\multicolumn{1}{c}{B$_{\rm cyc}$} \\
\multicolumn{1}{l}{} &
\multicolumn{1}{c}{(eV)} &
\multicolumn{1}{c}{(eV)} &
\multicolumn{1}{c}{($10^{13}$\,G)} &
\multicolumn{1}{c}{(eV)} &
\multicolumn{1}{c}{(eV)} &
\multicolumn{1}{c}{($10^{13}$\,G)} \\ 
\noalign{\smallskip}\hline\noalign{\smallskip}
 \object{RX\,J0420.0$-$5022}\tablefootmark{b} & 329:\tablefootmark{c}      & $-$43:       & 6.5:                           & & & \\
 \object{RX\,J0720.4$-$3125} & 300                           & $-$10 to $-$75 ($\pm$20) & 6.0                            & 750 & $\sim$ $-$30 & 14.9 \\
 \object{RX\,J0806.4$-$4123} & 410-490                       & $\sim$ $-$45             & 8.1-9.7                        & & & \\
 \object{RX\,J1308.6+2127}   & 300                           & $-$150 ($\pm$20)         & 6.0                            & 750 & $\sim$ $-$15 & 14.9 \\
 \object{RX\,J1605.3+3249}   & 443/828                       & $-$31/$-$13              & 8.8/16.4                       & & & \\
 \object{RX\,J1856.5$-$3754} & $\lesssim$250 or $\gtrsim$800 &  ?                       & $\lesssim$5.0 or $\gtrsim$15.9 & & & \\
 \object{RX\,J2143.0+0654}   & 750                           & $-$25                    & 14.9                           & & & \\
\noalign{\smallskip}\hline                                                                                        
\end{tabular}
\tablefoot{B is calculated as B/(10$^{13}$\,G) = (1+z)E/(63\,eV) with  
           (1 + z) = (1 - 2GM/Rc$^2$)$^{-1/2}$ assumed as 1.25 \citep[e.g.,][]{2001ApJ...560..384Z}.\\
\tablefoottext{a}{Equivalent width of broad absorption line. Numbers in parentheses indicate variations with pulse phase.}\\
\tablefoottext{b}{Line parameters subject to possible large systematic uncertainties (see text).} \\
\tablefoottext{c}{The colon indicates that the parameters are subject to  larger calibration uncertainties (see text).}}
\label{tab_M7lines}
\end{table*}

\subsection{Proton absorption features}
\label{m7-abs}

The ROSAT PSPC spectra of the M7 were modeled with blackbody emission attenuated by small amounts of absorption, 
which is interpreted to be induced by the interstellar medium. The energy resolution of the PSPC was not sufficient to 
discern additional absorption features. Moreover, the entrance window of the detector caused a deep carbon absorption 
edge with strong depression of the effective area between 300\,eV and 500\,eV, the energy band in which features 
were later found in the 
XMM-\textsl{Newton} and Chandra spectra. The line properties are described in the following and summarized in 
Table~\ref{tab_M7lines}.

The first absorption feature was discovered by \citet{2003A&A...403L..19H} in data collected from RX\,J1308.6+2127 by 
the EPIC-pn instrument of XMM-\textsl{Newton}.
The X-ray spectra show strong deviations from a Planckian energy distribution at 
energies below 500\,eV which can be modeled by a Gaussian-shaped  absorption line centered at an energy of 
$\sim$300\,eV and with $\sigma$ width of $\sim$100\,eV, although even broader lines at energies down to 100\,eV or 
several unresolved lines can not be excluded. The depth of the line relative to the continuum (the equivalent width, eqw) 
was measured to $-$150\,eV (we use negative values to indicate that the line is in absorption), the deepest absorption 
line seen from the M7. Moreover, pulse-phase spectroscopy revealed the line to vary in strength with a full amplitude of 
$\sim$40\,eV in equivalent width \citep{2005A&A...441..597S}. The line is deepest at the intensity maxima of the 
double-peaked pulse profile and shallowest at the minima.  

A similar behavior was found for RX\,J0720.4$-$3125. \citet{2004A&A...419.1077H} reported a phase-dependent 
absorption line in the spectrum which varies in depth with pulse phase. This line is broad ($\sim$64\,eV 1$\sigma$ width 
when modeled with a Gaussian profile) and centered at $\sim$300\,eV \citep{2004A&A...419.1077H,2006A&A...451L..17H}. 
The equivalent width varies with pulse phase from $-$31\,eV around intensity maximum and $-$58 eV at the declining part 
of the pulse. While the line is much weaker on average, the full amplitude variation with pulse phase is almost as large as 
for RX\,J1308.6+2127. Additionally, the line depth was found to change on timescales of years. Parallel to the increase 
in temperature with a maximum seen in May 2004 the line depth also increased from eqw = $-$40 eV to $-$74\,eV in 
the phase-averaged spectra \citep{2006A&A...451L..17H,2012MNRAS.423.1194H}. It is remarkable that the amplitude 
in the variations with pulse phase stayed constant at $\sim$40\,eV \citep{2009A&A...498..811H}.

While for RX\,J1308.6+2127 and RX\,J0720.4$-$3125 the variability of the absorption lines with pulse-phase excludes 
all possible doubt that they are real spectral features, the detection of a broad absorption line in the high resolution 
X-ray spectra of RX\,J1605.3+3249 obtained by the RGS instruments of 
XMM-\textsl{Newton} constrains the width of the line. 
Relative to CCD detectors, the superior energy resolution of the RGS revealed that the deviation from a blackbody 
spectrum can be well modeled with a single broad absorption line centered at 450\,eV with a 1$\sigma$ width of 
59\,eV \citep{2004ApJ...608..432V}. On the other hand, the high statistical quality of the EPIC-pn spectra of 
RX\,J1605.3+3249 revealed strong evidence for the existence of multiple lines. An acceptable fit was only reached 
when including three Gaussian absorption lines at energies of E$_1$ = 403$\pm$2\,eV, E$_2$ = 589$\pm$4\,eV and 
E$_3$ = 780$\pm$24 \citep{2007Ap&SS.308..181H}. It is remarkable that the ratios of the line energies are consistent 
with 1:1.5:2. The existing data at the time of that work did not allow the fitting of the width of the lines individually and a single 
common fit parameter was derived with $\sigma$ of 87\,eV. Including the spectra from new 
XMM-\textsl{Newton} observations required 
an even more complex modeling with two blackbody components for the continuum (a cool one with kT around 30\,eV 
and the dominant hot one with kT $\sim$110\,eV) and three absorption lines \citep{2014A&A...563A..50P}.
From RGS spectra these authors determined 
E$_1$ = 443$^{+13}_{-20}$\,eV ($\sigma_1$ = 74$^{+13}_{-11}$\,eV and eqw = $-$31\,eV) and
E$_2$ = 828$\pm$5 eV ($\sigma_2$ = 15 +/-4 eV and eqw $-$13 eV).
A somewhat narrower line with central energy of E = 576$\pm$8\,eV and Gaussian width $\sigma$ = 16$^{+7}_{-5}$\,eV 
(eqw = $-$5\,eV) was already reported by \citet{2004ApJ...608..432V} and might be explained by the presence of highly 
ionized oxygen (\ion{O}{VII}) in the interstellar medium and/or absorption in the neutron star atmosphere by \ion{O}{VIII}, 
gravitationally red-shifted \citep{2012MNRAS.419.1525H}. A narrow line at similar energy (569\,eV) was also reported 
from RX\,J0720.4$-$3125 \citep{2009A&A...497L...9H,2012MNRAS.419.1525H} and at 535\,eV from RX\,J1308.6+2127 
\citep{2012MNRAS.419.1525H}. The new RGS results for the lines at 576\,eV and 828\,eV makes it less clear if 
they are related to the broad absorption feature at 440\,eV.

The properties of absorption features in the spectra of RX\,J2143.0+0654 and RX\,J0806.4$-$4123 are less clear. 
A feature found in the first 
XMM-\textsl{Newton} observation of RX\,J2143.0+0654 was reported by \citet{2005ApJ...627..397Z} at 
an energy of 700\,eV (when modeled with an absorption edge) or 750\,eV (multiplicative Gaussian line). New 
XMM-\textsl{Newton} observations with a factor of 2.5 more exposure confirmed this and \citet{2009ApJ...692L..62K} inferred a 
significant improvement in fit quality when including a multiplicative Gaussian line with a width of 73\,eV, centered at 
756\,eV in the spectral model. This is the highest energy measured for an M7 star and yields a magnetic field strength 
more than a factor of seven higher than the field strength inferred from the dipole model, while for the other M7 stars this 
factor is more typical 1.2 to 4.
Also for RX\,J0806.4$-$4123 the spectral modeling is formally improved when considering an absorption line. Fixing 
the $\sigma$ width of the line at 70\,eV, \citet{2004A&A...424..635H} derive line parameters of 460$\pm$15\,eV for the 
central line energy and eqw = $-$33$\pm$6\,eV from the merged EPIC-pn spectrum of two 
XMM-\textsl{Newton} observations. For the 
corresponding RGS spectra the authors report 413$\pm$19\,eV and $-$56$\pm$13\,eV, respectively. With the errors at 
90\% confidence level, the parameters are formally inconsistent and indicate large systematic uncertainties. 
Moreover, \citet{2009ApJ...705..798K} report a line energy of 486$\pm$5\,eV from EPIC-pn observations performed in 
2008 and 2009 (a factor of four more exposure than available from the older observations). 
These authors also find further improvement in fit quality when using two Gaussian lines 
(E$_1$ = 460$\pm$5\,eV, eqw$_1$ = $-$89\,eV, E$_2$ = 693$\pm$12\,eV, eqw$_2$ = $-$55\,eV). It should be noted 
that they also fix all line widths at the same value (85\,eV) while experience with RX\,J1605.3+3249 shows that they can 
be considerably different. 

RX\,J0420.0$-$5022 is the X-ray faintest of the M7 and exhibits the softest spectrum, which makes it difficult to confirm 
the existence of absorption features due to possible systematic calibration uncertainties caused by insufficient energy 
resolution at low energies ($<$300\,eV) in modern CCD detectors. Formally, the fit to the EPIC-pn phase-averaged 
spectra improves when a Gaussian absorption line is added to the absorbed blackbody model \citep{2004A&A...424..635H}. 
Because the line parameters are subject to larger calibration uncertainties, we list them in Table~\ref{tab_M7lines} with colon.\\

The discovery of a second strongly phase-variable narrow absorption feature at $\sim$750\,eV was recently reported 
for RX\,J0720.4$-$3125 and RX\,J1308.6+2127 \citep{2015ApJ...807L..20B,2017MNRAS.468.2975B}. 
The fact that the features are significantly detected for only $\sim$20\% of the pulsar rotation suggests that they are 
formed very close to the neutron star's surface. 

\subsection{Origin of proton cyclotron lines}
\label{m7-discus}

The absence of non-thermal emission components in the X-ray spectra of the M7, no confirmed radio detection and 
their insufficient spin-down energy losses to power the X-ray emission lead to the generally accepted picture of cooling 
isolated neutron stars. The M7 provide the unique opportunity to obtain information about their magnetic fields in two 
independent ways. One method is based on the measurement of the spin down rate and assumes magnetic dipole 
breaking according to the model of a magnetic dipole rotating in vacuum.
A more direct determination of the magnetic field strength is possible when cyclotron resonance lines can be observed 
in the X-ray spectra. Unlike for accreting X-ray pulsars the absorption features observed from M7 stars can not be 
caused by electrons because the inferred magnetic field strength is by far incompatible with that derived from magnetic 
dipole breaking. For protons the derived field strengths are generally higher but more consistent with the dipole model. 
However: which particles contribute, is also a matter of the composition of the neutron star atmosphere. Moreover, 
bound-bound and bound-free transitions of neutral hydrogen are also expected at similar energies for magnetic fields 
above 10$^{13}$ G \citep[see Fig.\,7 in][]{2007Ap&SS.308..191V}.

However, whether or not transitions of free protons or magnetized hydrogen atoms contribute to the observed lines, is a 
matter of the temperature distribution in the neutron star atmosphere. According to Fig.\,1 in \citet{2003ApJ...599.1293H} 
the atomic fraction of hydrogen at temperatures of kT = 50 $-$ 100 eV typical for the M7 is rather low ($<$=0.3 \%). 
As a consequence the proton cyclotron lines dominate the spectrum \citep[Figs. 2 and 3 of][]{2003ApJ...599.1293H}, 
while atomic hydrogen transitions may produce additional features. That may explain some of the observed multiple 
lines. In view of the limited significance of the line detections it appears difficult to make specific assignments. 

Another way to explain multiple lines is to invoke -- beyond the radiation from the hot cap at the magnetic dipole -- 
separate hot spots on the stellar surface having higher multipolar magnetic fields. The first strong evidence for this comes 
from the discovery of strongly phase-variable absorption lines which are visible only for about 20\% of the neutron 
star rotation. The lines were detected in the X-ray spectra of RX\,J0720.4$-$3125 and RX\,J1308.6+2127 
\citep[][respectively]{2015ApJ...807L..20B,2017MNRAS.468.2975B} and have energies a factor of $\sim$2.5 higher 
than the broad lines (Table~\ref{tab_M7lines}). Such multipole fields must be quite young compared to the age of 
the NS. They may be formed by extracting magnetic energy from the toroidal field that resides in deep crustal 
layers, via Hall drift \citep[e.g.,][]{2014MNRAS.444.3198G}. 


We conclude that proton cyclotron lines produced in strong magnetic fields of isolated neutron 
stars can explain the absorption-type features observed in the soft X-ray spectra of young isolated neutron stars of the 
M7 class.  More accurate measurements in the future in combination with further theoretical work on magnetized 
atmospheres and the magneto-thermal evolution of neutron stars are required to make further progress in this exciting field. 

 \section{Summary}
\label{sec:summary}

With this contribution we attempt to provide a review on the current state of our knowledge about
X-ray sources that show cyclotron line features in their spectra -- both for electron cyclotron lines (this is by far 
the larger part) and for proton cyclotron lines. We concentrate on the observational aspects,
providing a number of Tables that contain detailed informations on all sources that we consider
reasonably well secured regarding the existence of cyclotron lines. In the case of electron cyclotron lines --
observed from binary X-ray pulsars, highly magnetized accreting neutron stars in binaries --
we also give a Table with candidate sources, mostly objects for which a detection of a cyclotron line was
claimed some time ago, but had not been confirmed in subsequent observations, or sources that were
observed rather recently, but only once and with somewhat peculiar properties. On the other hand: for some of
the objects listed in the main Tables, the existence of a cyclotron line still needs confirmation (or not)
through re-measurements (with a finite probability that they have to be moved into the candidates  
category). We try to address the basic physics underlying the cyclotron line phenomenon, 
including a large number of references to theoretical work, with some coverage of specific theoretical modeling.

Since the discovery of the first CRSF in Her~X-1 in 1976, a new field of Cyclotron Line Research has evolved. 
With the rather fast confirmation of the line in Her~X-1 after 1977 by a number of different instruments and the detection 
of the same phenomenon in other X-ray binary pulsars, an impressive wave of theoretical work was initiated, leading 
to a basic understanding of the underlying physics. Observational activities have lead to the discovery of many
more objects until today (at an approximate mean rate of one per year). In the meantime the emphasis has, however, 
shifted from a pure discovery of new cyclotron line objects to the investigation of physical details like the width, depth, 
line profile, phase dependence, etc., of the lines and to new associated phenomena like the dependence of the cyclotron 
line energy on luminosity or its long-term decay (so far found in only two sources). This has been possible through modern
instruments (like \textsl{NuSTAR}), with good energy resolution, broad spectral coverage and low background and
high sensitivity. On the theoretical front, progress has been made in the understanding and modeling of 
the phenomenon, leading for example, to a procedure to fit observed spectra to theoretical spectra that are generated by 
highly complex Monte Carlo modeling, allowing us to find solid physical parameters. Overall, however, our feeling is that 
theoretical work is somewhat lagging behind observations, and that there are many open questions  to be answered.

\begin{acknowledgements}
We thank the anonymous referee for very useful comments that allowed us to improve the text.
\end{acknowledgements}

\newpage
\bibliographystyle{aa}
\bibliography{refs_herx1_Sep2018,general}

\begin{thebibliography}{465}
\expandafter\ifx\csname natexlab\endcsname\relax\def\natexlab#1{#1}\fi

\bibitem[{{Ag{\"u}eros} {et~al.}(2011){Ag{\"u}eros}, {Posselt}, {Anderson},
  {Rosenfield}, {Haberl}, {Homer}, {Margon}, {Newsom}, \&
  {Voges}}]{2011AJ....141..176A}
{Ag{\"u}eros}, M.~A., {Posselt}, B., {Anderson}, S.~F., {et~al.} 2011, \aj,
  141, 176

\bibitem[{{Alford} \& {Halpern}(2016)}]{AlfordHalpern_16}
{Alford}, J.~A.~J. \& {Halpern}, J.~P. 2016, \apj, 818, 122

\bibitem[{{Antoniou} {et~al.}(2018){Antoniou}, {Zezas}, {Hong}, {Kennea},
  {Tomsick}, \& {Haberl}}]{Antoniou_etal18}
{Antoniou}, V., {Zezas}, A., {Hong}, J., {et~al.} 2018, The Astronomer's
  Telegram, 12234

\bibitem[{Araya \& Harding(1999)}]{ArayaHarding_99}
Araya, R. \& Harding, A. 1999, ApJ, 517, 334

\bibitem[{{Araya-G{\'o}chez} \& {Harding}(2000)}]{ArayaGochezHarding_00}
{Araya-G{\'o}chez}, R.~A. \& {Harding}, A.~K. 2000, \apj, 544, 1067

\bibitem[{{Argyle} {et~al.}(1983){Argyle}, {Eldridge}, {King}, {Honeycutt}, \&
  {Schlegel}}]{Argyle_etal83}
{Argyle}, R.~W., {Eldridge}, P., {King}, D.~L., {Honeycutt}, R.~K., \&
  {Schlegel}, E.~M. 1983, \iaucirc, 3899

\bibitem[{{Arons} {et~al.}(1987){Arons}, {Klein}, \& {Lea}}]{Arons_etal87}
{Arons}, J., {Klein}, R.~I., \& {Lea}, S.~M. 1987, \apj, 312, 666

\bibitem[{{Augello} {et~al.}(2003){Augello}, {Iaria}, {Robba}, {Di Salvo},
  {Burderi}, {Lavagetto}, \& {Stella}}]{Augello_etal03}
{Augello}, G., {Iaria}, R., {Robba}, N.~R., {et~al.} 2003, \apjl, 596, L63

\bibitem[{{Bailer-Jones} {et~al.}(2018){Bailer-Jones}, {Rybizki}, {Fouesneau},
  {Mantelet}, \& {Andrae}}]{BailerJones_etal18}
{Bailer-Jones}, C.~A.~L., {Rybizki}, J., {Fouesneau}, M., {Mantelet}, G., \&
  {Andrae}, R. 2018, \aj, 156, 58

\bibitem[{{Ballhausen} {et~al.}(2016){Ballhausen}, {K{\"u}hnel}, {Pottschmidt},
  {F{\"u}rst}, {Hemphill}, {Falkner}, {Gottlieb}, {Grinberg}, {Kretschmar},
  {Kreykenbohm}, {Rothschild}, \& {Wilms}}]{Ballhausen_etal16}
{Ballhausen}, R., {K{\"u}hnel}, M., {Pottschmidt}, K., {et~al.} 2016, \aap,
  591, A65

\bibitem[{{Ballhausen} {et~al.}(2017){Ballhausen}, {Pottschmidt}, {F{\"u}rst},
  {Wilms}, {Tomsick}, {Schwarm}, {Stern}, {Kretschmar}, {Caballero},
  {Harrison}, {Boggs}, {Christensen}, {Craig}, {Hailey}, \&
  {Zhang}}]{Ballhausen_etal17}
{Ballhausen}, R., {Pottschmidt}, K., {F{\"u}rst}, F., {et~al.} 2017, \aap, 608,
  A105

\bibitem[{{Barnstedt} {et~al.}(2008){Barnstedt}, {Staubert}, {Santangelo},
  {Ferrigno}, {Horns}, {Klochkov}, {Kretschmar}, {Kreykenbohm}, {Segreto}, \&
  {Wilms}}]{Barnstedt_etal08}
{Barnstedt}, J., {Staubert}, R., {Santangelo}, A., {et~al.} 2008, \aap, 486,
  293

\bibitem[{{Basko} \& {Sunyaev}(1975)}]{BaskoSunyaev_75}
{Basko}, M.~M. \& {Sunyaev}, R.~A. 1975, \aap, 42, 311

\bibitem[{{Basko} \& {Sunyaev}(1976)}]{BaskoSunyaev_76}
{Basko}, M.~M. \& {Sunyaev}, R.~A. 1976, \mnras, 175, 395

\bibitem[{{Baykal} {et~al.}(2010){Baykal}, {G{\"o}{\v g}{\"u}{\c s}}, {{\c
  C}a{\v g}da{\c s} {\.I}nam}, \& {Belloni}}]{Baykal_etal10}
{Baykal}, A., {G{\"o}{\v g}{\"u}{\c s}}, E., {{\c C}a{\v g}da{\c s} {\.I}nam},
  S., \& {Belloni}, T. 2010, \apj, 711, 1306

\bibitem[{{Baykal} {et~al.}(2007){Baykal}, {Inam}, {Stark}, {Heffner},
  {Erkoca}, \& {Swank}}]{Baykal_etal07}
{Baykal}, A., {Inam}, S.~{\c C}., {Stark}, M.~J., {et~al.} 2007, \mnras, 374,
  1108

\bibitem[{{Baykal} {et~al.}(2000){Baykal}, {Stark}, \& {Swank}}]{Baykal_etal00}
{Baykal}, A., {Stark}, M.~J., \& {Swank}, J. 2000, \apjl, 544, L129

\bibitem[{{Becker} {et~al.}(2012){Becker}, {Klochkov}, {Sch{\"o}nherr},
  {Nishimura}, {Ferrigno}, {Caballero}, {Kretschmar}, {Wolff}, {Wilms}, \&
  {Staubert}}]{Becker_etal12}
{Becker}, P.~A., {Klochkov}, D., {Sch{\"o}nherr}, G., {et~al.} 2012, \aap, 544,
  A123

\bibitem[{{Becker} \& {Wolff}(2007)}]{BeckerWolff_07}
{Becker}, P.~A. \& {Wolff}, M.~T. 2007, \apj, 654, 435

\bibitem[{{Bellm} {et~al.}(2014){Bellm}, {F{\"u}rst}, {Pottschmidt}, {Tomsick},
  {Boggs}, {Chakrabarty}, {Christensen}, {Craig}, {Hailey}, {Harrison},
  {Stern}, {Walton}, {Wilms}, \& {Zhang}}]{Bellm_etal14}
{Bellm}, E.~C., {F{\"u}rst}, F., {Pottschmidt}, K., {et~al.} 2014, \apj, 792,
  108

\bibitem[{{Bhalerao} {et~al.}(2015){Bhalerao}, {Romano}, {Tomsick},
  {Natalucci}, {Smith}, {Bellm}, {Boggs}, {Chakrabarty}, {Christensen},
  {Craig}, {Fuerst}, {Hailey}, {Harrison}, {Krivonos}, {Lu}, {Madsen}, {Stern},
  {Younes}, \& {Zhang}}]{Bhalerao_etal15}
{Bhalerao}, V., {Romano}, P., {Tomsick}, J., {et~al.} 2015, \mnras, 447, 2274

\bibitem[{{Bildsten} {et~al.}(1997){Bildsten}, {Chakrabarty}, {Chiu}, {Finger},
  {Koh}, {Nelson}, {Prince}, {Rubin}, {Scott}, {Stollberg}, {Vaughan},
  {Wilson}, \& {Wilson}}]{Bildsten_etal97}
{Bildsten}, L., {Chakrabarty}, D., {Chiu}, J., {et~al.} 1997, ApJS, 113, 367

\bibitem[{{Bisnovatyi-Kogan} \& {Fridman}(1970)}]{BisnovatyiFridman_70}
{Bisnovatyi-Kogan}, G.~S. \& {Fridman}, A.~M. 1970, \sovast, 13, 566

\bibitem[{{Blay} {et~al.}(2006){Blay}, {Negueruela}, {Reig}, {Coe}, {Corbet},
  {Fabregat}, \& {Tarasov}}]{Blay_etal06}
{Blay}, P., {Negueruela}, I., {Reig}, P., {et~al.} 2006, \aap, 446, 1095

\bibitem[{{Blay} {et~al.}(2005){Blay}, {Rib{\'o}}, {Negueruela},
  {Torrej{\'o}n}, {Reig}, {Camero}, {Mirabel}, \& {Reglero}}]{Blay_etal05}
{Blay}, P., {Rib{\'o}}, M., {Negueruela}, I., {et~al.} 2005, \aap, 438, 963

\bibitem[{{Bodaghee} {et~al.}(2016){Bodaghee}, {Tomsick}, {Fornasini},
  {Krivonos}, {Stern}, {Mori}, {Rahoui}, {Boggs}, {Christensen}, {Craig},
  {Hailey}, {Harrison}, \& {Zhang}}]{Bodaghee_etal16}
{Bodaghee}, A., {Tomsick}, J.~A., {Fornasini}, F.~M., {et~al.} 2016, \apj, 823,
  146

\bibitem[{{Bodaghee} {et~al.}(2006){Bodaghee}, {Walter}, {Zurita Heras},
  {Bird}, {Courvoisier}, {Malizia}, {Terrier}, \& {Ubertini}}]{Bodaghee_etal06}
{Bodaghee}, A., {Walter}, R., {Zurita Heras}, J.~A., {et~al.} 2006, \aap, 447,
  1027

\bibitem[{{Boldin} {et~al.}(2013){Boldin}, {Tsygankov}, \&
  {Lutovinov}}]{Boldin_etal13}
{Boldin}, P.~A., {Tsygankov}, S.~S., \& {Lutovinov}, A.~A. 2013, Astronomy
  Letters, 39, 375

\bibitem[{{Bonazzola} {et~al.}(1979){Bonazzola}, {Heyvaerts}, \&
  {Puget}}]{Bonazzola_etal79}
{Bonazzola}, S., {Heyvaerts}, J., \& {Puget}, J.~L. 1979, \aap, 78, 53

\bibitem[{{Bonnet-Bidaud} \& {Mouchet}(1998)}]{Bonnet-BidaudMouchet_98}
{Bonnet-Bidaud}, J.~M. \& {Mouchet}, M. 1998, \aap, 332, L9

\bibitem[{{Bonning} \& {Falanga}(2005)}]{BonningFalanga_05}
{Bonning}, E.~W. \& {Falanga}, M. 2005, \aap, 436, L31

\bibitem[{{Borghese} {et~al.}(2015){Borghese}, {Rea}, {Coti Zelati}, {Tiengo},
  \& {Turolla}}]{2015ApJ...807L..20B}
{Borghese}, A., {Rea}, N., {Coti Zelati}, F., {Tiengo}, A., \& {Turolla}, R.
  2015, \apjl, 807, L20

\bibitem[{{Borghese} {et~al.}(2017){Borghese}, {Rea}, {Coti Zelati}, {Tiengo},
  {Turolla}, \& {Zane}}]{2017MNRAS.468.2975B}
{Borghese}, A., {Rea}, N., {Coti Zelati}, F., {et~al.} 2017, \mnras, 468, 2975

\bibitem[{{Bozzo} {et~al.}(2011){Bozzo}, {Ferrigno}, {Falanga}, \&
  {Walter}}]{Bozzo_etal11}
{Bozzo}, E., {Ferrigno}, C., {Falanga}, M., \& {Walter}, R. 2011, \aap, 531,
  A65

\bibitem[{{Braun} \& {Yahel}(1984)}]{BraunYahel_84}
{Braun}, A. \& {Yahel}, R.~Z. 1984, \apj, 278, 349

\bibitem[{{Brightman} {et~al.}(2018){Brightman}, {Harrison}, {F{\"u}rst},
  {Middleton}, {Walton}, {Stern}, {Fabian}, {Heida}, {Barret}, \&
  {Bachetti}}]{Brightman_etal18}
{Brightman}, M., {Harrison}, F.~A., {F{\"u}rst}, F., {et~al.} 2018, Nature
  Astronomy, 2, 312

\bibitem[{{Brumback} {et~al.}(2018){Brumback}, {Hickox}, {F{\"u}rst},
  {Pottschmidt}, {Hemphill}, {Tomsick}, {Wilms}, \&
  {Ballhausen}}]{Brumback_etal18}
{Brumback}, M.~C., {Hickox}, R.~C., {F{\"u}rst}, F.~S., {et~al.} 2018, \apj,
  852, 132

\bibitem[{{Burderi} {et~al.}(2000){Burderi}, {Di Salvo}, {Robba}, {La Barbera},
  \& {Guainazzi}}]{Burderi_etal00}
{Burderi}, L., {Di Salvo}, T., {Robba}, N.~R., {La Barbera}, A., \&
  {Guainazzi}, M. 2000, \apj, 530, 429

\bibitem[{{Burnard} {et~al.}(1991){Burnard}, {Arons}, \&
  {Klein}}]{Burnard_etal91}
{Burnard}, D.~J., {Arons}, J., \& {Klein}, R.~I. 1991, ApJ, 367, 575

\bibitem[{{Burwitz} {et~al.}(2003){Burwitz}, {Haberl}, {Neuh{\"a}user},
  {Predehl}, {Tr{\"u}mper}, \& {Zavlin}}]{2003A&A...399.1109B}
{Burwitz}, V., {Haberl}, F., {Neuh{\"a}user}, R., {et~al.} 2003, \aap, 399,
  1109

\bibitem[{{Burwitz} {et~al.}(2001){Burwitz}, {Zavlin}, {Neuh{\"a}user},
  {Predehl}, {Tr{\"u}mper}, \& {Brinkman}}]{2001A&A...379L..35B}
{Burwitz}, V., {Zavlin}, V.~E., {Neuh{\"a}user}, R., {et~al.} 2001, \aap, 379,
  L35

\bibitem[{{Bussard}(1980)}]{Bussard_80}
{Bussard}, R.~W. 1980, \apj, 237, 970

\bibitem[{{Bykov} \& {Krasil'shchikov}(2004)}]{BykovKrasil_04}
{Bykov}, A.~M. \& {Krasil'shchikov}, A.~M. 2004, Astronomy Letters, 30, 309

\bibitem[{{Caballero} {et~al.}(2007){Caballero}, {Kretschmar}, {Santangelo},
  {Staubert}, {Klochkov}, {Camero}, {Ferrigno}, {Finger}, {Kreykenbohm},
  {McBride}, {Pottschmidt}, {Rothschild}, {Sch{\"o}nherr}, {Segreto}, {Suchy},
  {Wilms}, \& {Wilson}}]{Caballero_etal07}
{Caballero}, I., {Kretschmar}, P., {Santangelo}, A., {et~al.} 2007, A\&A, 465,
  L21

\bibitem[{{Caballero} {et~al.}(2013{\natexlab{a}}){Caballero}, {K{\"u}hnel},
  {Pottschmidt}, {Zurita Heras}, {Marcu}, {M{\"u}ller}, {Laurent}, {Klochkov},
  {Kretschmar}, {Ferrigno}, {Kreykenbohm}, {Wilms}, {Rothschild}, {Santangelo},
  {Staubert}, \& {Suchy}}]{Caballero_etal13b}
{Caballero}, I., {K{\"u}hnel}, M., {Pottschmidt}, K., {et~al.}
  2013{\natexlab{a}}, in AAS/High Energy Astrophysics Division, Vol.~13,
  AAS/High Energy Astrophysics Division, 126.32

\bibitem[{{Caballero} {et~al.}(2013{\natexlab{b}}){Caballero}, {Pottschmidt},
  {Marcu}, {Barragan}, {Ferrigno}, {Klochkov}, {Zurita Heras}, {Suchy},
  {Wilms}, {Kretschmar}, {Santangelo}, {Kreykenbohm}, {F{\"u}rst},
  {Rothschild}, {Staubert}, {Finger}, {Camero-Arranz}, {Makishima}, {Enoto},
  {Iwakiri}, \& {Terada}}]{Caballero_etal13}
{Caballero}, I., {Pottschmidt}, K., {Marcu}, D.~M., {et~al.}
  2013{\natexlab{b}}, \apjl, 764, L23

\bibitem[{{Caballero} {et~al.}(2008){Caballero}, {Santangelo}, {Kretschmar},
  {Staubert}, {Postnov}, {Klochkov}, {Camero-Arranz}, {Finger}, {Kreykenbohm},
  {Pottschmidt}, {Rothschild}, {Suchy}, {Wilms}, \&
  {Wilson}}]{Caballero_etal08}
{Caballero}, I., {Santangelo}, A., {Kretschmar}, P., {et~al.} 2008, A\&A, 480,
  L17

\bibitem[{{Caballero} \& {Wilms}(2012)}]{CaballeroWilms_12}
{Caballero}, I. \& {Wilms}, J. 2012, \memsai, 83, 230

\bibitem[{{Caballero-Garc{\'{\i}}a} {et~al.}(2016){Caballero-Garc{\'{\i}}a},
  {Camero-Arranz}, {{\"O}zbey Arabac{\i}}, {Zurita}, {Suso},
  {Guti{\'e}rrez-Soto}, {Beklen}, {Kiaeerad}, {Garrido}, \&
  {Hudec}}]{CaballeroGarcia_etal16}
{Caballero-Garc{\'{\i}}a}, M.~D., {Camero-Arranz}, A., {{\"O}zbey Arabac{\i}},
  M., {et~al.} 2016, \aap, 589, A9

\bibitem[{{Camero-Arranz} {et~al.}(2012{\natexlab{a}}){Camero-Arranz},
  {Finger}, {Wilson-Hodge}, {Jenke}, {Steele}, {Coe}, {Gutierrez-Soto},
  {Kretschmar}, {Caballero}, {Yan}, {Rodr{\'{\i}}guez}, {Suso}, {Case},
  {Cherry}, {Guiriec}, \& {McBride}}]{Camero_etal12}
{Camero-Arranz}, A., {Finger}, M.~H., {Wilson-Hodge}, C.~A., {et~al.}
  2012{\natexlab{a}}, ApJ, 754, 20

\bibitem[{{Camero-Arranz} {et~al.}(2012{\natexlab{b}}){Camero-Arranz},
  {Pottschmidt}, {Finger}, {Ikhsanov}, {Wilson-Hodge}, \&
  {Marcu}}]{Camero-Arranz_etal12}
{Camero-Arranz}, A., {Pottschmidt}, K., {Finger}, M.~H., {et~al.}
  2012{\natexlab{b}}, \aap, 546, A40

\bibitem[{{Campana} {et~al.}(2002){Campana}, {Stella}, {Israel}, {Moretti},
  {Parmar}, \& {Orlandini}}]{Campana_etal02}
{Campana}, S., {Stella}, L., {Israel}, G.~L., {et~al.} 2002, \apj, 580, 389

\bibitem[{{Carpano} {et~al.}(2018){Carpano}, {Haberl}, {Maitra}, \&
  {Vasilopoulos}}]{Carpano_etal18}
{Carpano}, S., {Haberl}, F., {Maitra}, C., \& {Vasilopoulos}, G. 2018, \mnras,
  476, L45

\bibitem[{{Chieregato} {et~al.}(2005){Chieregato}, {Campana}, {Treves},
  {Moretti}, {Mignani}, \& {Tagliaferri}}]{2005A&A...444...69C}
{Chieregato}, M., {Campana}, S., {Treves}, A., {et~al.} 2005, \aap, 444, 69

\bibitem[{Chodil {et~al.}(1967)Chodil, Mark, Rodrigues, Seward, \&
  Swift}]{Chodil_etal67}
Chodil, G., Mark, H., Rodrigues, R., Seward, F.~D., \& Swift, C.~D. 1967, \apj,
  150, 57

\bibitem[{{Chou} {et~al.}(2016){Chou}, {Hsieh}, {Hu}, {Yang}, \&
  {Su}}]{Chou_etal16}
{Chou}, Y., {Hsieh}, H.-E., {Hu}, C.-P., {Yang}, T.-C., \& {Su}, Y.-H. 2016,
  \apj, 831, 29

\bibitem[{{Christodoulou} {et~al.}(2017){Christodoulou}, {Laycock}, {Yang}, \&
  {Fingerman}}]{Christodoulou_etal17}
{Christodoulou}, D.~M., {Laycock}, S.~G.~T., {Yang}, J., \& {Fingerman}, S.
  2017, Research in Astronomy and Astrophysics, 17, 059

\bibitem[{{Clark} {et~al.}(1978){Clark}, {Doxsey}, {Li}, {Jernigan}, \& {van
  Paradijs}}]{Clark_etal78}
{Clark}, G., {Doxsey}, R., {Li}, F., {Jernigan}, J.~G., \& {van Paradijs}, J.
  1978, \apjl, 221, L37

\bibitem[{{Clark} {et~al.}(1979){Clark}, {Li}, \& {van
  Paradijs}}]{Clark_etal79}
{Clark}, G., {Li}, F., \& {van Paradijs}, J. 1979, \apj, 227, 54

\bibitem[{{Clark} {et~al.}(1990){Clark}, {Woo}, {Nagase}, {Makishima}, \&
  {Sakao}}]{Clark_etal90}
{Clark}, G.~W., {Woo}, J.~W., {Nagase}, F., {Makishima}, K., \& {Sakao}, T.
  1990, \apj, 353, 274

\bibitem[{{Coburn} {et~al.}(2001){Coburn}, {Heindl}, {Gruber}, {Rothschild},
  {Staubert}, {Wilms}, \& {Kreykenbohm}}]{Coburn_etal01}
{Coburn}, W., {Heindl}, W.~A., {Gruber}, D.~E., {et~al.} 2001, \apj, 552, 738

\bibitem[{{Coburn} {et~al.}(2002){Coburn}, {Heindl}, {Rothschild}, {Gruber},
  {Kreykenbohm}, {Wilms}, {Kretschmar}, \& {Staubert}}]{Coburn_etal02}
{Coburn}, W., {Heindl}, W.~A., {Rothschild}, R.~E., {et~al.} 2002, ApJ, 580,
  394

\bibitem[{{Coburn} {et~al.}(2006){Coburn}, {Pottschmidt}, {Rothschild},
  {Kretschmar}, {Kreykenbohm}, {Wilms}, \& {McBride}}]{Coburn_etal06}
{Coburn}, W., {Pottschmidt}, K., {Rothschild}, R., {et~al.} 2006, in Bulletin
  of the American Astronomical Society, Vol.~38, AAS/High Energy Astrophysics
  Division \#9, 340

\bibitem[{{Coe} {et~al.}(2007){Coe}, {Bird}, {Hill}, {McBride}, {Schurch},
  {Galache}, {Wilson}, {Finger}, {Buckley}, \& {Romero-Colmenero}}]{Coe_etal07}
{Coe}, M.~J., {Bird}, A.~J., {Hill}, A.~B., {et~al.} 2007, \mnras, 378, 1427

\bibitem[{{Cominsky} {et~al.}(1978){Cominsky}, {Clark}, {Li}, {Mayer}, \&
  {Rappaport}}]{Cominsky_etal78}
{Cominsky}, L., {Clark}, G.~W., {Li}, F., {Mayer}, W., \& {Rappaport}, S. 1978,
  \nat, 273, 367

\bibitem[{{Corbet} {et~al.}(1986){Corbet}, {Charles}, \& {van der
  Klis}}]{Corbet_etal86}
{Corbet}, R.~H.~D., {Charles}, P.~A., \& {van der Klis}, M. 1986, \aap, 162,
  117

\bibitem[{{Corbet} {et~al.}(2018){Corbet}, {Coley}, {Krim}, \&
  {Pottschmidt}}]{Corbet_etal18}
{Corbet}, R.~H.~D., {Coley}, J.~B.~C., {Krim}, H.~A., \& {Pottschmidt}, K.
  2018, The Astronomer's Telegram 11918

\bibitem[{{Corbet} \& {Krimm}(2009)}]{CorbetKrimm_09}
{Corbet}, R.~H.~D. \& {Krimm}, H.~A. 2009, The Astronomer's Telegram, 2008

\bibitem[{{Corbet} {et~al.}(2010{\natexlab{a}}){Corbet}, {Krimm}, {Barthelmy},
  {Baumgartner}, {Markwardt}, {Skinner}, \& {Tueller}}]{Corbet_etal10a}
{Corbet}, R.~H.~D., {Krimm}, H.~A., {Barthelmy}, S.~D., {et~al.}
  2010{\natexlab{a}}, The Astronomer's Telegram, 2570

\bibitem[{{Corbet} {et~al.}(2007){Corbet}, {Markwardt}, \&
  {Tueller}}]{Corbet_etal07}
{Corbet}, R.~H.~D., {Markwardt}, C.~B., \& {Tueller}, J. 2007, \apj, 655, 458

\bibitem[{{Corbet} {et~al.}(2001){Corbet}, {Marshall}, {Coe}, {Laycock}, \&
  {Handler}}]{Corbet_etal01}
{Corbet}, R.~H.~D., {Marshall}, F.~E., {Coe}, M.~J., {Laycock}, S., \&
  {Handler}, G. 2001, \apjl, 548, L41

\bibitem[{{Corbet} {et~al.}(2010{\natexlab{b}}){Corbet}, {Pearlman}, \&
  {Pottschmidt}}]{Corbet_etal10}
{Corbet}, R.~H.~D., {Pearlman}, A.~B., \& {Pottschmidt}, K. 2010{\natexlab{b}},
  The Astronomer's Telegram, 2766

\bibitem[{{Crampton} {et~al.}(1978){Crampton}, {Hutchings}, \&
  {Cowley}}]{Crampton_etal78}
{Crampton}, D., {Hutchings}, J.~B., \& {Cowley}, A.~P. 1978, \apjl, 223, L79

\bibitem[{{Crampton} {et~al.}(1985){Crampton}, {Hutchings}, \&
  {Cowley}}]{Crampton_etal85}
{Crampton}, D., {Hutchings}, J.~B., \& {Cowley}, A.~P. 1985, \apj, 299, 839

\bibitem[{{Cropper} {et~al.}(2001){Cropper}, {Zane}, {Ramsay}, {Haberl}, \&
  {Motch}}]{2001A&A...365L.302C}
{Cropper}, M., {Zane}, S., {Ramsay}, G., {Haberl}, F., \& {Motch}, C. 2001,
  \aap, 365, L302

\bibitem[{{Cropper} {et~al.}(2007){Cropper}, {Zane}, {Turolla}, {Zampieri},
  {Chieregato}, {Drake}, \& {Treves}}]{2007Ap&SS.308..161C}
{Cropper}, M., {Zane}, S., {Turolla}, R., {et~al.} 2007, \apss, 308, 161

\bibitem[{{Cusumano} {et~al.}(2016){Cusumano}, {La Parola}, {D'A{\`i}},
  {Segreto}, {Tagliaferri}, {Barthelmy}, \& {Gehrels}}]{Cusumano_etal16}
{Cusumano}, G., {La Parola}, V., {D'A{\`i}}, A., {et~al.} 2016, \mnras, 460,
  L99

\bibitem[{{Cusumano} {et~al.}(2010){Cusumano}, {La Parola}, {Romano},
  {Segreto}, {Vercellone}, \& {Chincarini}}]{Cusumano_etal10}
{Cusumano}, G., {La Parola}, V., {Romano}, P., {et~al.} 2010, \mnras, 406, L16

\bibitem[{{D'A{\`i}} {et~al.}(2017){D'A{\`i}}, {Cusumano}, {Del Santo}, {La
  Parola}, \& {Segreto}}]{D'Ai_etal17}
{D'A{\`i}}, A., {Cusumano}, G., {Del Santo}, M., {La Parola}, V., \& {Segreto},
  A. 2017, \mnras, 470, 2457

\bibitem[{{D'A{\`i}} {et~al.}(2011){D'A{\`i}}, {Cusumano}, {La Parola},
  {Segreto}, {di Salvo}, {Iaria}, \& {Robba}}]{D'Ai_etal11}
{D'A{\`i}}, A., {Cusumano}, G., {La Parola}, V., {et~al.} 2011, \aap, 532, A73

\bibitem[{{D'A{\`i}} {et~al.}(2015){D'A{\`i}}, {Di Salvo}, {Iaria},
  {Garc{\'{\i}}a}, {Sanna}, {Pintore}, {Riggio}, {Burderi}, {Bozzo}, {Dauser},
  {Matranga}, {Galiano}, \& {Robba}}]{D'ai_etal15}
{D'A{\`i}}, A., {Di Salvo}, T., {Iaria}, R., {et~al.} 2015, \mnras, 449, 4288

\bibitem[{{Dal Fiume} {et~al.}(1998){Dal Fiume}, {Orlandini}, {Cusumano}, {del
  Sordo}, {Feroci}, {Frontera}, {Oosterbroek}, {Palazzi}, {Parmar},
  {Santangelo}, \& {Segreto}}]{DalFiume_etal98}
{Dal Fiume}, D., {Orlandini}, M., {Cusumano}, G., {et~al.} 1998, A\&A, 329, L41

\bibitem[{{dal Fiume} {et~al.}(2000){dal Fiume}, {Orlandini}, {del Sordo},
  {Frontera}, {Oosterbroek}, {Palazzi}, {Parmar}, {Piraino}, {Santangelo}, \&
  {Segreto}}]{DalFiume_etal00}
{dal Fiume}, D., {Orlandini}, M., {del Sordo}, S., {et~al.} 2000, in Broad Band
  X-ray Spectra of Cosmic Sources, ed. K.~{Makishima}, L.~{Piro}, \&
  T.~{Takahashi}, 399

\bibitem[{{Daugherty} \& {Ventura}(1977)}]{DaughertyVentura_77}
{Daugherty}, J.~K. \& {Ventura}, J. 1977, \aap, 61, 723

\bibitem[{{Davidson}(1973)}]{Davidson_73}
{Davidson}, K. 1973, Nature Physical Science, 246, 1

\bibitem[{{de Vries} {et~al.}(2004){de Vries}, {Vink}, {M{\'e}ndez}, \&
  {Verbunt}}]{2004A&A...415L..31D}
{de Vries}, C.~P., {Vink}, J., {M{\'e}ndez}, M., \& {Verbunt}, F. 2004, \aap,
  415, L31

\bibitem[{{DeCesar} {et~al.}(2013){DeCesar}, {Boyd}, {Pottschmidt}, {Wilms},
  {Suchy}, \& {Miller}}]{DeCesar_etal13}
{DeCesar}, M.~E., {Boyd}, P.~T., {Pottschmidt}, K., {et~al.} 2013, \apj, 762,
  61

\bibitem[{{Degenaar} {et~al.}(2014){Degenaar}, {Miller}, {Harrison}, {Kennea},
  {Kouveliotou}, \& {Younes}}]{Degenaar_etal14}
{Degenaar}, N., {Miller}, J.~M., {Harrison}, F.~A., {et~al.} 2014, \apjl, 796,
  L9

\bibitem[{{den Hartog} {et~al.}(2006){den Hartog}, {Hermsen}, {Kuiper}, {Vink},
  {in't Zand}, \& {Collmar}}]{DenHartog_etal06}
{den Hartog}, P.~R., {Hermsen}, W., {Kuiper}, L., {et~al.} 2006, \aap, 451, 587

\bibitem[{{Doroshenko}(2017)}]{DoroshenkoA_17}
{Doroshenko}, A. 2017, private comm., 1, yyy

\bibitem[{Doroshenko(2017)}]{DoroR_17}
Doroshenko, R. 2017, PhD thesis, Univ. of T\"ubingen

\bibitem[{{Doroshenko} {et~al.}(2015){Doroshenko}, {Santangelo}, {Doroshenko},
  {Suleimanov}, \& {Piraino}}]{Doroshenko_etal15}
{Doroshenko}, R., {Santangelo}, A., {Doroshenko}, V., {Suleimanov}, V., \&
  {Piraino}, S. 2015, \mnras, 452, 2490

\bibitem[{{Doroshenko} {et~al.}(2012){Doroshenko}, {Santangelo}, {Kreykenbohm},
  \& {Doroshenko}}]{Doroshenko_etal12}
{Doroshenko}, V., {Santangelo}, A., {Kreykenbohm}, I., \& {Doroshenko}, R.
  2012, \aap, 540, L1

\bibitem[{{Doroshenko} {et~al.}(2010){Doroshenko}, {Suchy}, {Santangelo},
  {Staubert}, {Kreykenbohm}, {Rothschild}, {Pottschmidt}, \&
  {Wilms}}]{Doroshenko_etal10}
{Doroshenko}, V., {Suchy}, S., {Santangelo}, A., {et~al.} 2010, \aap, 515, L1

\bibitem[{{Doroshenko} {et~al.}(2017){Doroshenko}, {Tsygankov}, {Mushtukov},
  {Lutovinov}, {Santangelo}, {Suleimanov}, \& {Poutanen}}]{Doroshenko_etal17}
{Doroshenko}, V., {Tsygankov}, S.~S., {Mushtukov}, A.~A., {et~al.} 2017,
  \mnras, 466, 2143

\bibitem[{{Enoto} {et~al.}(2008){Enoto}, {Makishima}, {Terada}, {Mihara},
  {Nakazawa}, {Ueda}, {Dotani}, {Kokubun}, {Nagase}, {Naik}, {Suzuki},
  {Nakajima}, \& {Takahashi}}]{Enoto_etal08}
{Enoto}, T., {Makishima}, K., {Terada}, Y., {et~al.} 2008, \pasj, 60, 57

\bibitem[{{Epili} {et~al.}(2016){Epili}, {Naik}, \& {Jaisawal}}]{Epili_etal16}
{Epili}, P., {Naik}, S., \& {Jaisawal}, G.~K. 2016, Research in Astronomy and
  Astrophysics, 16, 77

\bibitem[{{Falkner} {et~al.}(2018){Falkner}, {Schwarm}, {Dauser}, \&
  et~al.}]{Falkner_etal18}
{Falkner}, S., {Schwarm}, F.-W., {Dauser}, T., \& et~al. 2018, submitted

\bibitem[{{Farinelli} {et~al.}(2012){Farinelli}, {Ceccobello}, {Romano}, \&
  {Titarchuk}}]{Farinelli_etal12}
{Farinelli}, R., {Ceccobello}, C., {Romano}, P., \& {Titarchuk}, L. 2012, \aap,
  538, A67

\bibitem[{{Farinelli} {et~al.}(2016){Farinelli}, {Ferrigno}, {Bozzo}, \&
  {Becker}}]{Farinelli_etal16}
{Farinelli}, R., {Ferrigno}, C., {Bozzo}, E., \& {Becker}, P.~A. 2016, \aap,
  591, A29

\bibitem[{{Farrell} {et~al.}(2006){Farrell}, {Sood}, \&
  {O'Neill}}]{Farrell_etal06}
{Farrell}, S.~A., {Sood}, R.~K., \& {O'Neill}, P.~M. 2006, \mnras, 367, 1457

\bibitem[{{Farrell} {et~al.}(2008){Farrell}, {Sood}, {O'Neill}, \&
  {Dieters}}]{Farrell_etal08}
{Farrell}, S.~A., {Sood}, R.~K., {O'Neill}, P.~M., \& {Dieters}, S. 2008,
  \mnras, 389, 608

\bibitem[{{Ferrigno} {et~al.}(2009){Ferrigno}, {Becker}, {Segreto}, {Mineo}, \&
  {Santangelo}}]{Ferrigno_etal09}
{Ferrigno}, C., {Becker}, P.~A., {Segreto}, A., {Mineo}, T., \& {Santangelo},
  A. 2009, \aap, 498, 825

\bibitem[{{Ferrigno} {et~al.}(2016{\natexlab{a}}){Ferrigno}, {Ducci}, {Bozzo},
  {Kretschmar}, {K{\"u}hnel}, {Malacaria}, {Pottschmidt}, {Santangelo},
  {Savchenko}, \& {Wilms}}]{Ferrigno_etal16a}
{Ferrigno}, C., {Ducci}, L., {Bozzo}, E., {et~al.} 2016{\natexlab{a}}, \aap,
  595, A17

\bibitem[{{Ferrigno} {et~al.}(2011){Ferrigno}, {Falanga}, {Bozzo}, {Becker},
  {Klochkov}, \& {Santangelo}}]{Ferrigno_etal11}
{Ferrigno}, C., {Falanga}, M., {Bozzo}, E., {et~al.} 2011, \aap, 532, A76

\bibitem[{{Ferrigno} {et~al.}(2013){Ferrigno}, {Farinelli}, {Bozzo},
  {Pottschmidt}, {Klochkov}, \& {Kretschmar}}]{Ferrigno_etal13}
{Ferrigno}, C., {Farinelli}, R., {Bozzo}, E., {et~al.} 2013, \aap, 553, A103

\bibitem[{{Ferrigno} {et~al.}(2016{\natexlab{b}}){Ferrigno}, {Pjanka}, {Bozzo},
  {Klochkov}, {Ducci}, \& {Zdziarski}}]{Ferrigno_etal16b}
{Ferrigno}, C., {Pjanka}, P., {Bozzo}, E., {et~al.} 2016{\natexlab{b}}, \aap,
  593, A105

\bibitem[{{Ferrigno} {et~al.}(2007){Ferrigno}, {Segreto}, {Santangelo},
  {Wilms}, {Kreykenbohm}, {Denis}, \& {Staubert}}]{Ferrigno_etal07}
{Ferrigno}, C., {Segreto}, A., {Santangelo}, A., {et~al.} 2007, A\&A, 462, 995

\bibitem[{{Finger} {et~al.}(1996){Finger}, {Wilson}, \&
  {Harmon}}]{Finger_etal96}
{Finger}, M.~H., {Wilson}, R.~B., \& {Harmon}, B.~A. 1996, ApJ, 459, 288

\bibitem[{Freeman {et~al.}(1996)}]{Freeman_96}
Freeman, P. {et~al.} 1996, in Gamma Ray Bursts, Proc. 3rd Huntsville Symp., ed.
  Koveliotou/Briggs/Fishman, 172

\bibitem[{{Fuerst} {et~al.}(2018){Fuerst}, {Falkner}, {Marcu-Cheatham},
  {Grefenstette}, {Tomsick}, {Pottschmidt}, {Walton}, {Natalucci}, \&
  {Kretschmar}}]{Fuerst_etal18}
{Fuerst}, F., {Falkner}, S., {Marcu-Cheatham}, D., {et~al.} 2018,
  ArXiv:1809.05691 [\eprint[arXiv:1809.05691]{1809.05691}]

\bibitem[{{F{\"u}rst} {et~al.}(2013){F{\"u}rst}, {Grefenstette}, {Staubert},
  {Tomsick}, {Bachetti}, {Barret}, {Bellm}, {Boggs}, {Chenevez}, {Christensen},
  {Craig}, {Hailey}, {Harrison}, {Klochkov}, {Madsen}, {Pottschmidt}, {Stern},
  {Walton}, {Wilms}, \& {Zhang}}]{Fuerst_etal13}
{F{\"u}rst}, F., {Grefenstette}, B.~W., {Staubert}, R., {et~al.} 2013, \apj,
  779, 69

\bibitem[{{F{\"u}rst} {et~al.}(2017){F{\"u}rst}, {Kretschmar}, {Kajava},
  {Alfonso-Garz{\'o}n}, {K{\"u}hnel}, {Sanchez-Fernandez}, {Blay},
  {Wilson-Hodge}, {Jenke}, {Kreykenbohm}, {Pottschmidt}, {Wilms}, \&
  {Rothschild}}]{Fuerst_etal17}
{F{\"u}rst}, F., {Kretschmar}, P., {Kajava}, J.~J.~E., {et~al.} 2017, \aap,
  606, A89

\bibitem[{{F{\"u}rst} {et~al.}(2011){F{\"u}rst}, {Kreykenbohm}, {Suchy},
  {Barrag{\'a}n}, {Wilms}, {Rothschild}, \& {Pottschmidt}}]{Fuerst_etal11}
{F{\"u}rst}, F., {Kreykenbohm}, I., {Suchy}, S., {et~al.} 2011, \aap, 525, A73

\bibitem[{{F{\"u}rst} {et~al.}(2012){F{\"u}rst}, {Pottschmidt}, {Kreykenbohm},
  {M{\"u}ller}, {K{\"u}hnel}, {Wilms}, \& {Rothschild}}]{Fuerst_etal12}
{F{\"u}rst}, F., {Pottschmidt}, K., {Kreykenbohm}, I., {et~al.} 2012, \aap,
  547, A2

\bibitem[{{F{\"u}rst} {et~al.}(2015){F{\"u}rst}, {Pottschmidt}, {Miyasaka},
  {Bhalerao}, {Bachetti}, {Boggs}, {Christensen}, {Craig}, {Grinberg},
  {Hailey}, {Harrison}, {Kennea}, {Rahoui}, {Stern}, {Tendulkar}, {Tomsick},
  {Walton}, {Wilms}, \& {Zhang}}]{Fuerst_etal15}
{F{\"u}rst}, F., {Pottschmidt}, K., {Miyasaka}, H., {et~al.} 2015, \apjl, 806,
  L24

\bibitem[{{F{\"u}rst} {et~al.}(2014{\natexlab{a}}){F{\"u}rst}, {Pottschmidt},
  {Wilms}, {Kennea}, {Bachetti}, {Bellm}, {Boggs}, {Chakrabarty},
  {Christensen}, {Craig}, {Hailey}, {Harrison}, {Stern}, {Tomsick}, {Walton},
  \& {Zhang}}]{Fuerst_etal14a}
{F{\"u}rst}, F., {Pottschmidt}, K., {Wilms}, J., {et~al.} 2014{\natexlab{a}},
  \apjl, 784, L40

\bibitem[{{F{\"u}rst} {et~al.}(2014{\natexlab{b}}){F{\"u}rst}, {Pottschmidt},
  {Wilms}, {Tomsick}, {Bachetti}, {Boggs}, {Christensen}, {Craig},
  {Grefenstette}, {Hailey}, {Harrison}, {Madsen}, {Miller}, {Stern}, {Walton},
  \& {Zhang}}]{Fuerst_etal14b}
{F{\"u}rst}, F., {Pottschmidt}, K., {Wilms}, J., {et~al.} 2014{\natexlab{b}},
  \apj, 780, 133

\bibitem[{{Galloway} {et~al.}(2005){Galloway}, {Wang}, \&
  {Morgan}}]{Galloway_etal05}
{Galloway}, D.~K., {Wang}, Z., \& {Morgan}, E.~H. 2005, \apj, 635, 1217

\bibitem[{{Geppert} \& {Vigan{\`o}}(2014)}]{2014MNRAS.444.3198G}
{Geppert}, U. \& {Vigan{\`o}}, D. 2014, \mnras, 444, 3198

\bibitem[{{Ghosh} \& {Lamb}(1979{\natexlab{a}})}]{GhoshLamb_79a}
{Ghosh}, P. \& {Lamb}, F.~K. 1979{\natexlab{a}}, \apj, 232, 259

\bibitem[{{Ghosh} \& {Lamb}(1979{\natexlab{b}})}]{GhoshLamb_79b}
{Ghosh}, P. \& {Lamb}, F.~K. 1979{\natexlab{b}}, ApJ, 234, 296

\bibitem[{{Ghosh} {et~al.}(1977){Ghosh}, {Lamb}, \& {Pethick}}]{GhoshLamb_77}
{Ghosh}, P., {Lamb}, F.~K., \& {Pethick}, C.~J. 1977, \apj, 217, 578

\bibitem[{{Giacconi} {et~al.}(1972){Giacconi}, {Murray}, {Gursky}, {Kellogg},
  {Schreier}, \& {Tananbaum}}]{Giacconi_etal72}
{Giacconi}, R., {Murray}, S., {Gursky}, H., {et~al.} 1972, \apj, 178, 281

\bibitem[{{Giangrande} {et~al.}(1980){Giangrande}, {Giovannelli}, {Bartolini},
  {Guarnieri}, \& {Piccioni}}]{Giangrande_etal80}
{Giangrande}, A., {Giovannelli}, F., {Bartolini}, C., {Guarnieri}, A., \&
  {Piccioni}, A. 1980, \aaps, 40, 289

\bibitem[{{Gnedin} \& {Sunyaev}(1974)}]{GnedinSunyaev_74}
{Gnedin}, I.~N. \& {Sunyaev}, R.~A. 1974, \aap, 36, 379

\bibitem[{{Gold}(1968)}]{Gold_68}
{Gold}, T. 1968, \nat, 218, 731

\bibitem[{{Goldreich} \& {Julian}(1969)}]{GoldreichJulian_69}
{Goldreich}, P. \& {Julian}, W.~H. 1969, \apj, 157, 869

\bibitem[{{Grove} {et~al.}(1995){Grove}, {Strickman}, {Johnson}, {Kurfess},
  {Kinzer}, {Starr}, {Jung}, {Kendziorra}, {Kretschmar}, {Maisack}, \&
  {Staubert}}]{Grove_etal95}
{Grove}, J.~E., {Strickman}, M.~S., {Johnson}, W.~N., {et~al.} 1995, \apjl,
  438, L25

\bibitem[{{Gruber} {et~al.}(2001){Gruber}, {Heindl}, {Rothschild}, {Coburn},
  {Staubert}, {Kreykenbohm}, \& {Wilms}}]{Gruber_etal01}
{Gruber}, D.~E., {Heindl}, W.~A., {Rothschild}, R.~E., {et~al.} 2001, ApJ, 562,
  499

\bibitem[{{Gruber} {et~al.}(1999){Gruber}, {Heindl}, {Rothschild}, {Staubert},
  {Wilms}, \& {Scott}}]{Gruber_etal99}
{Gruber}, D.~E., {Heindl}, W.~A., {Rothschild}, R.~E., {et~al.} 1999, in
  Highlights in X-ray Astronomy, ed. B.~{Aschenbach} \& M.~J. {Freyberg}, Vol.
  272, 33

\bibitem[{{Gruber} {et~al.}(1980){Gruber}, {Matteson}, {Nolan}, \&
  et~al.}]{Gruber_etal80}
{Gruber}, D.~E., {Matteson}, J.~L., {Nolan}, P.~L., \& et~al. 1980, ApJ, 240,
  L127

\bibitem[{{Haberl}(2007)}]{2007Ap&SS.308..181H}
{Haberl}, F. 2007, \apss, 308, 181

\bibitem[{{Haberl} {et~al.}(1997){Haberl}, {Motch}, {Buckley}, {Zickgraf}, \&
  {Pietsch}}]{1997A&A...326..662H}
{Haberl}, F., {Motch}, C., {Buckley}, D.~A.~H., {Zickgraf}, F.-J., \&
  {Pietsch}, W. 1997, \aap, 326, 662

\bibitem[{{Haberl} {et~al.}(1998){Haberl}, {Motch}, \&
  {Pietsch}}]{1998AN....319...97H}
{Haberl}, F., {Motch}, C., \& {Pietsch}, W. 1998, Astronomische Nachrichten,
  319, 97

\bibitem[{{Haberl} {et~al.}(2004{\natexlab{a}}){Haberl}, {Motch}, {Zavlin},
  {Reinsch}, {G{\"a}nsicke}, {Cropper}, {Schwope}, {Turolla}, \&
  {Zane}}]{2004A&A...424..635H}
{Haberl}, F., {Motch}, C., {Zavlin}, V.~E., {et~al.} 2004{\natexlab{a}}, \aap,
  424, 635

\bibitem[{{Haberl} {et~al.}(1999){Haberl}, {Pietsch}, \&
  {Motch}}]{1999A&A...351L..53H}
{Haberl}, F., {Pietsch}, W., \& {Motch}, C. 1999, \aap, 351, L53

\bibitem[{{Haberl} {et~al.}(2003){Haberl}, {Schwope}, {Hambaryan}, {Hasinger},
  \& {Motch}}]{2003A&A...403L..19H}
{Haberl}, F., {Schwope}, A.~D., {Hambaryan}, V., {Hasinger}, G., \& {Motch}, C.
  2003, \aap, 403, L19

\bibitem[{{Haberl} {et~al.}(2006){Haberl}, {Turolla}, {de Vries}, {Zane},
  {Vink}, {M{\'e}ndez}, \& {Verbunt}}]{2006A&A...451L..17H}
{Haberl}, F., {Turolla}, R., {de Vries}, C.~P., {et~al.} 2006, \aap, 451, L17

\bibitem[{{Haberl} \& {Zavlin}(2002)}]{2002A&A...391..571H}
{Haberl}, F. \& {Zavlin}, V.~E. 2002, \aap, 391, 571

\bibitem[{{Haberl} {et~al.}(2004{\natexlab{b}}){Haberl}, {Zavlin},
  {Tr{\"u}mper}, \& {Burwitz}}]{2004A&A...419.1077H}
{Haberl}, F., {Zavlin}, V.~E., {Tr{\"u}mper}, J., \& {Burwitz}, V.
  2004{\natexlab{b}}, \aap, 419, 1077

\bibitem[{{Haigh} {et~al.}(2004){Haigh}, {Coe}, \& {Fabregat}}]{Haigh_etal04}
{Haigh}, N.~J., {Coe}, M.~J., \& {Fabregat}, J. 2004, \mnras, 350, 1457

\bibitem[{{Halpern}(2012)}]{Halpern_12}
{Halpern}, J.~P. 2012, The Astronomer's Telegram, 3949

\bibitem[{{Halpern} \& {Gotthelf}(2007)}]{HalpernGotthelf_07}
{Halpern}, J.~P. \& {Gotthelf}, E.~V. 2007, \apj, 669, 579

\bibitem[{{Hambaryan} {et~al.}(2002){Hambaryan}, {Hasinger}, {Schwope}, \&
  {Schulz}}]{2002A&A...381...98H}
{Hambaryan}, V., {Hasinger}, G., {Schwope}, A.~D., \& {Schulz}, N.~S. 2002,
  \aap, 381, 98

\bibitem[{{Hambaryan} {et~al.}(2009){Hambaryan}, {Neuh{\"a}user}, {Haberl},
  {Hohle}, \& {Schwope}}]{2009A&A...497L...9H}
{Hambaryan}, V., {Neuh{\"a}user}, R., {Haberl}, F., {Hohle}, M.~M., \&
  {Schwope}, A.~D. 2009, \aap, 497, L9

\bibitem[{{Hambaryan} {et~al.}(2017){Hambaryan}, {Suleimanov}, {Haberl},
  {Schwope}, {Neuh{\"a}user}, {Hohle}, \& {Werner}}]{2017A&A...601A.108H}
{Hambaryan}, V., {Suleimanov}, V., {Haberl}, F., {et~al.} 2017, \aap, 601, A108

\bibitem[{{Hambaryan} {et~al.}(2011){Hambaryan}, {Suleimanov}, {Schwope},
  {Neuh{\"a}user}, {Werner}, \& {Potekhin}}]{2011A&A...534A..74H}
{Hambaryan}, V., {Suleimanov}, V., {Schwope}, A.~D., {et~al.} 2011, \aap, 534,
  A74

\bibitem[{{Heindl} {et~al.}(2003){Heindl}, {Coburn}, {Kreykenbohm}, \&
  {Wilms}}]{Heindl_etal03}
{Heindl}, W., {Coburn}, W., {Kreykenbohm}, I., \& {Wilms}, J. 2003, The
  Astronomer's Telegram, 200

\bibitem[{{Heindl} {et~al.}(2000){Heindl}, {Coburn}, {Gruber}, {Pelling},
  {Rothschild}, {Kretschmar}, {Kreykenbohm}, {Wilms}, {Pottschmidt}, \&
  {Staubert}}]{Heindl_etal00}
{Heindl}, W.~A., {Coburn}, W., {Gruber}, D.~E., {et~al.} 2000, in American
  Institute of Physics Conference Series, Vol. 510, American Institute of
  Physics Conference Series, ed. M.~L. {McConnell} \& J.~M. {Ryan}, 178--182

\bibitem[{{Heindl} {et~al.}(1999){Heindl}, {Coburn}, {Gruber}, {Pelling},
  {Rothschild}, {Wilms}, {Pottschmidt}, \& {Staubert}}]{Heindl_etal99a}
{Heindl}, W.~A., {Coburn}, W., {Gruber}, D.~E., {et~al.} 1999, \apjl, 521, L49

\bibitem[{{Heindl} {et~al.}(2001){Heindl}, {Coburn}, {Gruber}, {Rothschild},
  {Kreykenbohm}, {Wilms}, \& {Staubert}}]{Heindl_etal01}
{Heindl}, W.~A., {Coburn}, W., {Gruber}, D.~E., {et~al.} 2001, \apjl, 563, L35

\bibitem[{{Heindl} {et~al.}(2004){Heindl}, {Rothschild}, {Coburn}, {Staubert},
  {Wilms}, {Kreykenbohm}, \& {Kretschmar}}]{Heindl_etal04}
{Heindl}, W.~A., {Rothschild}, R.~E., {Coburn}, W., {et~al.} 2004, in American
  Institute of Physics Conference Series, Vol. 714, X-ray Timing 2003: Rossi
  and Beyond, ed. P.~{Kaaret}, F.~K. {Lamb}, \& J.~H. {Swank}, 323--330

\bibitem[{{Hemphill} {et~al.}(2013){Hemphill}, {Rothschild}, {Caballero},
  {Pottschmidt}, {K{\"u}hnel}, {F{\"u}rst}, \& {Wilms}}]{Hemphill_etal13}
{Hemphill}, P.~B., {Rothschild}, R.~E., {Caballero}, I., {et~al.} 2013, \apj,
  777, 61

\bibitem[{{Hemphill} {et~al.}(2016){Hemphill}, {Rothschild}, {F{\"u}rst},
  {Grinberg}, {Klochkov}, {Kretschmar}, {Pottschmidt}, {Staubert}, \&
  {Wilms}}]{Hemphill_etal16}
{Hemphill}, P.~B., {Rothschild}, R.~E., {F{\"u}rst}, F., {et~al.} 2016, \mnras,
  458, 2745

\bibitem[{{Hemphill} {et~al.}(2014){Hemphill}, {Rothschild}, {Markowitz},
  {F{\"u}rst}, {Pottschmidt}, \& {Wilms}}]{Hemphill_etal14}
{Hemphill}, P.~B., {Rothschild}, R.~E., {Markowitz}, A., {et~al.} 2014, \apj,
  792, 14

\bibitem[{{Ho} {et~al.}(2003){Ho}, {Lai}, {Potekhin}, \&
  {Chabrier}}]{2003ApJ...599.1293H}
{Ho}, W.~C.~G., {Lai}, D., {Potekhin}, A.~Y., \& {Chabrier}, G. 2003, \apj,
  599, 1293

\bibitem[{{Hohle} {et~al.}(2012{\natexlab{a}}){Hohle}, {Haberl}, {Vink}, {de
  Vries}, \& {Neuh{\"a}user}}]{2012MNRAS.419.1525H}
{Hohle}, M.~M., {Haberl}, F., {Vink}, J., {de Vries}, C.~P., \&
  {Neuh{\"a}user}, R. 2012{\natexlab{a}}, \mnras, 419, 1525

\bibitem[{{Hohle} {et~al.}(2012{\natexlab{b}}){Hohle}, {Haberl}, {Vink}, {de
  Vries}, {Turolla}, {Zane}, \& {M{\'e}ndez}}]{2012MNRAS.423.1194H}
{Hohle}, M.~M., {Haberl}, F., {Vink}, J., {et~al.} 2012{\natexlab{b}}, \mnras,
  423, 1194

\bibitem[{{Hohle} {et~al.}(2009){Hohle}, {Haberl}, {Vink}, {Turolla},
  {Hambaryan}, {Zane}, {de Vries}, \& {M{\'e}ndez}}]{2009A&A...498..811H}
{Hohle}, M.~M., {Haberl}, F., {Vink}, J., {et~al.} 2009, \aap, 498, 811

\bibitem[{{Hulleman} {et~al.}(1998){Hulleman}, {in 't Zand}, \&
  {Heise}}]{Hulleman_etal98}
{Hulleman}, F., {in 't Zand}, J.~J.~M., \& {Heise}, J. 1998, \aap, 337, L25

\bibitem[{{Iaria} {et~al.}(2015){Iaria}, {Di Salvo}, {Matranga}, {Galiano},
  {D'A{\'{\i}}}, {Riggio}, {Burderi}, {Sanna}, {Ferrigno}, {Del Santo},
  {Pintore}, \& {Robba}}]{Iaria_etal15}
{Iaria}, R., {Di Salvo}, T., {Matranga}, M., {et~al.} 2015, \aap, 577, A63

\bibitem[{{Ibrahim} {et~al.}(2002){Ibrahim}, {Safi-Harb}, {Swank}, {Parke},
  {Zane}, \& {Turolla}}]{Ibrahim_etal02}
{Ibrahim}, A.~I., {Safi-Harb}, S., {Swank}, J.~H., {et~al.} 2002, \apjl, 574,
  L51

\bibitem[{{Ibrahim} {et~al.}(2003){Ibrahim}, {Swank}, \&
  {Parke}}]{Ibrahim_etal03}
{Ibrahim}, A.~I., {Swank}, J.~H., \& {Parke}, W. 2003, \apjl, 584, L17

\bibitem[{{Ikhsanov} \& {Beskrovnaya}(2013)}]{Ikhsarov_etal13}
{Ikhsanov}, N.~R. \& {Beskrovnaya}, N.~G. 2013, Astronomy Reports, 57, 287

\bibitem[{{Illarionov} \& {Sunyaev}(1975)}]{IllarionovSunyaev_75}
{Illarionov}, A.~F. \& {Sunyaev}, R.~A. 1975, \aap, 39, 185

\bibitem[{{Inoue} {et~al.}(2005){Inoue}, {Kunieda}, {White}, {Kelley},
  {Mihara}, {Terada}, {Takahashi}, {Kokubun}, {Makishima}, \& {Suzaku
  Team}}]{Inoue_etal05}
{Inoue}, H., {Kunieda}, H., {White}, N., {et~al.} 2005, IAU Circ., 613

\bibitem[{{Isenberg} {et~al.}(1998){Isenberg}, {Lamb}, \&
  {Wang}}]{Isenberg_etal98}
{Isenberg}, M., {Lamb}, D.~Q., \& {Wang}, J.~C.~L. 1998, \apj, 493, 154

\bibitem[{{Iwakiri} {et~al.}(2012){Iwakiri}, {Terada}, {Mihara}, {Angelini},
  {Tashiro}, {Enoto}, {Yamada}, {Makishima}, {Nakajima}, \&
  {Yoshida}}]{Iwakiri_etal12}
{Iwakiri}, W.~B., {Terada}, Y., {Mihara}, T., {et~al.} 2012, \apj, 751, 35

\bibitem[{{Iyer} {et~al.}(2015){Iyer}, {Mukherjee}, {Dewangan}, {Bhattacharya},
  \& {Seetha}}]{Iyer_etal15}
{Iyer}, N., {Mukherjee}, D., {Dewangan}, G.~C., {Bhattacharya}, D., \&
  {Seetha}, S. 2015, \mnras, 454, 741

\bibitem[{{Jain} {et~al.}(2010){Jain}, {Paul}, \& {Dutta}}]{Jain_etal10}
{Jain}, C., {Paul}, B., \& {Dutta}, A. 2010, \mnras, 409, 755

\bibitem[{{Jaisawal} \& {Naik}(2015{\natexlab{a}})}]{JaisawalNaik_15}
{Jaisawal}, G.~K. \& {Naik}, S. 2015{\natexlab{a}}, \mnras, 448, 620

\bibitem[{{Jaisawal} \& {Naik}(2015{\natexlab{b}})}]{Jaisawal_etal15}
{Jaisawal}, G.~K. \& {Naik}, S. 2015{\natexlab{b}}, \mnras, 453, L21

\bibitem[{{Jaisawal} \& {Naik}(2016)}]{JaisawalNaik_16}
{Jaisawal}, G.~K. \& {Naik}, S. 2016, \mnras, 461, L97

\bibitem[{{Jaisawal} \& {Naik}(2017)}]{JaisawalNaik_17}
{Jaisawal}, G.~K. \& {Naik}, S. 2017, in 7 years of MAXI: monitoring X-ray
  Transients, held 5-7 December 2016 at RIKEN. Online at <A
  href=``https://indico2.riken.jp/indico/conferenceDisplay.py?confId=2357''>
  https://indico2.riken.jp/indico/conferenceDisplay.py?confId=2357</A>, p.153,
  ed. M.~{Serino}, M.~{Shidatsu}, W.~{Iwakiri}, \& T.~{Mihara}, 153

\bibitem[{{Jaisawal} {et~al.}(2016){Jaisawal}, {Naik}, \&
  {Epili}}]{Jaisawal_etal16}
{Jaisawal}, G.~K., {Naik}, S., \& {Epili}, P. 2016, \mnras, 457, 2749

\bibitem[{{Jaisawal} {et~al.}(2013){Jaisawal}, {Naik}, \&
  {Paul}}]{Jaisawal_etal13}
{Jaisawal}, G.~K., {Naik}, S., \& {Paul}, B. 2013, \apj, 779, 54

\bibitem[{{Janot-Pacheco} {et~al.}(1981){Janot-Pacheco}, {Ilovaisky}, \&
  {Chevalier}}]{Janot-Pacheco_etal81}
{Janot-Pacheco}, E., {Ilovaisky}, S.~A., \& {Chevalier}, C. 1981, \aap, 99, 274

\bibitem[{{Johns} {et~al.}(1978){Johns}, {Koski}, {Canizares}, {McClintock},
  {Rappaport}, {Clark}, {Cominsky}, \& {Li}}]{Johns_etal78}
{Johns}, M., {Koski}, A., {Canizares}, C., {et~al.} 1978, \iaucirc, 3171

\bibitem[{{Jones} {et~al.}(1973){Jones}, {Forman}, \& {Liller}}]{Jones_etal73}
{Jones}, C.~A., {Forman}, W., \& {Liller}, W. 1973, \apjl, 182, L109

\bibitem[{{Kaper} {et~al.}(2006){Kaper}, {van der Meer}, {van Kerkwijk}, \&
  {van den Heuvel}}]{Kaper_etal06}
{Kaper}, L., {van der Meer}, A., {van Kerkwijk}, M., \& {van den Heuvel}, E.
  2006, The Messenger, 126, 27

\bibitem[{{Kaplan}(2008)}]{2008AIPC..968..129K}
{Kaplan}, D.~L. 2008, in American Institute of Physics Conference Series, Vol.
  968, Astrophysics of Compact Objects, ed. Y.-F. {Yuan}, X.-D. {Li}, \&
  D.~{Lai}, 129--136

\bibitem[{{Kaplan} \& {van Kerkwijk}(2005{\natexlab{a}})}]{2005ApJ...628L..45K}
{Kaplan}, D.~L. \& {van Kerkwijk}, M.~H. 2005{\natexlab{a}}, \apjl, 628, L45

\bibitem[{{Kaplan} \& {van Kerkwijk}(2005{\natexlab{b}})}]{2005ApJ...635L..65K}
{Kaplan}, D.~L. \& {van Kerkwijk}, M.~H. 2005{\natexlab{b}}, \apjl, 635, L65

\bibitem[{{Kaplan} \& {van Kerkwijk}(2009{\natexlab{a}})}]{2009ApJ...705..798K}
{Kaplan}, D.~L. \& {van Kerkwijk}, M.~H. 2009{\natexlab{a}}, \apj, 705, 798

\bibitem[{{Kaplan} \& {van Kerkwijk}(2009{\natexlab{b}})}]{2009ApJ...692L..62K}
{Kaplan}, D.~L. \& {van Kerkwijk}, M.~H. 2009{\natexlab{b}}, \apjl, 692, L62

\bibitem[{{Kaplan} \& {van Kerkwijk}(2011)}]{2011ApJ...740L..30K}
{Kaplan}, D.~L. \& {van Kerkwijk}, M.~H. 2011, \apjl, 740, L30

\bibitem[{{Kendziorra} {et~al.}(1994){Kendziorra}, {Kretschmar}, {Pan}, {Kunz},
  {Maisack}, {Staubert}, {Pietsch}, {Truemper}, {Efremov}, \&
  {Sunyaev}}]{Kendziorra_etal94}
{Kendziorra}, E., {Kretschmar}, P., {Pan}, H.~C., {et~al.} 1994, \aap, 291, L31

\bibitem[{{Kendziorra} {et~al.}(1992){Kendziorra}, {Mony}, {Kretschmar},
  {Maisack}, {Staubert}, {D{\"o}bereiner}, {Englhauser}, {Pietsch}, {Reppin},
  {Tr{\"u}mper}, {Efremov}, {Kaniovsky}, \& {Sunyaev}}]{Kendziorra_etal92}
{Kendziorra}, E., {Mony}, B., {Kretschmar}, P., {et~al.} 1992, in Frontiers
  Science Series, ed. Y.~{Tanaka} \& K.~{Koyama}, 51

\bibitem[{{Kendziorra} {et~al.}(1977){Kendziorra}, {Staubert}, {Pietsch},
  {Reppin}, {Sacco}, \& {Truemper}}]{Kendziorra_etal77}
{Kendziorra}, E., {Staubert}, R., {Pietsch}, W., {et~al.} 1977, \apjl, 217, L93

\bibitem[{{Kirk} \& {Meszaros}(1980)}]{KirkMeszaros_80}
{Kirk}, J.~G. \& {Meszaros}, P. 1980, \apj, 241, 1153

\bibitem[{{Klochkov} {et~al.}(2012){Klochkov}, {Doroshenko}, {Santangelo},
  {Staubert}, {Ferrigno}, {Kretschmar}, {Caballero}, {Wilms}, {Kreykenbohm},
  {Pottschmidt}, {Rothschild}, {Wilson-Hodge}, \&
  {P{\"u}hlhofer}}]{Klochkov_etal12}
{Klochkov}, D., {Doroshenko}, V., {Santangelo}, A., {et~al.} 2012, \aap, 542,
  L28

\bibitem[{{Klochkov} {et~al.}(2008{\natexlab{a}}){Klochkov}, {Santangelo},
  {Staubert}, \& {Ferrigno}}]{Klochkov_etal08c}
{Klochkov}, D., {Santangelo}, A., {Staubert}, R., \& {Ferrigno}, C.
  2008{\natexlab{a}}, \aap, 491, 833

\bibitem[{{Klochkov} {et~al.}(2008{\natexlab{b}}){Klochkov}, {Staubert},
  {Postnov}, {Shakura}, {Santangelo}, {Tsygankov}, {Lutovinov}, {Kreykenbohm},
  \& {Wilms}}]{Klochkov_etal08}
{Klochkov}, D., {Staubert}, R., {Postnov}, K., {et~al.} 2008{\natexlab{b}},
  A\&A, 482, 907

\bibitem[{{Klochkov} {et~al.}(2015){Klochkov}, {Staubert}, {Postnov}, {Wilms},
  {Rothschild}, \& {Santangelo}}]{Klochkov_etal15}
{Klochkov}, D., {Staubert}, R., {Postnov}, K., {et~al.} 2015, \aap, 578, A88

\bibitem[{{Klochkov} {et~al.}(2011){Klochkov}, {Staubert}, {Santangelo},
  {Rothschild}, \& {Ferrigno}}]{Klochkov_etal11}
{Klochkov}, D., {Staubert}, R., {Santangelo}, A., {Rothschild}, R.~E., \&
  {Ferrigno}, C. 2011, \aap, 532, A126

\bibitem[{{Klus} {et~al.}(2014){Klus}, {Ho}, {Coe}, {Corbet}, \&
  {Townsend}}]{Klus_etal14}
{Klus}, H., {Ho}, W.~C.~G., {Coe}, M.~J., {Corbet}, R.~H.~D., \& {Townsend},
  L.~J. 2014, \mnras, 437, 3863

\bibitem[{{Klu{\'z}niak} \& {Rappaport}(2007)}]{KluzniakRappaport_07}
{Klu{\'z}niak}, W. \& {Rappaport}, S. 2007, \apj, 671, 1990

\bibitem[{{Kretschmar} {et~al.}(2005){Kretschmar}, {Kreykenbohm},
  {Pottschmidt}, {Wilms}, {Coburn}, {Boggs}, {Staubert}, {Santangelo},
  {Kendziorra}, {Segreto}, {Orlandini}, {Bildsten}, \&
  {Araya-Gochez}}]{Kretschmar_etal05}
{Kretschmar}, P., {Kreykenbohm}, I., {Pottschmidt}, K., {et~al.} 2005, IAU
  Circ., 601

\bibitem[{{Kretschmar} {et~al.}(1996){Kretschmar}, {Pan}, {Kendziorra}, {Kunz},
  {Maisack}, {Staubert}, {Pietsch}, {Truemper}, {Efremov}, \&
  {Sunyaev}}]{Kretschmar_etal96}
{Kretschmar}, P., {Pan}, H.~C., {Kendziorra}, E., {et~al.} 1996, \aaps, 120,
  175

\bibitem[{{Kretschmar} {et~al.}(1997){Kretschmar}, {Pan}, {Kendziorra},
  {Maisack}, {Staubert}, {Skinner}, {Pietsch}, {Truemper}, {Efremov}, \&
  {Sunyaev}}]{Kretschmar_etal97}
{Kretschmar}, P., {Pan}, H.~C., {Kendziorra}, E., {et~al.} 1997, \aap, 325, 623

\bibitem[{{Kreykenbohm} {et~al.}(2002){Kreykenbohm}, {Coburn}, {Wilms},
  {Kretschmar}, {Staubert}, {Heindl}, \& {Rothschild}}]{Kreykenbohm_etal02}
{Kreykenbohm}, I., {Coburn}, W., {Wilms}, J., {et~al.} 2002, \aap, 395, 129

\bibitem[{{Kreykenbohm} {et~al.}(2005){Kreykenbohm}, {Mowlavi}, {Produit},
  {Soldi}, {Walter}, {Dubath}, {Lubi{\'n}ski}, {T{\"u}rler}, {Coburn},
  {Santangelo}, {Rothschild}, \& {Staubert}}]{Kreykenbohm_etal05}
{Kreykenbohm}, I., {Mowlavi}, N., {Produit}, N., {et~al.} 2005, \aap, 433, L45

\bibitem[{{Kreykenbohm} {et~al.}(2004){Kreykenbohm}, {Wilms}, {Coburn},
  {Kuster}, {Rothschild}, {Heindl}, {Kretschmar}, \&
  {Staubert}}]{Kreykenbohm_etal04}
{Kreykenbohm}, I., {Wilms}, J., {Coburn}, W., {et~al.} 2004, A\&A, 427, 975

\bibitem[{{K{\"u}hnel} {et~al.}(2013){K{\"u}hnel}, {M{\"u}ller}, {Kreykenbohm},
  {F{\"u}rst}, {Pottschmidt}, {Rothschild}, {Caballero}, {Grinberg},
  {Sch{\"o}nherr}, {Shrader}, {Klochkov}, {Staubert}, {Ferrigno},
  {Torrej{\'o}n}, {Mart{\'{\i}}nez-N{\'u}{\~n}ez}, \& {Wilms}}]{Kuehnel_etal13}
{K{\"u}hnel}, M., {M{\"u}ller}, S., {Kreykenbohm}, I., {et~al.} 2013, \aap,
  555, A95

\bibitem[{{Kulkarni} \& {Romanova}(2013)}]{KulkarniRomanova_13}
{Kulkarni}, A.~K. \& {Romanova}, M.~M. 2013, \mnras, 433, 3048

\bibitem[{{La Barbera} {et~al.}(2001){La Barbera}, {Burderi}, {Di Salvo},
  {Iaria}, \& {Robba}}]{LaBarbera_etal01}
{La Barbera}, A., {Burderi}, L., {Di Salvo}, T., {Iaria}, R., \& {Robba}, N.~R.
  2001, \apj, 553, 375

\bibitem[{{La Barbera} {et~al.}(2003){La Barbera}, {Santangelo}, {Orlandini},
  \& {Segreto}}]{LaBarbera_etal03}
{La Barbera}, A., {Santangelo}, A., {Orlandini}, M., \& {Segreto}, A. 2003,
  \aap, 400, 993

\bibitem[{{La Barbera} {et~al.}(2005){La Barbera}, {Segreto}, {Santangelo},
  {Kreykenbohm}, \& {Orlandini}}]{LaBarbera_etal05}
{La Barbera}, A., {Segreto}, A., {Santangelo}, A., {Kreykenbohm}, I., \&
  {Orlandini}, M. 2005, \aap, 438, 617

\bibitem[{{La Parola} {et~al.}(2016){La Parola}, {Cusumano}, {Segreto}, \&
  {D'A{\`i}}}]{LaParola_etal16}
{La Parola}, V., {Cusumano}, G., {Segreto}, A., \& {D'A{\`i}}, A. 2016, \mnras,
  463, 185

\bibitem[{{Langer} {et~al.}(1980){Langer}, {McCray}, \& {Baan}}]{Langer_etal80}
{Langer}, S.~H., {McCray}, R., \& {Baan}, W.~A. 1980, \apj, 238, 731

\bibitem[{{Langer} \& {Rappaport}(1982)}]{LangerRappaport_82}
{Langer}, S.~H. \& {Rappaport}, S. 1982, \apj, 257, 733

\bibitem[{{Levine} {et~al.}(1988){Levine}, {Ma}, {McClintock}, {Rappaport},
  {van der Klis}, \& {Verbunt}}]{Levine_etal88}
{Levine}, A., {Ma}, C.~P., {McClintock}, J., {et~al.} 1988, \apj, 327, 732

\bibitem[{{Li} {et~al.}(2012{\natexlab{a}}){Li}, {Wang}, \& {Zhao}}]{Li_etal12}
{Li}, J., {Wang}, W., \& {Zhao}, Y. 2012{\natexlab{a}}, \mnras, 423, 2854

\bibitem[{{Li} {et~al.}(2012{\natexlab{b}}){Li}, {Zhang}, {Torres}, {Papitto},
  {Chen}, \& {Wang}}]{LiWang_etal12}
{Li}, J., {Zhang}, S., {Torres}, D.~F., {et~al.} 2012{\natexlab{b}}, \mnras,
  426, L16

\bibitem[{{Liller}(1975)}]{Liller_75}
{Liller}, W. 1975, IAU Circ., 2780

\bibitem[{{Lin} {et~al.}(2010){Lin}, {Takata}, {Kong}, \& {Hwang}}]{Lin_etal10}
{Lin}, L.~C.-C., {Takata}, J., {Kong}, A.~K.~H., \& {Hwang}, C.-Y. 2010,
  \mnras, 409, 1127

\bibitem[{{Lipunov}(1992)}]{Lipunov_92}
{Lipunov}, V.~M. 1992, {Astrophysics of neutron stars} (Berlin ; New York :
  Springer-Verlag, c1992.), 5713

\bibitem[{{Litwin} {et~al.}(2001){Litwin}, {Brown}, \&
  {Rosner}}]{Litwin_etal01}
{Litwin}, C., {Brown}, E.~F., \& {Rosner}, R. 2001, \apj, 553, 788

\bibitem[{{Liu} \& {Mirabel}(2005)}]{LiuMirabel_05}
{Liu}, Q.~Z. \& {Mirabel}, I.~F. 2005, \aap, 429, 1125

\bibitem[{{Lodenquai} {et~al.}(1974){Lodenquai}, {Canuto}, {Ruderman}, \&
  {Tsuruta}}]{Lodenquai_etal74}
{Lodenquai}, J., {Canuto}, V., {Ruderman}, M., \& {Tsuruta}, S. 1974, \apj,
  190, 141

\bibitem[{{Lovelace} {et~al.}(1995){Lovelace}, {Romanova}, \&
  {Bisnovatyi-Kogan}}]{Lovelace_etal95}
{Lovelace}, R.~V.~E., {Romanova}, M.~M., \& {Bisnovatyi-Kogan}, G.~S. 1995,
  \mnras, 275, 244

\bibitem[{{Lutovinov} {et~al.}(2012){Lutovinov}, {Tsygankov}, \&
  {Chernyakova}}]{Lutovinov_etal12}
{Lutovinov}, A., {Tsygankov}, S., \& {Chernyakova}, M. 2012, \mnras, 423, 1978

\bibitem[{{Lutovinov} {et~al.}(2016){Lutovinov}, {Buckley}, {Townsend},
  {Tsygankov}, \& {Kennea}}]{Lutovinov_etal16}
{Lutovinov}, A.~A., {Buckley}, D.~A.~H., {Townsend}, L.~J., {Tsygankov}, S.~S.,
  \& {Kennea}, J. 2016, \mnras, 462, 3823

\bibitem[{{Lutovinov} {et~al.}(2017{\natexlab{a}}){Lutovinov}, {Tsygankov},
  {Krivonos}, {Molkov}, \& {Poutanen}}]{Lutovinov_etal17a}
{Lutovinov}, A.~A., {Tsygankov}, S.~S., {Krivonos}, R.~A., {Molkov}, S.~V., \&
  {Poutanen}, J. 2017{\natexlab{a}}, \apj, 834, 209

\bibitem[{{Lutovinov} {et~al.}(2017{\natexlab{b}}){Lutovinov}, {Tsygankov},
  {Postnov}, {Krivonos}, {Molkov}, \& {Tomsick}}]{Lutovinov_etal17b}
{Lutovinov}, A.~A., {Tsygankov}, S.~S., {Postnov}, K.~A., {et~al.}
  2017{\natexlab{b}}, \mnras, 466, 593

\bibitem[{{Lutovinov} {et~al.}(2015){Lutovinov}, {Tsygankov}, {Suleimanov},
  {Mushtukov}, {Doroshenko}, {Nagirner}, \& {Poutanen}}]{Lutovinov_etal15}
{Lutovinov}, A.~A., {Tsygankov}, S.~S., {Suleimanov}, V.~F., {et~al.} 2015,
  \mnras, 448, 2175

\bibitem[{{Lyubarskii} \& {Sunyaev}(1982)}]{LyubarskiiSunyaev_82}
{Lyubarskii}, Y.~E. \& {Sunyaev}, R.~A. 1982, Soviet Astronomy Letters, 8, 330

\bibitem[{{Maisack} {et~al.}(1996){Maisack}, {Grove}, {Johnson}, {Jung},
  {Kendziorra}, {Kinzer}, {Kretschmar}, {Kurfess}, {Starr}, {Staubert}, \&
  {Strickman}}]{Maisack_etal96}
{Maisack}, M., {Grove}, J.~E., {Johnson}, W.~N., {et~al.} 1996, \aaps, 120, 179

\bibitem[{{Maitra}(2017)}]{Maitra_17}
{Maitra}, C. 2017, Journal of Astrophysics and Astronomy, 38, 50

\bibitem[{{Maitra} \& {Paul}(2013{\natexlab{a}})}]{MaitraPaul_13a}
{Maitra}, C. \& {Paul}, B. 2013{\natexlab{a}}, \apj, 771, 96

\bibitem[{{Maitra} \& {Paul}(2013{\natexlab{b}})}]{MaitraPaul_13b}
{Maitra}, C. \& {Paul}, B. 2013{\natexlab{b}}, \apj, 763, 79

\bibitem[{{Maitra} {et~al.}(2018){Maitra}, {Paul}, {Haberl}, \&
  {Vasilopoulos}}]{Maitra_etal18}
{Maitra}, C., {Paul}, B., {Haberl}, F., \& {Vasilopoulos}, G. 2018, \mnras,
  480, L136

\bibitem[{{Maitra} {et~al.}(2012){Maitra}, {Paul}, \& {Naik}}]{Maitra_etal12}
{Maitra}, C., {Paul}, B., \& {Naik}, S. 2012, \mnras, 420, 2307

\bibitem[{{Makishima} \& {Mihara}(1992)}]{MakishimaMihara_92}
{Makishima}, K. \& {Mihara}. 1992, {Frontiers O X-ray Astronomy} (Universal
  Academy Press Inc, Tokyo), 23

\bibitem[{{Makishima} {et~al.}(1990{\natexlab{a}}){Makishima}, {Mihara},
  {Ishida}, {Ohashi}, {Sakao}, {Tashiro}, {Tsuru}, {Kii}, {Makino}, {Murakami},
  {Nagase}, {Tanaka}, {Kunieda}, {Tawara}, {Kitamoto}, {Miyamoto}, {Yoshida},
  \& {Turner}}]{Makishima_etal90}
{Makishima}, K., {Mihara}, T., {Ishida}, M., {et~al.} 1990{\natexlab{a}},
  \apjl, 365, L59

\bibitem[{{Makishima} {et~al.}(1999){Makishima}, {Mihara}, {Nagase}, \&
  {Tanaka}}]{Makishima_etal99}
{Makishima}, K., {Mihara}, T., {Nagase}, F., \& {Tanaka}, Y. 1999, \apj, 525,
  978

\bibitem[{{Makishima} {et~al.}(1990{\natexlab{b}}){Makishima}, {Ohashi},
  {Kawai}, {Matsuoka}, {Koyama}, {Kunieda}, {Tawara}, {Ushimaru}, {Corbet},
  {Inoue}, {Kii}, {Makino}, {Mitsuda}, {Murakami}, {Nagase}, {Ogawara},
  {Tanaka}, {Kitamoto}, {Miyamoto}, {Tsunemi}, \&
  {Yamashita}}]{Makishima_etal90a}
{Makishima}, K., {Ohashi}, T., {Kawai}, N., {et~al.} 1990{\natexlab{b}}, \pasj,
  42, 295

\bibitem[{{Malacaria} {et~al.}(2015){Malacaria}, {Klochkov}, {Santangelo}, \&
  {Staubert}}]{Malacaria_etal15}
{Malacaria}, C., {Klochkov}, D., {Santangelo}, A., \& {Staubert}, R. 2015,
  \aap, 581, A121

\bibitem[{{Manousakis} {et~al.}(2008){Manousakis}, {Beckmann}, {Bianchin},
  {Brandt}, {Chenevez}, {Hermsen}, {von Kienlin}, {Krivonos}, {Mas-Hesse},
  {Parmar}, \& {Reglero}}]{Manousakis_etal08}
{Manousakis}, A., {Beckmann}, V., {Bianchin}, V., {et~al.} 2008, The
  Astronomer's Telegram, 1613

\bibitem[{{Manousakis} {et~al.}(2009){Manousakis}, {Walter}, {Audard}, \&
  {Lanz}}]{Manousakis_etal09}
{Manousakis}, A., {Walter}, R., {Audard}, M., \& {Lanz}, T. 2009, \aap, 498,
  217

\bibitem[{{Maravelias} {et~al.}(2018){Maravelias}, {Antoniou}, {Zezas},
  {Strantzalis}, {Hatzidimitriou}, \& {Haberl}}]{Maravelias_etal18}
{Maravelias}, G., {Antoniou}, V., {Zezas}, A., {et~al.} 2018, The Astronomer's
  Telegram, 12224

\bibitem[{{Marcu-Cheatham} {et~al.}(2015){Marcu-Cheatham}, {Pottschmidt},
  {K{\"u}hnel}, {M{\"u}ller}, {Falkner}, {Caballero}, {Finger}, {Jenke},
  {Wilson-Hodge}, {F{\"u}rst}, {Grinberg}, {Hemphill}, {Kreykenbohm},
  {Klochkov}, {Rothschild}, {Terada}, {Enoto}, {Iwakiri}, {Wolff}, {Becker},
  {Wood}, \& {Wilms}}]{Marcu-Cheatham_etal15}
{Marcu-Cheatham}, D.~M., {Pottschmidt}, K., {K{\"u}hnel}, M., {et~al.} 2015,
  \apj, 815, 44

\bibitem[{{Marcu-Cheatham} {et~al.}(2018){Marcu-Cheatham}, {Pottschmidt},
  {Wolff}, {Gottlieb}, {Hemphill}, \& {Ballhausen}}]{Marcu-Cheatham_etal18}
{Marcu-Cheatham}, D.~M., {Pottschmidt}, K., {Wolff}, M.~T., {et~al.} 2018,
  \mnras, 00, 00

\bibitem[{{Markwardt} {et~al.}(2007){Markwardt}, {Swank}, \&
  {Corbet}}]{Markwardt_etal07}
{Markwardt}, C.~B., {Swank}, J.~H., \& {Corbet}, R. 2007, The Astronomer's
  Telegram, 1176

\bibitem[{{Mart{\'{\i}}nez-N{\'u}{\~n}ez}
  {et~al.}(2015){Mart{\'{\i}}nez-N{\'u}{\~n}ez}, {Sander},
  {G{\'{\i}}menez-Garc{\'{\i}}a}, {G{\'o}nzalez-Gal{\'a}n}, {Torrej{\'o}n},
  {G{\'o}nzalez-Fern{\'a}ndez}, \& {Hamann}}]{Martinez-Nunez_etal15}
{Mart{\'{\i}}nez-N{\'u}{\~n}ez}, S., {Sander}, A.,
  {G{\'{\i}}menez-Garc{\'{\i}}a}, A., {et~al.} 2015, \aap, 578, A107

\bibitem[{{Masetti} {et~al.}(2004){Masetti}, {Dal Fiume}, {Amati}, {Del Sordo},
  {Frontera}, {Orlandini}, \& {Palazzi}}]{Masetti_etal04}
{Masetti}, N., {Dal Fiume}, D., {Amati}, L., {et~al.} 2004, \aap, 423, 311

\bibitem[{{Masetti} {et~al.}(2006){Masetti}, {Orlandini}, {dal Fiume}, {del
  Sordo}, {Amati}, {Frontera}, {Palazzi}, \& {Santangelo}}]{Masetti_etal06}
{Masetti}, N., {Orlandini}, M., {dal Fiume}, D., {et~al.} 2006, \aap, 445, 653

\bibitem[{{Masetti} {et~al.}(2014){Masetti}, {Orlandini}, {Parisi}, {Fiocchi},
  {Sanchez-Fernandez}, \& {Kuulkers}}]{Masetti_etal14}
{Masetti}, N., {Orlandini}, M., {Parisi}, P., {et~al.} 2014, The Astronomer's
  Telegram, 5997

\bibitem[{{Mason} \& {Cordova}(1982)}]{MasonCordova_82}
{Mason}, K.~O. \& {Cordova}, F.~A. 1982, \apj, 255, 603

\bibitem[{{Mason} {et~al.}(1978){Mason}, {Murdin}, {Parkes}, \&
  {Visvanathan}}]{Mason_etal78}
{Mason}, K.~O., {Murdin}, P.~G., {Parkes}, G.~E., \& {Visvanathan}, N. 1978,
  \mnras, 184, 45P

\bibitem[{{McBride} {et~al.}(2008){McBride}, {Coe}, {Negueruela}, {Schurch}, \&
  {McGowan}}]{McBride_etal08}
{McBride}, V.~A., {Coe}, M.~J., {Negueruela}, I., {Schurch}, M.~P.~E., \&
  {McGowan}, K.~E. 2008, \mnras, 388, 1198

\bibitem[{{McBride} {et~al.}(2006){McBride}, {Wilms}, {Coe}, {Kreykenbohm},
  {Rothschild}, {Coburn}, {Galache}, {Kretschmar}, {Edge}, \&
  {Staubert}}]{McBride_etal06}
{McBride}, V.~A., {Wilms}, J., {Coe}, M.~J., {et~al.} 2006, \aap, 451, 267

\bibitem[{{McBride} {et~al.}(2007){McBride}, {Wilms}, {Kreykenbohm}, {Coe},
  {Rothschild}, {Kretschmar}, {Pottschmidt}, {Fisher}, \&
  {Hamson}}]{McBride_etal07}
{McBride}, V.~A., {Wilms}, J., {Kreykenbohm}, I., {et~al.} 2007, \aap, 470,
  1065

\bibitem[{{McClintock} {et~al.}(1977){McClintock}, {Rappaport}, {Nugent}, \&
  {Li}}]{McClintock_etal77}
{McClintock}, J.~E., {Rappaport}, S.~A., {Nugent}, J.~J., \& {Li}, F.~K. 1977,
  \apjl, 216, L15

\bibitem[{{McClintock} {et~al.}(1971){McClintock}, {Ricker}, \&
  {Lewin}}]{McClintock_etal71}
{McClintock}, J.~E., {Ricker}, G.~R., \& {Lewin}, W.~H.~G. 1971, \apjl, 166,
  L73

\bibitem[{{Melrose} \& {Zhelezniyakov}(1981)}]{MelroseZheleznyakov_81}
{Melrose}, D.~B. \& {Zhelezniyakov}, V.~V. 1981, \aap, 95, 86

\bibitem[{{Meszaros}(1978)}]{Meszaros_78}
{Meszaros}, P. 1978, \aap, 63, L19

\bibitem[{{Meszaros}(1992)}]{Meszaros_92}
{Meszaros}, P. 1992, Journal of the British Astronomical Association, 102, 287

\bibitem[{{Meszaros} {et~al.}(1983){Meszaros}, {Harding}, {Kirk}, \&
  {Galloway}}]{Meszaros_etal83}
{Meszaros}, P., {Harding}, A.~K., {Kirk}, J.~G., \& {Galloway}, D.~J. 1983,
  \apjl, 266, L33

\bibitem[{{Mignani} {et~al.}(2013){Mignani}, {Vande Putte}, {Cropper},
  {Turolla}, {Zane}, {Pellizza}, {Bignone}, {Sartore}, \&
  {Treves}}]{2013MNRAS.429.3517M}
{Mignani}, R.~P., {Vande Putte}, D., {Cropper}, M., {et~al.} 2013, \mnras, 429,
  3517

\bibitem[{Mihara(1995)}]{Mihara_95}
Mihara, T. 1995, PhD thesis, Thesis, Univ. of Tokyo

\bibitem[{{Mihara} {et~al.}(1991){Mihara}, {Makishima}, {Kamijo}, {Ohashi},
  {Nagase}, {Tanaka}, \& {Koyama}}]{Mihara_etal91a}
{Mihara}, T., {Makishima}, K., {Kamijo}, S., {et~al.} 1991, \apjl, 379, L61

\bibitem[{{Mihara} {et~al.}(1995){Mihara}, {Makishima}, \&
  {Nagase}}]{Mihara_etal95}
{Mihara}, T., {Makishima}, K., \& {Nagase}, F. 1995, in Bulletin of the
  American Astronomical Society, Vol.~27, American Astronomical Society Meeting
  Abstracts, 1434

\bibitem[{{Mihara} {et~al.}(2004){Mihara}, {Makishima}, \&
  {Nagase}}]{Mihara_etal04}
{Mihara}, T., {Makishima}, K., \& {Nagase}, F. 2004, \apj, 610, 390

\bibitem[{Mihara {et~al.}(1990)}]{Mihara_90}
Mihara, T. {et~al.} 1990, \nat, 346, 250

\bibitem[{Mihara {et~al.}(1991)}]{Mihara_etal91}
Mihara, T. {et~al.} 1991, Adv. Space Res., 22(7), 987

\bibitem[{{Miller} {et~al.}(1987){Miller}, {Salpeter}, \&
  {Wasserman}}]{Miller_etal87}
{Miller}, G.~S., {Salpeter}, E.~E., \& {Wasserman}, I. 1987, \apj, 314, 215

\bibitem[{{Miyasaka} {et~al.}(2013){Miyasaka}, {Bachetti}, {Harrison},
  {F{\"u}rst}, {Barret}, {Bellm}, {Boggs}, {Chakrabarty}, {Chenevez},
  {Christensen}, {Craig}, {Grefenstette}, {Hailey}, {Madsen}, {Natalucci},
  {Pottschmidt}, {Stern}, {Tomsick}, {Walton}, {Wilms}, \&
  {Zhang}}]{Miyasaka_etal13}
{Miyasaka}, H., {Bachetti}, M., {Harrison}, F.~A., {et~al.} 2013, \apj, 775, 65

\bibitem[{{Motch} {et~al.}(1997){Motch}, {Haberl}, {Dennerl}, {Pakull}, \&
  {Janot-Pacheco}}]{Motch_etal97}
{Motch}, C., {Haberl}, F., {Dennerl}, K., {Pakull}, M., \& {Janot-Pacheco}, E.
  1997, \aap, 323, 853

\bibitem[{{Motch} {et~al.}(1999){Motch}, {Haberl}, {Zickgraf}, {Hasinger}, \&
  {Schwope}}]{1999A&A...351..177M}
{Motch}, C., {Haberl}, F., {Zickgraf}, F.-J., {Hasinger}, G., \& {Schwope},
  A.~D. 1999, \aap, 351, 177

\bibitem[{{Motch} {et~al.}(2009){Motch}, {Pires}, {Haberl}, {Schwope}, \&
  {Zavlin}}]{2009A&A...497..423M}
{Motch}, C., {Pires}, A.~M., {Haberl}, F., {Schwope}, A., \& {Zavlin}, V.~E.
  2009, \aap, 497, 423

\bibitem[{{Mowlavi} {et~al.}(2006){Mowlavi}, {Kreykenbohm}, {Shaw},
  {Pottschmidt}, {Wilms}, {Rodriguez}, {Produit}, {Soldi}, {Larsson}, \&
  {Dubath}}]{Mowlavi_etal06}
{Mowlavi}, N., {Kreykenbohm}, I., {Shaw}, S.~E., {et~al.} 2006, \aap, 451, 187

\bibitem[{{Mukherjee} \& {Bhattacharya}(2012)}]{MukhBhatt_12}
{Mukherjee}, D. \& {Bhattacharya}, D. 2012, \mnras, 420, 720

\bibitem[{{Mukherjee} {et~al.}(2013){Mukherjee}, {Bhattacharya}, \&
  {Mignone}}]{Mukherjee_etal13b}
{Mukherjee}, D., {Bhattacharya}, D., \& {Mignone}, A. 2013, \mnras, 435, 718

\bibitem[{{Mukherjee} {et~al.}(2014){Mukherjee}, {Bhattacharya}, \&
  {Mignone}}]{Mukherjee_etal14}
{Mukherjee}, D., {Bhattacharya}, D., \& {Mignone}, A. 2014, in European
  Physical Journal Web of Conferences, Vol.~64, European Physical Journal Web
  of Conferences, 2004

\bibitem[{{M{\"u}ller} {et~al.}(2013{\natexlab{a}}){M{\"u}ller}, {Klochkov},
  {Caballero}, \& {Santangelo}}]{DMueller_etal13}
{M{\"u}ller}, D., {Klochkov}, D., {Caballero}, I., \& {Santangelo}, A.
  2013{\natexlab{a}}, \aap, 552, A81

\bibitem[{{M{\"u}ller} {et~al.}(2013{\natexlab{b}}){M{\"u}ller}, {Ferrigno},
  {K{\"u}hnel}, {Sch{\"o}nherr}, {Becker}, {Wolff}, {Hertel}, {Schwarm},
  {Grinberg}, {Obst}, {Caballero}, {Pottschmidt}, {F{\"u}rst}, {Kreykenbohm},
  {Rothschild}, {Hemphill}, {N{\'u}{\~n}ez}, {Torrej{\'o}n}, {Klochkov},
  {Staubert}, \& {Wilms}}]{SMueller_etal13}
{M{\"u}ller}, S., {Ferrigno}, C., {K{\"u}hnel}, M., {et~al.}
  2013{\natexlab{b}}, \aap, 551, A6

\bibitem[{{M{\"u}ller} {et~al.}(2013{\natexlab{c}}){M{\"u}ller}, {Ferrigno},
  {K{\"u}hnel}, {Sch{\"o}nherr}, {Becker}, {Wolff}, {Hertel}, {Schwarm},
  {Grinberg}, {Obst}, {Caballero}, {Pottschmidt}, {F{\"u}rst}, {Kreykenbohm},
  {Rothschild}, {Hemphill}, {N{\'u}{\~n}ez}, {Torrej{\'o}n}, {Klochkov},
  {Staubert}, \& {Wilms}}]{Mueller_etal13}
{M{\"u}ller}, S., {Ferrigno}, C., {K{\"u}hnel}, M., {et~al.}
  2013{\natexlab{c}}, \aap, 551, A6

\bibitem[{{Mushtukov} {et~al.}(2015{\natexlab{a}}){Mushtukov}, {Suleimanov},
  {Tsygankov}, \& {Poutanen}}]{Mushtukov_etal15a}
{Mushtukov}, A.~A., {Suleimanov}, V.~F., {Tsygankov}, S.~S., \& {Poutanen}, J.
  2015{\natexlab{a}}, \mnras, 447, 1847

\bibitem[{{Mushtukov} {et~al.}(2015{\natexlab{b}}){Mushtukov}, {Tsygankov},
  {Serber}, {Suleimanov}, \& {Poutanen}}]{Mushtukov_etal15b}
{Mushtukov}, A.~A., {Tsygankov}, S.~S., {Serber}, A.~V., {Suleimanov}, V.~F.,
  \& {Poutanen}, J. 2015{\natexlab{b}}, \mnras, 454, 2714

\bibitem[{{Nagase}(1989)}]{Nagase_89}
{Nagase}, F. 1989, PASJ, 41, 1

\bibitem[{{Nagase} {et~al.}(1991){Nagase}, {Dotani}, {Tanaka}, {Makishima},
  {Mihara}, {Sakao}, {Tsunemi}, {Kitamoto}, {Tamura}, {Yoshida}, \&
  {Nakamura}}]{Nagase_etal91}
{Nagase}, F., {Dotani}, T., {Tanaka}, Y., {et~al.} 1991, \apjl, 375, L49

\bibitem[{{Nagel}(1980)}]{Nagel_80}
{Nagel}, W. 1980, \apj, 236, 904

\bibitem[{{Nagel}(1981{\natexlab{a}})}]{Nagel_81a}
{Nagel}, W. 1981{\natexlab{a}}, \apj, 251, 288

\bibitem[{{Nagel}(1981{\natexlab{b}})}]{Nagel_81b}
{Nagel}, W. 1981{\natexlab{b}}, \apj, 251, 278

\bibitem[{{Naik} {et~al.}(2006){Naik}, {Callanan}, {Paul}, \&
  {Dotani}}]{Naik_etal06}
{Naik}, S., {Callanan}, P.~J., {Paul}, B., \& {Dotani}, T. 2006, \apj, 647,
  1293

\bibitem[{{Naik} {et~al.}(2008){Naik}, {Dotani}, {Terada}, {Nakajima},
  {Mihara}, {Suzuki}, {Makishima}, {Sudoh}, {Kitamoto}, {Nagase}, {Enoto}, \&
  {Takahashi}}]{Naik_etal08}
{Naik}, S., {Dotani}, T., {Terada}, Y., {et~al.} 2008, \apj, 672, 516

\bibitem[{{Nakajima} {et~al.}(2010){Nakajima}, {Mihara}, \&
  {Makishima}}]{Nakajima_etal10}
{Nakajima}, M., {Mihara}, T., \& {Makishima}, K. 2010, \apj, 710, 1755

\bibitem[{{Nakajima} {et~al.}(2006){Nakajima}, {Mihara}, {Makishima}, \&
  {Niko}}]{Nakajima_etal06}
{Nakajima}, M., {Mihara}, T., {Makishima}, K., \& {Niko}, H. 2006, ApJ, 646,
  1125

\bibitem[{{Negueruela} \& {Marco}(2006)}]{NegueruelaMarco_06}
{Negueruela}, I. \& {Marco}, A. 2006, The Astronomer's Telegram, 739

\bibitem[{{Negueruela} \& {Okazaki}(2001)}]{NegueruelaOkazaki_01}
{Negueruela}, I. \& {Okazaki}, A.~T. 2001, \aap, 369, 108

\bibitem[{{Negueruela} {et~al.}(1999){Negueruela}, {Roche}, {Fabregat}, \&
  {Coe}}]{Negueruela_etal99}
{Negueruela}, I., {Roche}, P., {Fabregat}, J., \& {Coe}, M.~J. 1999, \mnras,
  307, 695

\bibitem[{{Nelson} {et~al.}(1993){Nelson}, {Salpeter}, \&
  {Wasserman}}]{Nelson_etal93}
{Nelson}, R.~W., {Salpeter}, E.~E., \& {Wasserman}, I. 1993, \apj, 418, 874

\bibitem[{{Nishimura}(2008)}]{Nishimura_08}
{Nishimura}, O. 2008, \apj, 672, 1127

\bibitem[{{Nishimura}(2011)}]{Nishimura_11}
{Nishimura}, O. 2011, \apj, 730, 106

\bibitem[{{Nishimura}(2013)}]{Nishimura_13}
{Nishimura}, O. 2013, \pasj, 65, 84

\bibitem[{{Nishimura}(2014)}]{Nishimura_14}
{Nishimura}, O. 2014, \apj, 781, 30

\bibitem[{{Nishimura}(2015)}]{Nishimura_15}
{Nishimura}, O. 2015, \apj, 807, 164

\bibitem[{{Nobili} {et~al.}(2008){Nobili}, {Turolla}, \&
  {Zane}}]{Nobili_etal08}
{Nobili}, L., {Turolla}, R., \& {Zane}, S. 2008, \mnras, 389, 989

\bibitem[{{Nowak} {et~al.}(2012){Nowak}, {Paizis}, {Rodriguez}, {Chaty}, {Del
  Santo}, {Grinberg}, {Wilms}, {Ubertini}, \& {Chini}}]{Nowak_etal12}
{Nowak}, M.~A., {Paizis}, A., {Rodriguez}, J., {et~al.} 2012, \apj, 757, 143

\bibitem[{{Orlandini} {et~al.}(1999){Orlandini}, {dal Fiume}, {del Sordo},
  {Frontera}, {Parmar}, {Santangelo}, \& {Segreto}}]{Orlandini_etal99}
{Orlandini}, M., {dal Fiume}, D., {del Sordo}, S., {et~al.} 1999, \aap, 349, L9

\bibitem[{{Orlandini} {et~al.}(1998){Orlandini}, {Dal Fiume}, {Frontera}, {Del
  Sordo}, {Piraino}, {Santangelo}, {Segreto}, {Oosterbroek}, \&
  {Parmar}}]{Orlandini_etal98}
{Orlandini}, M., {Dal Fiume}, D., {Frontera}, F., {et~al.} 1998, \apjl, 500,
  L163

\bibitem[{{Orlandini} {et~al.}(2012){Orlandini}, {Frontera}, {Masetti},
  {Sguera}, \& {Sidoli}}]{Orlandini_etal12}
{Orlandini}, M., {Frontera}, F., {Masetti}, N., {Sguera}, V., \& {Sidoli}, L.
  2012, \apj, 748, 86

\bibitem[{{Ostriker} \& {Gunn}(1969)}]{OstrikerGunn_69}
{Ostriker}, J.~P. \& {Gunn}, J.~E. 1969, \apj, 157, 1395

\bibitem[{{Palmer} {et~al.}(2005){Palmer}, {Barthelmy}, {Cummings}, {Gehrels},
  {Kennea}, {Krimm}, {Markwardt}, \& {Tueller}}]{Palmer_etal05}
{Palmer}, D., {Barthelmy}, S., {Cummings}, J., {et~al.} 2005, The Astronomer's
  Telegram, 678

\bibitem[{{Parfrey} {et~al.}(2016){Parfrey}, {Spitkovsky}, \&
  {Beloborodov}}]{Parfrey_etal16}
{Parfrey}, K., {Spitkovsky}, A., \& {Beloborodov}, A.~M. 2016, \apj, 822, 33

\bibitem[{{Parkes} {et~al.}(1980){Parkes}, {Murdin}, \&
  {Mason}}]{Parkes_etal80}
{Parkes}, G.~E., {Murdin}, P.~G., \& {Mason}, K.~O. 1980, \mnras, 190, 537

\bibitem[{{Payne} \& {Melatos}(2007)}]{PayneMelatos_07}
{Payne}, D.~J.~B. \& {Melatos}, A. 2007, \mnras, 376, 609

\bibitem[{{Pearlman} {et~al.}(2013){Pearlman}, {Corbet}, \&
  {Pottschmidt}}]{Pearlman_etal13}
{Pearlman}, A.~B., {Corbet}, R., \& {Pottschmidt}, K. 2013, in American
  Astronomical Society Meeting Abstracts, Vol. 221, American Astronomical
  Society Meeting Abstracts \#221, 142.38

\bibitem[{{Pearlman} {et~al.}(2011){Pearlman}, {Corbet}, {Pottschmidt}, \&
  {Skinner}}]{Pearlman_etal11}
{Pearlman}, A.~B., {Corbet}, R.~H.~D., {Pottschmidt}, K., \& {Skinner}, G.~K.
  2011, in AAS/High Energy Astrophysics Division, Vol.~12, AAS/High Energy
  Astrophysics Division \#12, 42.06

\bibitem[{{Pellizza} {et~al.}(2006){Pellizza}, {Chaty}, \&
  {Negueruela}}]{Pellizza_etal06}
{Pellizza}, L.~J., {Chaty}, S., \& {Negueruela}, I. 2006, \aap, 455, 653

\bibitem[{{Pietsch} {et~al.}(1986){Pietsch}, {Oegelman}, {Kahabka}, {Collmar},
  \& {Gottwald}}]{Pietsch_etal86}
{Pietsch}, W., {Oegelman}, H., {Kahabka}, P., {Collmar}, W., \& {Gottwald}, M.
  1986, \aap, 163, 93

\bibitem[{{Piraino} {et~al.}(2000){Piraino}, {Santangelo}, {Segreto},
  {Giarrusso}, {Cusumano}, {Del Sordo}, {Robba}, {Dal Fiume}, {Orlandini},
  {Oosterbroek}, \& {Parmar}}]{Piraino_etal00}
{Piraino}, S., {Santangelo}, A., {Segreto}, A., {et~al.} 2000, \aap, 357, 501

\bibitem[{{Pires} {et~al.}(2014){Pires}, {Haberl}, {Zavlin}, {Motch}, {Zane},
  \& {Hohle}}]{2014A&A...563A..50P}
{Pires}, A.~M., {Haberl}, F., {Zavlin}, V.~E., {et~al.} 2014, \aap, 563, A50

\bibitem[{{Pires} {et~al.}(2017){Pires}, {Schwope}, {Haberl}, {Motch}, {Zane},
  \& {Zavlin}}]{2017xru.Pires}
{Pires}, A.~M., {Schwope}, A., {Haberl}, F., {et~al.} 2017, in The X-ray
  Universe 2017, Poster

\bibitem[{{Posselt} {et~al.}(2007){Posselt}, {Popov}, {Haberl}, {Tr{\"u}mper},
  {Turolla}, \& {Neuh{\"a}user}}]{2007Ap&SS.308..171P}
{Posselt}, B., {Popov}, S.~B., {Haberl}, F., {et~al.} 2007, \apss, 308, 171

\bibitem[{{Posselt} {et~al.}(2008){Posselt}, {Popov}, {Haberl}, {Tr{\"u}mper},
  {Turolla}, \& {Neuh{\"a}user}}]{2008A&A...482..617P}
{Posselt}, B., {Popov}, S.~B., {Haberl}, F., {et~al.} 2008, \aap, 482, 617

\bibitem[{{Postnov} {et~al.}(2013){Postnov}, {Shakura}, {Staubert},
  {Kochetkova}, {Klochkov}, \& {Wilms}}]{Postnov_etal13}
{Postnov}, K., {Shakura}, N., {Staubert}, R., {et~al.} 2013, \mnras, 435, 1147

\bibitem[{{Postnov} {et~al.}(2011){Postnov}, {Shakura}, {Kochetkova}, \&
  {Hjalmarsdotter}}]{Postnov_etal11}
{Postnov}, K., {Shakura}, N.~I., {Kochetkova}, A.~Y., \& {Hjalmarsdotter}, L.
  2011, in Extreme and Variable High Energy Sky (Extremesky 2011), 17

\bibitem[{{Postnov} {et~al.}(2015){Postnov}, {Gornostaev}, {Klochkov},
  {Laplace}, {Lukin}, \& {Shakura}}]{Postnov_etal15}
{Postnov}, K.~A., {Gornostaev}, M.~I., {Klochkov}, D., {et~al.} 2015, \mnras,
  452, 1601

\bibitem[{{Potekhin}(2010)}]{2010A&A...518A..24P}
{Potekhin}, A.~Y. 2010, \aap, 518, A24

\bibitem[{{Potekhin} {et~al.}(2015){Potekhin}, {De Luca}, \&
  {Pons}}]{2015SSRv..191..171P}
{Potekhin}, A.~Y., {De Luca}, A., \& {Pons}, J.~A. 2015, \ssr, 191, 171

\bibitem[{{Pottschmidt} {et~al.}(2005){Pottschmidt}, {Kreykenbohm}, {Wilms},
  {Coburn}, {Rothschild}, {Kretschmar}, {McBride}, {Suchy}, \&
  {Staubert}}]{Pottschmidt_etal05}
{Pottschmidt}, K., {Kreykenbohm}, I., {Wilms}, J., {et~al.} 2005, \apjl, 634,
  L97

\bibitem[{{Poutanen} {et~al.}(2013){Poutanen}, {Mushtukov}, {Suleimanov},
  {Tsygankov}, {Nagirner}, {Doroshenko}, \& {Lutovinov}}]{Poutanen_etal13}
{Poutanen}, J., {Mushtukov}, A.~A., {Suleimanov}, V.~F., {et~al.} 2013, \apj,
  777, 115

\bibitem[{{Pravdo} \& {Bussard}(1981)}]{PravdoBussard_81}
{Pravdo}, S.~H. \& {Bussard}, R.~W. 1981, \apjl, 246, L115

\bibitem[{{Qu} {et~al.}(2005){Qu}, {Zhang}, {Song}, \& {Falanga}}]{Qu_etal05}
{Qu}, J.~L., {Zhang}, S., {Song}, L.~M., \& {Falanga}, M. 2005, \apjl, 629, L33

\bibitem[{{Rappaport} {et~al.}(1978){Rappaport}, {Clark}, {Cominsky}, {Li}, \&
  {Joss}}]{Rappaport_etal78}
{Rappaport}, S., {Clark}, G.~W., {Cominsky}, L., {Li}, F., \& {Joss}, P.~C.
  1978, \apjl, 224, L1

\bibitem[{{Rawls} {et~al.}(2011){Rawls}, {Orosz}, {McClintock}, {Torres},
  {Bailyn}, \& {Buxton}}]{Rawls_etal11}
{Rawls}, M.~L., {Orosz}, J.~A., {McClintock}, J.~E., {et~al.} 2011, \apj, 730,
  25

\bibitem[{{Reig}(2004)}]{Reig_04}
{Reig}, P. 2004, in ESA Special Publication, Vol. 552, 5th INTEGRAL Workshop on
  the INTEGRAL Universe, ed. V.~{Schoenfelder}, G.~{Lichti}, \& C.~{Winkler},
  373

\bibitem[{{Reig} \& {Coe}(1999)}]{ReigCoe_99}
{Reig}, P. \& {Coe}, M.~J. 1999, \mnras, 302, 700

\bibitem[{{Reig} \& {Milonaki}(2016)}]{ReigMilonaki_16}
{Reig}, P. \& {Milonaki}, F. 2016, \aap, 594, A45

\bibitem[{{Reig} {et~al.}(2005){Reig}, {Negueruela}, {Fabregat}, {Chato}, \&
  {Coe}}]{Reig_etal05}
{Reig}, P., {Negueruela}, I., {Fabregat}, J., {Chato}, R., \& {Coe}, M.~J.
  2005, \aap, 440, 1079

\bibitem[{{Reig} {et~al.}(2016){Reig}, {Nersesian}, {Zezas}, {Gkouvelis}, \&
  {Coe}}]{Reig_etal16}
{Reig}, P., {Nersesian}, A., {Zezas}, A., {Gkouvelis}, L., \& {Coe}, M.~J.
  2016, \aap, 590, A122

\bibitem[{{Reig} \& {Nespoli}(2013)}]{ReigNespoli_13}
{Reig}, P. \& {Nespoli}, E. 2013, \aap, 551, A1

\bibitem[{{Revnivtsev} \& {Mereghetti}(2016)}]{RevnivtsevMereghetti_16}
{Revnivtsev}, M. \& {Mereghetti}, S. 2016, {Magnetic Fields of Neutron Stars in
  X-Ray Binaries}, ed. V.~S. {Beskin}, A.~{Balogh}, M.~{Falanga},
  M.~{Lyutikov}, S.~{Mereghetti}, T.~{Piran}, \& R.~A. {Treumann}, 299

\bibitem[{{Reynolds} {et~al.}(1997){Reynolds}, {Quaintrell}, {Still}, {Roche},
  {Chakrabarty}, \& {Levine}}]{Reynolds_etal97}
{Reynolds}, A.~P., {Quaintrell}, H., {Still}, M.~D., {et~al.} 1997, \mnras,
  288, 43

\bibitem[{{Rivers} {et~al.}(2010){Rivers}, {Markowitz}, {Pottschmidt}, {Roth},
  {Barrag{\'a}n}, {F{\"u}rst}, {Suchy}, {Kreykenbohm}, {Wilms}, \&
  {Rothschild}}]{Rivers_etal10}
{Rivers}, E., {Markowitz}, A., {Pottschmidt}, K., {et~al.} 2010, \apj, 709, 179

\bibitem[{{Rodes-Roca} {et~al.}(2018){Rodes-Roca}, {Bernabeu}, {Magazz{\`u}},
  {Torrej{\'o}n}, \& {Solano}}]{Rodes-Roca_etal18}
{Rodes-Roca}, J.~J., {Bernabeu}, G., {Magazz{\`u}}, A., {Torrej{\'o}n}, J.~M.,
  \& {Solano}, E. 2018, \mnras, 476, 2110

\bibitem[{{Rodes-Roca} {et~al.}(2009){Rodes-Roca}, {Torrej{\'o}n},
  {Kreykenbohm}, {Mart{\'{\i}}nez N{\'u}{\~n}ez}, {Camero-Arranz}, \&
  {Bernab{\'e}u}}]{Rodes-Roca_etal09}
{Rodes-Roca}, J.~J., {Torrej{\'o}n}, J.~M., {Kreykenbohm}, I., {et~al.} 2009,
  \aap, 508, 395

\bibitem[{{Rodriguez} {et~al.}(2009){Rodriguez}, {Tomsick}, {Bodaghee}, {Zurita
  Heras}, {Chaty}, {Paizis}, \& {Corbel}}]{Rodriguez_etal09}
{Rodriguez}, J., {Tomsick}, J.~A., {Bodaghee}, A., {et~al.} 2009, \aap, 508,
  889

\bibitem[{{Romanova} {et~al.}(2008){Romanova}, {Kulkarni}, \&
  {Lovelace}}]{Romanova_etal08}
{Romanova}, M.~M., {Kulkarni}, A.~K., \& {Lovelace}, R.~V.~E. 2008, \apjl, 673,
  L171

\bibitem[{{Rosenberg} {et~al.}(1975){Rosenberg}, {Eyles}, {Skinner}, \&
  {Willmore}}]{Rosenberg_etal75}
{Rosenberg}, F.~D., {Eyles}, C.~J., {Skinner}, G.~K., \& {Willmore}, A.~P.
  1975, \nat, 256, 628

\bibitem[{{Rothschild} {et~al.}(2013){Rothschild}, {Markowitz}, {Hemphill},
  {Caballero}, {Pottschmidt}, {K{\"u}hnel}, {Wilms}, {F{\"u}rst}, {Doroshenko},
  \& {Camero-Arranz}}]{Rothschild_etal13}
{Rothschild}, R., {Markowitz}, A., {Hemphill}, P., {et~al.} 2013, \apj, 770, 19

\bibitem[{{Rothschild} {et~al.}(2016){Rothschild}, {K{\"u}hnel}, {Britton
  Hemphill}, {Markowitz}, {Pottschmidt}, {Wilms}, {Staubert}, {Klochkov},
  {Postnov}, \& {Goronostaev}}]{Rothschild_etal16}
{Rothschild}, R.~E., {K{\"u}hnel}, M., {Britton Hemphill}, P., {et~al.} 2016,
  in AAS/High Energy Astrophysics Division, Vol.~15, AAS/High Energy
  Astrophysics Division, 120.21

\bibitem[{{Rothschild} {et~al.}(2017){Rothschild}, {K{\"u}hnel}, {Pottschmidt},
  {Hemphill}, {Postnov}, {Gornostaev}, {Shakura}, {F{\"u}rst}, {Wilms},
  {Staubert}, \& {Klochkov}}]{Rothschild_etal17}
{Rothschild}, R.~E., {K{\"u}hnel}, M., {Pottschmidt}, K., {et~al.} 2017,
  \mnras, 466, 2752

\bibitem[{{Roy} {et~al.}(2017){Roy}, {Choudhury}, \& {Agrawal}}]{Roy_etal17}
{Roy}, J., {Choudhury}, M., \& {Agrawal}, P.~C. 2017, \apj, 848, 124

\bibitem[{{Rutledge} {et~al.}(2003){Rutledge}, {Fox}, {Bogosavljevic}, \&
  {Mahabal}}]{2003ApJ...598..458R}
{Rutledge}, R.~E., {Fox}, D.~W., {Bogosavljevic}, M., \& {Mahabal}, A. 2003,
  \apj, 598, 458

\bibitem[{{Santangelo} {et~al.}(1998){Santangelo}, {del Sordo}, {Segreto}, {dal
  Fiume}, {Orlandini}, \& {Piraino}}]{Santangelo_etal98}
{Santangelo}, A., {del Sordo}, S., {Segreto}, A., {et~al.} 1998, \aap, 340, L55

\bibitem[{{Santangelo} {et~al.}(1999){Santangelo}, {Segreto}, {Giarrusso}, {Dal
  Fiume}, {Orlandini}, {Parmar}, {Oosterbroek}, {Bulik}, {Mihara}, {Campana},
  {Israel}, \& {Stella}}]{Santangelo_etal99}
{Santangelo}, A., {Segreto}, A., {Giarrusso}, S., {et~al.} 1999, \apjl, 523,
  L85

\bibitem[{{Sartore} {et~al.}(2015){Sartore}, {Jourdain}, \&
  {Roques}}]{Sartore_etal15}
{Sartore}, N., {Jourdain}, E., \& {Roques}, J.~P. 2015, \apj, 806, 193

\bibitem[{{Sasano} {et~al.}(2014){Sasano}, {Makishima}, {Sakurai}, {Zhang}, \&
  {Enoto}}]{Sasano_etal14}
{Sasano}, M., {Makishima}, K., {Sakurai}, S., {Zhang}, Z., \& {Enoto}, T. 2014,
  \pasj, 66, 35

\bibitem[{{Sch{\"o}nherr} {et~al.}(2014){Sch{\"o}nherr}, {Schwarm}, {Falkner},
  {Dauser}, {Ferrigno}, {K{\"u}hnel}, {Klochkov}, {Kretschmar}, {Becker},
  {Wolff}, {Pottschmidt}, {Falanga}, {Kreykenbohm}, {F{\"u}rst}, {Staubert}, \&
  {Wilms}}]{Schoenherr_etal14}
{Sch{\"o}nherr}, G., {Schwarm}, F.-W., {Falkner}, S., {et~al.} 2014, \aap, 564,
  L8

\bibitem[{{Sch{\"o}nherr} {et~al.}(2007){Sch{\"o}nherr}, {Wilms}, {Kretschmar},
  {Kreykenbohm}, {Santangelo}, {Rothschild}, {Coburn}, \&
  {Staubert}}]{Schoenherr_etal07}
{Sch{\"o}nherr}, G., {Wilms}, J., {Kretschmar}, P., {et~al.} 2007, \aap, 472,
  353

\bibitem[{{Schurch} {et~al.}(2011){Schurch}, {Coe}, {McBride}, {Townsend},
  {Udalski}, {Haberl}, \& {Corbet}}]{Schurch_etal11}
{Schurch}, M.~P.~E., {Coe}, M.~J., {McBride}, V.~A., {et~al.} 2011, \mnras,
  412, 391

\bibitem[{{Schwarm} {et~al.}(2017{\natexlab{a}}){Schwarm}, {Ballhausen},
  {Falkner}, {Sch{\"o}nherr}, {Pottschmidt}, {Wolff}, {Becker}, {F{\"u}rst},
  {Marcu-Cheatham}, {Hemphill}, {Sokolova-Lapa}, {Dauser}, {Klochkov},
  {Ferrigno}, \& {Wilms}}]{Schwarm_etal17b}
{Schwarm}, F.-W., {Ballhausen}, R., {Falkner}, S., {et~al.} 2017{\natexlab{a}},
  \aap, 601, A99

\bibitem[{{Schwarm} {et~al.}(2017{\natexlab{b}}){Schwarm}, {Sch{\"o}nherr},
  {Falkner}, {Pottschmidt}, {Wolff}, {Becker}, {Sokolova-Lapa}, {Klochkov},
  {Ferrigno}, {F{\"u}rst}, {Hemphill}, {Marcu-Cheatham}, {Dauser}, \&
  {Wilms}}]{Schwarm_etal17a}
{Schwarm}, F.-W., {Sch{\"o}nherr}, G., {Falkner}, S., {et~al.}
  2017{\natexlab{b}}, \aap, 597, A3

\bibitem[{{Schwope} {et~al.}(2000){Schwope}, {Hasinger}, {Lehmann}, {Schwarz},
  {Brunner}, {Neizvestny}, {Ugryumov}, {Balega}, {Tr{\"u}mper}, \&
  {Voges}}]{2000AN....321....1S}
{Schwope}, A., {Hasinger}, G., {Lehmann}, I., {et~al.} 2000, Astronomische
  Nachrichten, 321, 1

\bibitem[{{Schwope} {et~al.}(2005){Schwope}, {Hambaryan}, {Haberl}, \&
  {Motch}}]{2005A&A...441..597S}
{Schwope}, A.~D., {Hambaryan}, V., {Haberl}, F., \& {Motch}, C. 2005, \aap,
  441, 597

\bibitem[{{Schwope} {et~al.}(2007){Schwope}, {Hambaryan}, {Haberl}, \&
  {Motch}}]{2007Ap&SS.308..619S}
{Schwope}, A.~D., {Hambaryan}, V., {Haberl}, F., \& {Motch}, C. 2007, \apss,
  308, 619

\bibitem[{{Schwope} {et~al.}(1999){Schwope}, {Hasinger}, {Schwarz}, {Haberl},
  \& {Schmidt}}]{1999A&A...341L..51S}
{Schwope}, A.~D., {Hasinger}, G., {Schwarz}, R., {Haberl}, F., \& {Schmidt}, M.
  1999, \aap, 341, L51

\bibitem[{{Shakura} {et~al.}(2012){Shakura}, {Postnov}, {Kochetkova}, \&
  {Hjalmarsdotter}}]{Shakura_etal12}
{Shakura}, N., {Postnov}, K., {Kochetkova}, A., \& {Hjalmarsdotter}, L. 2012,
  \mnras, 420, 216

\bibitem[{{Shapiro} \& {Salpeter}(1975)}]{ShapiroSalpeter_75}
{Shapiro}, S.~L. \& {Salpeter}, E.~E. 1975, \apj, 198, 671

\bibitem[{{Shi} {et~al.}(2015){Shi}, {Zhang}, \& {Li}}]{Shi_etal15}
{Shi}, C.-S., {Zhang}, S.-N., \& {Li}, X.-D. 2015, \apj, 813, 91

\bibitem[{{Shrader} {et~al.}(1999){Shrader}, {Sutaria}, {Singh}, \&
  {Macomb}}]{Shrader_etal99}
{Shrader}, C.~R., {Sutaria}, F.~K., {Singh}, K.~P., \& {Macomb}, D.~J. 1999,
  \apj, 512, 920

\bibitem[{{Shtykovsky} {et~al.}(2018){Shtykovsky}, {Lutovinov}, {Tsygankov}, \&
  {Molkov}}]{Shtykovsky_etal18}
{Shtykovsky}, A.~E., {Lutovinov}, A.~A., {Tsygankov}, S.~S., \& {Molkov}, S.~V.
  2018, ArXiv:1809.00880 [\eprint[arXiv:1809.00880]{1809.00880}]

\bibitem[{{Soong} {et~al.}(1990){Soong}, {Gruber}, {Peterson}, \&
  {Rothschild}}]{Soong_etal90b}
{Soong}, Y., {Gruber}, D.~E., {Peterson}, L.~E., \& {Rothschild}, R.~E. 1990,
  ApJ, 348, 641

\bibitem[{Staubert(2003)}]{Staubert_03}
Staubert, R. 2003, in Multifrequency behaviour of high energy cosmic sources,
  ed. L.S.-G.F. Giovanelli, Vol. ChJAA, Vol. 3, S270

\bibitem[{{Staubert}(2014)}]{Staubert_14}
{Staubert}, R. 2014, in PoS(INTEGRAL2014)024

\bibitem[{{Staubert} {et~al.}(1980){Staubert}, {Kendziorra}, {Pietsch},
  {Reppin}, {Truemper}, \& {Voges}}]{Staubert_etal80}
{Staubert}, R., {Kendziorra}, E., {Pietsch}, W., {et~al.} 1980, \apj, 239, 1010

\bibitem[{{Staubert} {et~al.}(2017){Staubert}, {Klochkov}, {F{\"u}rst},
  {Wilms}, {Rothschild}, \& {Harrison}}]{Staubert_etal17}
{Staubert}, R., {Klochkov}, D., {F{\"u}rst}, F., {et~al.} 2017, \aap, 606, L13

\bibitem[{{Staubert} {et~al.}(2009){Staubert}, {Klochkov}, {Postnov},
  {Shakura}, {Wilms}, \& {Rothschild}}]{Staubert_etal09}
{Staubert}, R., {Klochkov}, D., {Postnov}, K., {et~al.} 2009, A\&A, 494, 1025

\bibitem[{{Staubert} {et~al.}(2010{\natexlab{a}}){Staubert}, {Klochkov},
  {Postnov}, {Shakura}, {Wilms}, \& {Rothschild}}]{Staubert_etal10a}
{Staubert}, R., {Klochkov}, D., {Postnov}, K., {et~al.} 2010{\natexlab{a}},
  X-ray Astronomy 2009; Present Status, Multi-Wavelength Approach and Future
  Perspectives, 1248, 209

\bibitem[{{Staubert} {et~al.}(2013){Staubert}, {Klochkov}, {Vasco}, {Postnov},
  {Shakura}, {Wilms}, \& {Rothschild}}]{Staubert_etal13}
{Staubert}, R., {Klochkov}, D., {Vasco}, D., {et~al.} 2013, \aap, 550, A110

\bibitem[{{Staubert} {et~al.}(2010{\natexlab{b}}){Staubert}, {Klochkov},
  {Vasco}, \& {Wilms}}]{Staubert_etal10b}
{Staubert}, R., {Klochkov}, D., {Vasco}, D., \& {Wilms}, J. 2010{\natexlab{b}},
  in Proceedings of the 8th INTEGRAL Workshop ''The Restless Gamma-ray
  Universe'' (INTEGRAL 2010). September 27-30 2010. Dublin Castle, Dublin,
  Ireland. Published online at <A
  href=''http://pos.sissa.it/cgi-bin/reader/conf.cgi?confid=115''>http://pos.sissa.it/cgi-bin/reader/conf.cgi?confid=115</A>.,
  id.48

\bibitem[{{Staubert} {et~al.}(2010{\natexlab{c}}){Staubert}, {Klochkov},
  {Vasco}, \& {Wilms}}]{Staubert_etal10c}
{Staubert}, R., {Klochkov}, D., {Vasco}, D., \& {Wilms}, J. 2010{\natexlab{c}},
  in Proceedings of the 8th INTEGRAL Workshop ''The Restless Gamma-ray
  Universe'' (INTEGRAL 2010). September 27-30 2010. Dublin Castle, Dublin,
  Ireland. Published online at <A
  href=''http://pos.sissa.it/cgi-bin/reader/conf.cgi?confid=115''>http://pos.sissa.it/cgi-bin/reader/conf.cgi?confid=115</A>.,
  id.141

\bibitem[{{Staubert} {et~al.}(2016){Staubert}, {Klochkov}, {Vybornov}, {Wilms},
  \& {Harrison}}]{Staubert_etal16}
{Staubert}, R., {Klochkov}, D., {Vybornov}, V., {Wilms}, J., \& {Harrison},
  F.~A. 2016, \aap, 590, A91

\bibitem[{{Staubert} {et~al.}(2014){Staubert}, {Klochkov}, {Wilms}, {Postnov},
  {Shakura}, {Rothschild}, {F{\"u}rst}, \& {Harrison}}]{Staubert_etal14}
{Staubert}, R., {Klochkov}, D., {Wilms}, J., {et~al.} 2014, \aap, 572, A119

\bibitem[{{Staubert} {et~al.}(2011){Staubert}, {Pottschmidt}, {Doroshenko},
  {Wilms}, {Suchy}, {Rothschild}, \& {Santangelo}}]{Staubert_etal11a}
{Staubert}, R., {Pottschmidt}, K., {Doroshenko}, V., {et~al.} 2011, \aap, 527,
  A7

\bibitem[{{Staubert} {et~al.}(2007){Staubert}, {Shakura}, {Postnov}, {Wilms},
  {Rothschild}, {Coburn}, {Rodina}, \& {Klochkov}}]{Staubert_etal07}
{Staubert}, R., {Shakura}, N.~I., {Postnov}, K., {et~al.} 2007, A\&A, 465, L25

\bibitem[{{Steele} {et~al.}(1998){Steele}, {Negueruela}, {Coe}, \&
  {Roche}}]{Steele_etal98}
{Steele}, I.~A., {Negueruela}, I., {Coe}, M.~J., \& {Roche}, P. 1998, \mnras,
  297, L5

\bibitem[{{Stella} {et~al.}(1985){Stella}, {White}, {Davelaar}, {Parmar},
  {Blissett}, \& {van der Klis}}]{Stella_etal85}
{Stella}, L., {White}, N.~E., {Davelaar}, J., {et~al.} 1985, \apjl, 288, L45

\bibitem[{{Stella} {et~al.}(1986){Stella}, {White}, \&
  {Rosner}}]{Stella_etal86}
{Stella}, L., {White}, N.~E., \& {Rosner}, R. 1986, \apj, 308, 669

\bibitem[{{Stoyanov} {et~al.}(2014){Stoyanov}, {Zamanov}, {Latev}, {Abedin}, \&
  {Tomov}}]{Stoyanov_etal14}
{Stoyanov}, K.~A., {Zamanov}, R.~K., {Latev}, G.~Y., {Abedin}, A.~Y., \&
  {Tomov}, N.~A. 2014, Astronomische Nachrichten, 335, 1060

\bibitem[{{Suchy} {et~al.}(2012){Suchy}, {F{\"u}rst}, {Pottschmidt},
  {Caballero}, {Kreykenbohm}, {Wilms}, {Markowitz}, \&
  {Rothschild}}]{Suchy_etal12}
{Suchy}, S., {F{\"u}rst}, F., {Pottschmidt}, K., {et~al.} 2012, \apj, 745, 124

\bibitem[{{Suchy} {et~al.}(2011){Suchy}, {Pottschmidt}, {Rothschild}, {Wilms},
  {F{\"u}rst}, {Barragan}, {Caballero}, {Grinberg}, {Kreykenbohm},
  {Doroshenko}, {Santangelo}, {Staubert}, {Terada}, {Iwakari}, \&
  {Makishima}}]{Suchy_etal11}
{Suchy}, S., {Pottschmidt}, K., {Rothschild}, R.~E., {et~al.} 2011, \apj, 733,
  15

\bibitem[{{Suchy} {et~al.}(2008){Suchy}, {Pottschmidt}, {Wilms}, {Kreykenbohm},
  {Sch{\"o}nherr}, {Kretschmar}, {McBride}, {Caballero}, {Rothschild}, \&
  {Grinberg}}]{Suchy_etal08}
{Suchy}, S., {Pottschmidt}, K., {Wilms}, J., {et~al.} 2008, \apj, 675, 1487

\bibitem[{{Sugizaki} {et~al.}(2015){Sugizaki}, {Yamamoto}, {Mihara},
  {Nakajima}, \& {Makishima}}]{Sugizaki_etal15}
{Sugizaki}, M., {Yamamoto}, T., {Mihara}, T., {Nakajima}, M., \& {Makishima},
  K. 2015, \pasj, 67, 73

\bibitem[{{Takagi} {et~al.}(2016){Takagi}, {Mihara}, {Sugizaki}, {Makishima},
  \& {Morii}}]{Takagi_etal16}
{Takagi}, T., {Mihara}, T., {Sugizaki}, M., {Makishima}, K., \& {Morii}, M.
  2016, \pasj, 68, S13

\bibitem[{{Takeshima} {et~al.}(1994){Takeshima}, {Dotani}, {Mitsuda}, \&
  {Nagase}}]{Takeshima_etal94}
{Takeshima}, T., {Dotani}, T., {Mitsuda}, K., \& {Nagase}, F. 1994, \apj, 436,
  871

\bibitem[{{Tanaka}(1983)}]{Tanaka_83}
{Tanaka}, Y. 1983, \iaucirc, 3891

\bibitem[{{Tanaka}(1986)}]{Tanaka_86}
{Tanaka}, Y. 1986, in Lecture Notes in Physics, Berlin Springer Verlag, Vol.
  255, IAU Colloq. 89: Radiation Hydrodynamics in Stars and Compact Objects,
  ed. D.~{Mihalas} \& K.-H.~A. {Winkler}, 198

\bibitem[{{Tananbaum} {et~al.}(1972){Tananbaum}, {Gursky}, {Kellogg},
  {Levinson}, {Schreier}, \& {Giacconi}}]{Tananbaum_etal72}
{Tananbaum}, H., {Gursky}, H., {Kellogg}, E.~M., {et~al.} 1972, ApJ, 174, L143

\bibitem[{{Tendulkar} {et~al.}(2014){Tendulkar}, {F{\"u}rst}, {Pottschmidt},
  {Bachetti}, {Bhalerao}, {Boggs}, {Christensen}, {Craig}, {Hailey},
  {Harrison}, {Stern}, {Tomsick}, {Walton}, \& {Zhang}}]{Tendulkar_etal14}
{Tendulkar}, S.~P., {F{\"u}rst}, F., {Pottschmidt}, K., {et~al.} 2014, \apj,
  795, 154

\bibitem[{{Terada} {et~al.}(2007){Terada}, {Mihara}, {Nagase}, {Angelini},
  {Dotani}, {Enoto}, {Kitamoto}, {Kohmura}, {Kokubun}, {Kotani}, {Makishima},
  {Naik}, {Nakajima}, {Sugita}, {Sudoh}, {Suzuki}, {Takahashi}, {Yonetoku}, \&
  {Yoshida}}]{Terada_etal07}
{Terada}, Y., {Mihara}, T., {Nagase}, F., {et~al.} 2007, Adv. in Space
  Research, 40, 1485

\bibitem[{{Terada} {et~al.}(2006){Terada}, {Mihara}, {Nakajima}, {Suzuki},
  {Isobe}, {Makishima}, {Takahashi}, {Enoto}, {Kokubun}, {Kitaguchi}, {Naik},
  {Dotani}, {Nagase}, {Tanaka}, {Watanabe}, {Kitamoto}, {Sudoh}, {Yoshida},
  {Nakagawa}, {Sugita}, {Kohmura}, {Kotani}, {Yonetoku}, {Angelini}, {Cottam},
  {Mukai}, {Kelley}, {Soong}, {Bautz}, {Kissel}, \& {Doty}}]{Terada_etal06}
{Terada}, Y., {Mihara}, T., {Nakajima}, M., {et~al.} 2006, ApJ, 648, L139

\bibitem[{{Terrell} \& {Priedhorsky}(1983)}]{TerrellPriedhorsky_83}
{Terrell}, J. \& {Priedhorsky}, W.~C. 1983, in \baas, Vol.~15, Bulletin of the
  American Astronomical Society, 979

\bibitem[{{Terrell} \& {Priedhorsky}(1984)}]{TerrellPriedhorsky_84}
{Terrell}, J. \& {Priedhorsky}, W.~C. 1984, \apjl, 285, L15

\bibitem[{{Tetzlaff} {et~al.}(2011){Tetzlaff}, {Eisenbeiss}, {Neuh{\"a}user},
  \& {Hohle}}]{2011MNRAS.417..617T}
{Tetzlaff}, N., {Eisenbeiss}, T., {Neuh{\"a}user}, R., \& {Hohle}, M.~M. 2011,
  \mnras, 417, 617

\bibitem[{{Tetzlaff} {et~al.}(2010){Tetzlaff}, {Neuh{\"a}user}, {Hohle}, \&
  {Maciejewski}}]{2010MNRAS.402.2369T}
{Tetzlaff}, N., {Neuh{\"a}user}, R., {Hohle}, M.~M., \& {Maciejewski}, G. 2010,
  \mnras, 402, 2369

\bibitem[{{Tetzlaff} {et~al.}(2012){Tetzlaff}, {Schmidt}, {Hohle}, \&
  {Neuh{\"a}user}}]{2012PASA...29...98T}
{Tetzlaff}, N., {Schmidt}, J.~G., {Hohle}, M.~M., \& {Neuh{\"a}user}, R. 2012,
  \pasa, 29, 98

\bibitem[{{Tiengo} \& {Mereghetti}(2007)}]{2007ApJ...657L.101T}
{Tiengo}, A. \& {Mereghetti}, S. 2007, \apjl, 657, L101

\bibitem[{{Torii} {et~al.}(2000){Torii}, {Kohmura}, {Yokogawa}, \&
  {Koyama}}]{Torii_etal00}
{Torii}, K., {Kohmura}, T., {Yokogawa}, J., \& {Koyama}, K. 2000, \iaucirc,
  7441

\bibitem[{{Torrej{\'o}n} {et~al.}(2004){Torrej{\'o}n}, {Kreykenbohm}, {Orr},
  {Titarchuk}, \& {Negueruela}}]{Torrejon_etal04}
{Torrej{\'o}n}, J.~M., {Kreykenbohm}, I., {Orr}, A., {Titarchuk}, L., \&
  {Negueruela}, I. 2004, \aap, 423, 301

\bibitem[{{Torrej{\'o}n} {et~al.}(2010){Torrej{\'o}n}, {Negueruela}, {Smith},
  \& {Harrison}}]{Torrejon_etal10}
{Torrej{\'o}n}, J.~M., {Negueruela}, I., {Smith}, D.~M., \& {Harrison}, T.~E.
  2010, \aap, 510, A61

\bibitem[{{Townsend} {et~al.}(2011){Townsend}, {Coe}, {Corbet}, \&
  {Hill}}]{Townsend_etal11}
{Townsend}, L.~J., {Coe}, M.~J., {Corbet}, R.~H.~D., \& {Hill}, A.~B. 2011,
  \mnras, 416, 1556

\bibitem[{{Tr\"umper} {et~al.}(1986){Tr\"umper}, {Kahabka}, {Oegelman},
  {Pietsch}, \& {Voges}}]{Truemper_etal86}
{Tr\"umper}, J., {Kahabka}, P., {Oegelman}, H., {Pietsch}, W., \& {Voges}, W.
  1986, ApJ, 300, L63

\bibitem[{{Tr\"umper} {et~al.}(1978){Tr\"umper}, {Pietsch}, {Reppin}, {Voges},
  {Staubert}, \& {Kendziorra}}]{Truemper_etal78}
{Tr\"umper}, J., {Pietsch}, W., {Reppin}, C., {et~al.} 1978, ApJ, 219, L105

\bibitem[{Tr\"umper {et~al.}(1977)}]{Truemper_etal77}
Tr\"umper, J. {et~al.} 1977, Ann. N.Y. Acad. Sci., 302, 538

\bibitem[{{Tsygankov} {et~al.}(2018){Tsygankov}, {Doroshenko}, {Mushtukov},
  {Lutovinov}, \& {Poutanen}}]{Tsygankov_etal18}
{Tsygankov}, S.~S., {Doroshenko}, V., {Mushtukov}, A.~A., {Lutovinov}, A.~A.,
  \& {Poutanen}, J. 2018, ArXiv:1811.08912
  [\eprint[arXiv:1811.08912]{1811.08912}]

\bibitem[{{Tsygankov} {et~al.}(2012){Tsygankov}, {Krivonos}, \&
  {Lutovinov}}]{Tsygankov_etal12}
{Tsygankov}, S.~S., {Krivonos}, R.~A., \& {Lutovinov}, A.~A. 2012, \mnras, 421,
  2407

\bibitem[{{Tsygankov} \& {Lutovinov}(2005)}]{TsygankovLutovinov_05}
{Tsygankov}, S.~S. \& {Lutovinov}, A.~A. 2005, Astronomy Letters, 31, 88

\bibitem[{{Tsygankov} {et~al.}(2006){Tsygankov}, {Lutovinov}, {Churazov}, \&
  {Sunyaev}}]{Tsygankov_etal06}
{Tsygankov}, S.~S., {Lutovinov}, A.~A., {Churazov}, E.~M., \& {Sunyaev}, R.~A.
  2006, \mnras, 371, 19

\bibitem[{{Tsygankov} {et~al.}(2007){Tsygankov}, {Lutovinov}, {Churazov}, \&
  {Sunyaev}}]{Tsygankov_etal07}
{Tsygankov}, S.~S., {Lutovinov}, A.~A., {Churazov}, E.~M., \& {Sunyaev}, R.~A.
  2007, Astronomy Letters, 33, 368

\bibitem[{{Tsygankov} {et~al.}(2016{\natexlab{a}}){Tsygankov}, {Lutovinov},
  {Doroshenko}, {Mushtukov}, {Suleimanov}, \& {Poutanen}}]{Tsygankov_etal16a}
{Tsygankov}, S.~S., {Lutovinov}, A.~A., {Doroshenko}, V., {et~al.}
  2016{\natexlab{a}}, \aap, 593, A16

\bibitem[{{Tsygankov} {et~al.}(2016{\natexlab{b}}){Tsygankov}, {Lutovinov},
  {Krivonos}, {Molkov}, {Jenke}, {Finger}, \& {Poutanen}}]{Tsygankov_etal16b}
{Tsygankov}, S.~S., {Lutovinov}, A.~A., {Krivonos}, R.~A., {et~al.}
  2016{\natexlab{b}}, \mnras, 457, 258

\bibitem[{{Tsygankov} {et~al.}(2010){Tsygankov}, {Lutovinov}, \&
  {Serber}}]{Tsygankov_etal10}
{Tsygankov}, S.~S., {Lutovinov}, A.~A., \& {Serber}, A.~V. 2010, \mnras, 401,
  1628

\bibitem[{{Tsygankov} {et~al.}(2016{\natexlab{c}}){Tsygankov}, {Mushtukov},
  {Suleimanov}, \& {Poutanen}}]{Tsygankov_etal16c}
{Tsygankov}, S.~S., {Mushtukov}, A.~A., {Suleimanov}, V.~F., \& {Poutanen}, J.
  2016{\natexlab{c}}, \mnras, 457, 1101

\bibitem[{{Tuerler} {et~al.}(2012){Tuerler}, {Chenevez}, {Bozzo}, {Ferrigno},
  {Tramacere}, {Caballero}, {Rodriguez}, {Cadolle-Bel}, {Sanchez-Fernandez},
  {Del Santo}, {Fiocchi}, {Tarana}, {den Hartog}, {Kreykenbohm}, {Kuehnel},
  {Paizis}, {Puehlhofer}, {Watanabe}, {Weidenspointner}, \&
  {Zhang}}]{Tuerler_etal12}
{Tuerler}, M., {Chenevez}, J., {Bozzo}, E., {et~al.} 2012, The Astronomer's
  Telegram, 3947

\bibitem[{{Ulmer} {et~al.}(1973){Ulmer}, {Baity}, {Wheaton}, \&
  {Peterson}}]{Ulmer_etal73}
{Ulmer}, M.~P., {Baity}, W.~A., {Wheaton}, W.~A., \& {Peterson}, L.~E. 1973,
  \apjl, 184, L117

\bibitem[{{van Kerkwijk} \& {Kaplan}(2007)}]{2007Ap&SS.308..191V}
{van Kerkwijk}, M.~H. \& {Kaplan}, D.~L. 2007, \apss, 308, 191

\bibitem[{{van Kerkwijk} \& {Kaplan}(2008)}]{2008ApJ...673L.163V}
{van Kerkwijk}, M.~H. \& {Kaplan}, D.~L. 2008, \apjl, 673, L163

\bibitem[{{van Kerkwijk} {et~al.}(2004){van Kerkwijk}, {Kaplan}, {Durant},
  {Kulkarni}, \& {Paerels}}]{2004ApJ...608..432V}
{van Kerkwijk}, M.~H., {Kaplan}, D.~L., {Durant}, M., {Kulkarni}, S.~R., \&
  {Paerels}, F. 2004, \apj, 608, 432

\bibitem[{{Vasco} {et~al.}(2013){Vasco}, {Staubert}, {Klochkov}, {Santangelo},
  {Shakura}, \& {Postnov}}]{Vasco_etal13}
{Vasco}, D., {Staubert}, R., {Klochkov}, D., {et~al.} 2013, \aap, 550, A111

\bibitem[{{Ventura} {et~al.}(1979){Ventura}, {Nagel}, \&
  {Meszaros}}]{Ventura_etal79}
{Ventura}, J., {Nagel}, W., \& {Meszaros}, P. 1979, \apjl, 233, L125

\bibitem[{{Vigan{\`o}} {et~al.}(2014){Vigan{\`o}}, {Perna}, {Rea}, \&
  {Pons}}]{2014MNRAS.443...31V}
{Vigan{\`o}}, D., {Perna}, R., {Rea}, N., \& {Pons}, J.~A. 2014, \mnras, 443,
  31

\bibitem[{{Voges} {et~al.}(1982){Voges}, {Pietsch}, {Reppin}, {Tr\"umper},
  {Kendziorra}, \& {Staubert}}]{Voges_etal82}
{Voges}, W., {Pietsch}, W., {Reppin}, C., {et~al.} 1982, ApJ, 263, 803

\bibitem[{{Vurgun} {et~al.}(2018){Vurgun}, {Chakraborty}, {G{\"u}ver}, \&
  {G{\"o}{\u g}{\"u}{\c s}}}]{Vurgun_etal18}
{Vurgun}, E., {Chakraborty}, M., {G{\"u}ver}, T., \& {G{\"o}{\u g}{\"u}{\c s}},
  E. 2018, ArXiv:1809.04738 [\eprint[arXiv:1809.04738]{1809.04738}]

\bibitem[{{Vybornov} {et~al.}(2018){Vybornov}, {Doroshenko}, {Staubert}, \&
  {Santangelo}}]{Vybornov_etal18}
{Vybornov}, V., {Doroshenko}, V., {Staubert}, R., \& {Santangelo}, A. 2018,
  \aap, 610, A88

\bibitem[{{Vybornov} {et~al.}(2017){Vybornov}, {Klochkov}, {Gornostaev},
  {Postnov}, {Sokolova-Lapa}, {Staubert}, {Pottschmidt}, \&
  {Santangelo}}]{Vybornov_etal17}
{Vybornov}, V., {Klochkov}, D., {Gornostaev}, M., {et~al.} 2017, \aap, 601,
  A126

\bibitem[{{Walter}(2001)}]{2001ApJ...549..433W}
{Walter}, F.~M. 2001, \apj, 549, 433

\bibitem[{{Walter} {et~al.}(2010){Walter}, {Eisenbei{\ss}}, {Lattimer}, {Kim},
  {Hambaryan}, \& {Neuh{\"a}user}}]{2010ApJ...724..669W}
{Walter}, F.~M., {Eisenbei{\ss}}, T., {Lattimer}, J.~M., {et~al.} 2010, \apj,
  724, 669

\bibitem[{{Walter} {et~al.}(1996){Walter}, {Wolk}, \&
  {Neuh{\"a}user}}]{1996Natur.379..233W}
{Walter}, F.~M., {Wolk}, S.~J., \& {Neuh{\"a}user}, R. 1996, \nat, 379, 233

\bibitem[{{Walton} {et~al.}(2018){Walton}, {Bachetti}, {F{\"u}rst}, {Barret},
  {Brightman}, {Fabian}, {Grefenstette}, {Harrison}, {Heida}, {Kennea},
  {Kosec}, {Lau}, {Madsen}, {Middleton}, {Pinto}, {Steiner}, \&
  {Webb}}]{Walton_etal18}
{Walton}, D.~J., {Bachetti}, M., {F{\"u}rst}, F., {et~al.} 2018, \apjl, 857, L3

\bibitem[{{Wang} {et~al.}(1988){Wang}, {Wasserman}, \&
  {Salpeter}}]{Wang_etal88}
{Wang}, J.~C.~L., {Wasserman}, I.~M., \& {Salpeter}, E.~E. 1988, \apjs, 68, 735

\bibitem[{{Wang}(2009)}]{Wang_09}
{Wang}, W. 2009, \mnras, 398, 1428

\bibitem[{{Wang}(2013)}]{Wang_13}
{Wang}, W. 2013, \mnras, 432, 954

\bibitem[{{Wang}(1981)}]{Wang_81}
{Wang}, Y.-M. 1981, \aap, 102, 36

\bibitem[{{Wang}(1987)}]{Wang_87}
{Wang}, Y.-M. 1987, A\&A, 183, 257

\bibitem[{{Wang} \& {Frank}(1981)}]{WangFrank_81}
{Wang}, Y.-M. \& {Frank}, J. 1981, \aap, 93, 255

\bibitem[{{Wang} \& {Welter}(1981)}]{WangWelter_81}
{Wang}, Y.-M. \& {Welter}, G.~L. 1981, \aap, 102, 97

\bibitem[{{Wasserman} \& {Salpeter}(1980)}]{WassermanSalpeter_80}
{Wasserman}, I. \& {Salpeter}, E. 1980, \apj, 241, 1107

\bibitem[{{Wheaton} {et~al.}(1979){Wheaton}, {Doty}, {Primini}, {Cooke},
  {Dobson}, {Goldman}, {Hecht}, {Howe}, {Hoffman}, \&
  {Scheepmaker}}]{Wheaton_etal79}
{Wheaton}, W.~A., {Doty}, J.~P., {Primini}, F.~A., {et~al.} 1979, \nat, 282,
  240

\bibitem[{{White} {et~al.}(1983){White}, {Swank}, \& {Holt}}]{White_etal83}
{White}, N.~E., {Swank}, J.~H., \& {Holt}, S.~S. 1983, ApJ, 270, 711

\bibitem[{Wilms(2012)}]{Wilms_12}
Wilms, J. 2012, in Proceed. 39th COSPAR Sci. Assembly, 14-22 July 2012, Mysore,
  India, Vol.~39, 2159

\bibitem[{{Wilson} {et~al.}(2005){Wilson}, {Fabregat}, \&
  {Coburn}}]{Wilson_etal05}
{Wilson}, C.~A., {Fabregat}, J., \& {Coburn}, W. 2005, ApJ, 620, L99

\bibitem[{{Wilson} {et~al.}(2002){Wilson}, {Finger}, {Coe}, {Laycock}, \&
  {Fabregat}}]{Wilson_etal02}
{Wilson}, C.~A., {Finger}, M.~H., {Coe}, M.~J., {Laycock}, S., \& {Fabregat},
  J. 2002, ApJ, 570, 287

\bibitem[{{Wolff} {et~al.}(2016){Wolff}, {Becker}, {Gottlieb}, {F{\"u}rst},
  {Hemphill}, {Marcu-Cheatham}, {Pottschmidt}, {Schwarm}, {Wilms}, \&
  {Wood}}]{Wolff_etal16}
{Wolff}, M.~T., {Becker}, P.~A., {Gottlieb}, A.~M., {et~al.} 2016, \apj, 831,
  194

\bibitem[{{Yahel}(1979{\natexlab{a}})}]{Yahel_79b}
{Yahel}, R.~Z. 1979{\natexlab{a}}, \apjl, 229, L73

\bibitem[{{Yahel}(1979{\natexlab{b}})}]{Yahel_79a}
{Yahel}, R.~Z. 1979{\natexlab{b}}, \aap, 78, 136

\bibitem[{{Yahel}(1980{\natexlab{a}})}]{Yahel_80a}
{Yahel}, R.~Z. 1980{\natexlab{a}}, \aap, 90, 26

\bibitem[{{Yahel}(1980{\natexlab{b}})}]{Yahel_80b}
{Yahel}, R.~Z. 1980{\natexlab{b}}, \apj, 236, 911

\bibitem[{{Yamamoto} {et~al.}(2014){Yamamoto}, {Mihara}, {Sugizaki},
  {Nakajima}, {Makishima}, \& {Sasano}}]{Yamamoto_etal14}
{Yamamoto}, T., {Mihara}, T., {Sugizaki}, M., {et~al.} 2014, \pasj, 66, 59

\bibitem[{{Yamamoto} {et~al.}(2011){Yamamoto}, {Sugizaki}, {Mihara},
  {Nakajima}, {Yamaoka}, {Matsuoka}, {Morii}, \& {Makishima}}]{Yamamoto_etal11}
{Yamamoto}, T., {Sugizaki}, M., {Mihara}, T., {et~al.} 2011, \pasj, 63, 751

\bibitem[{{Yan} {et~al.}(2016){Yan}, {Zhang}, {Liu}, \& {Liu}}]{Yan_etal16}
{Yan}, J., {Zhang}, P., {Liu}, W., \& {Liu}, Q. 2016, \aj, 151, 104

\bibitem[{{Yan} {et~al.}(2012){Yan}, {Zurita Heras}, {Chaty}, {Li}, \&
  {Liu}}]{Yan_etal12}
{Yan}, J., {Zurita Heras}, J.~A., {Chaty}, S., {Li}, H., \& {Liu}, Q. 2012,
  \apj, 753, 73

\bibitem[{{Yokogawa} {et~al.}(2001){Yokogawa}, {Torii}, {Kohmura}, \&
  {Koyama}}]{Yokogawa_etal01}
{Yokogawa}, J., {Torii}, K., {Kohmura}, T., \& {Koyama}, K. 2001, \pasj, 53,
  227

\bibitem[{{Younes} {et~al.}(2015){Younes}, {Kouveliotou}, {Grefenstette},
  {Tomsick}, {Tennant}, {Finger}, {F{\"u}rst}, {Pottschmidt}, {Bhalerao},
  {Boggs}, {Boirin}, {Chakrabarty}, {Christensen}, {Craig}, {Degenaar},
  {Fabian}, {Gandhi}, {G{\"o}{\u g}{\"u}{\c s}}, {Hailey}, {Harrison},
  {Kennea}, {Miller}, {Stern}, \& {Zhang}}]{Younes_etal15}
{Younes}, G., {Kouveliotou}, C., {Grefenstette}, B.~W., {et~al.} 2015, \apj,
  804, 43

\bibitem[{{Zampieri} {et~al.}(2001){Zampieri}, {Campana}, {Turolla},
  {Chieregato}, {Falomo}, {Fugazza}, {Moretti}, \&
  {Treves}}]{2001A&A...378L...5Z}
{Zampieri}, L., {Campana}, S., {Turolla}, R., {et~al.} 2001, \aap, 378, L5

\bibitem[{{Zane} {et~al.}(2005){Zane}, {Cropper}, {Turolla}, {Zampieri},
  {Chieregato}, {Drake}, \& {Treves}}]{2005ApJ...627..397Z}
{Zane}, S., {Cropper}, M., {Turolla}, R., {et~al.} 2005, \apj, 627, 397

\bibitem[{{Zane} {et~al.}(2002){Zane}, {Haberl}, {Cropper}, {Zavlin}, {Lumb},
  {Sembay}, \& {Motch}}]{2002MNRAS.334..345Z}
{Zane}, S., {Haberl}, F., {Cropper}, M., {et~al.} 2002, \mnras, 334, 345

\bibitem[{{Zane} {et~al.}(2001){Zane}, {Turolla}, {Stella}, \&
  {Treves}}]{2001ApJ...560..384Z}
{Zane}, S., {Turolla}, R., {Stella}, L., \& {Treves}, A. 2001, \apj, 560, 384

\bibitem[{{Zel'dovich} \& {Shakura}(1969)}]{ZeldovichShakura_69}
{Zel'dovich}, Y.~B. \& {Shakura}, N.~I. 1969, \sovast, 13, 175

\bibitem[{{Zhang} {et~al.}(2005){Zhang}, {Qu}, {Song}, \&
  {Torres}}]{Zhang_etal05}
{Zhang}, S., {Qu}, J.-L., {Song}, L.-M., \& {Torres}, D.~F. 2005, \apjl, 630,
  L65

\bibitem[{{Zheleznyakov}(1996)}]{Zheleznyakov_96}
{Zheleznyakov}, V.~V., ed. 1996, Astrophysics and Space Science Library, Vol.
  204, {Radiation in Astrophysical Plasmas}

\end{thebibliography}

\newpage

\longtab{
\begin{longtable}{llclclllll}
 \caption[]{\label{tab:collection}Cyclotron line sources (listed first in alphabetical order, then according to right ascension).}\\
  \hline\noalign{\smallskip}
System                   & Type$^{c}$  & P$_{spin}$ & P$_{orb}$  & Ecl. & E$_{cyc}$     & Instr. of            & Ref.$^{f}$  & Line        & other \\
                               &                     &  (s)         &  (days)        &            & (keV)           & 1st Det.            & 1st Det.     & conf         & Refs.$^{f}$ \\
                               &                     &                &                   &            &                     &                          &                  &                &         \\
 \hline\noalign{\smallskip}                        
Cen~X-3                  & HMXB$^{b}$ & 4.84       &  2.09          & yes      & 28               & BeppoSAX         & 3            & yes        & 49,50,186,188  \\  
Cep~X-4                  & Be trans.        & 66           & 20.85?     &  no      & 30               & Ginga                 &14,46    & yes$^{e}$ & 48 \\  
      " " "                    &                       &                &                  &            & 28,45          & Suzaku               & 47           & yes       & 48,130,162 \\  
      " " "                    &                       &                &                  &            & 30,55          & NuSTAR             & 130         & yes       & 130  \\  
GX~301-2                & HMXB           & 681         & 41.5          & near    & 37/50          & Ginga/NuS$^{i}$ & 7,10/223  & yes        & 51,52 \\  
GX~304-1                & Be trans        & 275         & 132.2        & no       & 54               & RXTE                 & 20            & yes        & 23,24,28,188,213,214 \\ 
Her~X-1                   & LMXB$^{a}$  & 1.2377   & 1.70          & yes      & 37               & Balloon               & 1             & yes        & 25-27,186 \\  
NGC300 ULX1        & Be HMXB       & 20          & --               & no        & 13               & NuSTAR             & 190        & no         & 191 \\  
SMC~X-2                 & HM trans        & 2.37       & 18.4          & no        & 27               & NuSTAR             & 91          & no         & 149,155   \\  
SXP~15.3                 & Be trans         & 15.2       &                  & no        & 5-8        & AstroSat/NuS$^{i}$  & 215        & (yes)      & RX J0052.1-7319  \\ 
Vela~X-1                  & HMXB            & 283        &  8.96         & yes      & 25,53          & Mir-HEXE            & 9,6,7      & yes        & 147,32,70,186,188 \\ 
X~Persei                  & Be XRB         & 837        & 250.3        & no        & 29               & RXTE                   & 16          & yes        & 36 \\  
0115+63 (4U)          & Be trans.        & 3.61       &  24.3         & no        & 12,24,36,    & HEAO-1               & 2           & yes         & 148,41,42,118\\ 
       " " "                   &                       &              &                   &            & 48,62          & RXTE/SAX$^{g}$ & 39/40    & yes          & 68,186,188 \\              
0332+53 (V)            & Be trans         & 4.38      &  34             & no       & 28               & Tenma/Ginga       &120/13   & yes         & 63,64,65 \\ 
      " " "                    & Be pers          &              &                   &            & 51/74          & Ginga/RXTE         &13,62      & yes        & 99,100,121 \\  
0440.9+4431 (RX)  & Be trans         & 203        & 155            & no      & 32                & RXTE                   & 17          & no          &   \\ 
0520.5-6932 (RX)   & Be (LMC)       & 8.03       & 23.9          &  no      & 31               & NuSTAR               & 31          & no          &   \\ 
0535+26 (A)            & Be trans         & 104        & 110.6        &  no      & 50               & Mir-HEXE             & 11          & yes         & 6,72,73,94 \\ 
      " " "                    &                       &               &                  &            & 110             & OSSE                   & 12          & yes         & 74,75,188 \\ 
0658-073 (XTE)      & Be trans        & 161        & 101            & no       & 33               & RXTE                   & 18          & yes         & 60,61 \\ 
1008-57 (GRO)       & Be trans        & 93.5        & 249.5        & no       & 78               & GRO/Suzaku       & 76/77     & yes         & 78 \\ 
1118-616 (1A)         & Be trans        & 407         & 24             & no      & 55, 110?      & RXTE                   & 19,45     & yes         & 43,44,188 \\ 
1409-619 (MAXI)     & HMXB           & 500        & ?                & no      & 44,73,128    & BeppoSAX            & 79          & no           &   \\ 
1538-52 (4U)           & HMXB            & 526        & 3.73          & yes     & 22               & Ginga                   & 8            & yes          & 81,87,186 \\ 
       " " "                    &                      &               &                   &            & 47               & RXTE/INT            & 80          & yes         & 81,87 \\ 
1553-542 (2S)         & Be trans         & 9.28       & 30.6          & --         & 23-27          & NuSTAR              & 82          & no           &   \\ 
1626-67 (4U)           & LMXB            & 7.67       & 0.0289       & no       & 37,61?        & BeppoSAX           & 4,161     & y/n          & 83,95,186,188 \\ 
1626.6-5156 (Swift) & Be pers         & 15.36      & 132.9        & no       & 10,18         & RXTE                    & 84,15     & no           & 15 \\ 
16393-4643 (IGR)   & HMXB            & 904        & 4.2            & no       & 29               & NuSTAR               & 29           & no          & 194,195,196 \\ 
16493-4348 (IGR)   & SG HMXB      & 1093      & 6.78          & yes      & 31               & Swift/BAT/             & 30          & no          & 96,192,193 \\ 
      " " "                     &                       &               &                  &            &                    & Suzaku                 &               &                & --   \\
1744-28 (GRO)        & LMXB            & 0.467     & 11.83        & no       & 4.7?            & XMM/                    & 106        & yes         & 107,186 \\ 
        " " "                   &                       &               &                  &            & 10.4,15.8?  & INTEGRAL           &               & no           & 108-110 \\ 
17544-2619 (IGR)    & Be trans    & 71.5$^{d}$ & 4.93          & no       & 17,33?        & NuSTAR               & 22          & no            &  \\ 
18027-2016 (IGR)    & HMXB           & 139.9       & 4.6           & no       & 24               & NuSTAR               & 142        & no           &  \\ 
18179-1621 (IGR)    & HMXB            & 11.82      & ??            & ?         & 21               & INTEGRAL           & 111        & yes         & 112-114 \\ 
1822-371 (4U,X)      & LMXB            & 0.592       & 0.232       & yes      & 0.7?/33       & XMM/Suzaku        & 85/86    & no           &  \\ 
1829-098 (XTE)       & Be trans?       & 7.84        & 246?         & no       & 15               & NuS$^{i}$/RXTE     & 224       & yes          & 225 \\ 
1907+09 (4U)          & HMXB            & 440          & 8.37          & near    & 18,36          & Ginga                   & 7           & yes         & 87,72,172 \\ 
19294+1816 (IGR)  & Be trans         & 12.4         & 117            & no       & (36)43        & (RXTE)NuS$^{i}$  & (197)198  & (yes)   & 203-206 \\  
1946+274 (XTE)      & Be trans         & 15.83       & 169.2        & no       & 36               & RXTE                    & 5           & yes         & 72,97,186,188 \\ 
1947+300 (KS)        & Be trans         & 18.7         & 40.4           & no       & 12               & NuSTAR                & 88          & no          & 98,183,185 \\ 
2206+54 (4U)          & HMXB            & 5570        & 9.57           & no       & 29-35         & RXTE/SAX$^{g}$    & 34/35     & y/n         & 35,119 \\ 
      " " "                    &                       &                 & 19.25?       &            &                    & INTEGRAL             & 175       &               & 178,179 \\ 
 \noalign{\smallskip}\hline\\ 
\end{longtable}

\tablefoot{$^{a}$ Low Mass X-ray Binary; $^{b}$ High Mass X-ray Binary; $^{c}$ a more specific definition of Type is given in Table~\ref{tab:companion}; $^{d}$ still questionable; $^{e}$ but see Ref. 117; $^{g}$ read: BeppoSAX, $^{h}$ read: INTEGRAL; $^{i}$ read: NuSTAR; \\
$^{f}$ References:
1: \citet{Truemper_etal78}; 2: \citet{Wheaton_etal79}; 3: \citet{Santangelo_etal98}; 4: \citet{Orlandini_etal98};  5: \citet{Heindl_etal01}; 6: \citet{Kretschmar_etal96}; 
7: \citet{MakishimaMihara_92}; 8: \citet{Clark_etal90}; 9: \citet{Kendziorra_etal92}; 10: \citet{Mihara_95}; 11: \citet{Kendziorra_etal94}; 12: \citet{Grove_etal95}; 
13: \citet{Makishima_etal90}; 14: \citet{Mihara_etal91a}; 15: \citet{DeCesar_etal13}; 16: \citet{Coburn_etal01}; 17: \citet{Tsygankov_etal12}; 18: \citet{Heindl_etal03}; 
19: \citet{Doroshenko_etal10}; 20: \citet{Yamamoto_etal11}; 21: \citet{Jaisawal_etal13}; 22: \citet{Bhalerao_etal15}; 23: \citet{Klochkov_etal12}; 
24: \citet{Malacaria_etal15}; 25: \citet{Staubert_etal07}; 26: \citet{Staubert_etal14};  27: \citet{Staubert_etal16}; 28: \citet{Rothschild_etal16}; 
29: \citet{Bodaghee_etal16}; 30: \citet{D'Ai_etal11}; 31: \citet{Tendulkar_etal14}; 32: \citet{Fuerst_etal14b}; 33: \citet{Vasco_etal13}; 34: \citet{Torrejon_etal04}; 
35: \citet{Masetti_etal04}; 36: \citet{Lutovinov_etal12}; 37: \citet{BonningFalanga_05}; 38: \citet{Farrell_etal08}; 39: \citet{Heindl_etal99a}; 
40: \citet{Santangelo_etal99}; 41: \citet{Iyer_etal15}; 42: \citet{Mueller_etal13}; 43: \citet{Maitra_etal12}; 44: \citet{Staubert_etal11a}; 45: \citet{Suchy_etal11}; 
46: \citet{McBride_etal07}; 47: \citet{Jaisawal_etal15}; 48: \citet{Fuerst_etal15}; 49: \citet{Burderi_etal00}; 50: \citet{Suchy_etal08}; 
51: \citet{Kreykenbohm_etal04}; 52: \citet{Suchy_etal12}; 53: \citet{LaBarbera_etal05}; 54: \citet{Jaisawal_etal16}; 55: \citet{Rothschild_etal17}; 
56: \citet{Klochkov_etal11}; 57: \citet{Nakajima_etal06}; 58: \citet{Tsygankov_etal07}; 59: \citet{Postnov_etal15}; 60: \citet{McBride_etal06}; 61: \citet{Yan_etal12}; 
62: \citet{Tsygankov_etal06}; 63: \citet{Cusumano_etal16}; 64:\citet{Mowlavi_etal06}; 65: \citet{Pottschmidt_etal05}; 66: \citet{Mihara_etal04}; 
67: \citet{Lutovinov_etal15}; 68: \citet{Li_etal12}; 69: \citet{MaitraPaul_13a}; 70: \citet{Kreykenbohm_etal02}; 71: \citet{LaBarbera_etal03};  72: \citet{MaitraPaul_13b}; 
73: \citet{Caballero_etal07}; 74: \cite{Sartore_etal15}; 75: \citet{DMueller_etal13}; 76: \citet{Shrader_etal99}; 77: \citet{Yamamoto_etal14}; 78: \citet{Bellm_etal14}; 
79: \citet{Orlandini_etal12}; 80: \citet{Rodes-Roca_etal09}; 81: \citet{Hemphill_etal16}; 82: \citet{Tsygankov_etal16b}; 83: \citet{Iwakiri_etal12}; 84: \citet{Coburn_etal06}; 
85: \citet{Iaria_etal15}; 86: \citet{Sasano_etal14}; 87: \citet{Hemphill_etal13}; 88: \citet{Fuerst_etal14a}; 89: \citet{ReigCoe_99}; 90: \citet{Klochkov_etal08c}; 
91: \citet{JaisawalNaik_16}; 92: \citet{Mason_etal78}; 93: \citet{Parkes_etal80}; 94: \citet{Caballero_etal13}; 95: \citet{Camero-Arranz_etal12}; 96: \citet{Pearlman_etal13}; 
97: \citet{Marcu-Cheatham_etal15}; 98: \citet{Ballhausen_etal16}; 99: \citet{Doroshenko_etal17}; 100: \citet{Ferrigno_etal16a}; 101: \citet{McBride_etal08}; 
102: \citet{Motch_etal97}; 103: \citet{Reig_etal05}; 104: \citet{Rawls_etal11}; 105: \citet{Kaper_etal06}; 106: \citet{D'ai_etal15}; 107: \citet{Doroshenko_etal15}; 
108: \cite{Masetti_etal14}; 109: \citet{Younes_etal15}; 110: \citet{Degenaar_etal14}; 111: \citet{Tuerler_etal12}; 112: \citet{Halpern_12}; 113: \citet{LiWang_etal12}; 
114: \citet{Nowak_etal12}; 115: \citet{LaParola_etal16}; 116: \citet{MasonCordova_82}; 117: \citet{Doroshenko_etal12}; 118: \citet{Ferrigno_etal11}; 
119: \citet{Ikhsarov_etal13}; 120: \citet{Makishima_etal90a}; 121: \citet{Kreykenbohm_etal05}; 122:\citet{Negueruela_etal99}; 123: \citet{Tsygankov_etal10}; 
124: \citet{Haigh_etal04}; 125: \citet{Coe_etal07}; 126: \citet{Maisack_etal96}; 127: \citet{Naik_etal08}; 128: \citet{Coburn_etal02}; 129: \citet{Pellizza_etal06}; 
130: \citet{Vybornov_etal17}; 131: \citet{Janot-Pacheco_etal81}; 132: \citet{Levine_etal88}; 133: \citet{ReigNespoli_13}; 134: \citet{Wilson_etal02}; 
135: \citet{Reynolds_etal97}; 136: \citet{Klochkov_etal08}; 137: \citet{Fuerst_etal13}; 138: \cite{Kuehnel_etal13}; 139: \citet{Hemphill_etal14}; 
140: \citet{ReigMilonaki_16}; 141: \citet{Galloway_etal05}; 142: \citet{Lutovinov_etal17b}; 143: \citet{Torrejon_etal10}; 144: \citet{Augello_etal03}; 
145: \citet{Martinez-Nunez_etal15}; 146: \citet{Lutovinov_etal16}; 147: \citet{Kretschmar_etal97};  148: \citet{Nagase_etal91}; 149: \citet{Lutovinov_etal17a};
150: \citet{GhoshLamb_79b}; 151: \citet{Wang_87}; 152: \citet{Wang_81}; 153: \citet{WangWelter_81}; 154: \citet{Staubert_etal80}; 155: \citet{Klus_etal14}; 
156: \citet{Lipunov_92}; 157: \citet{Jain_etal10}; 158: \citet{Chou_etal16}; 159: \citet{Takagi_etal16}; 160: \citet{Lin_etal10}; 161: \citet{D'Ai_etal17}; 
162: \citet{JaisawalNaik_17}; 163: \citet{Mihara_etal95}; 164: \citet{Piraino_etal00}; 165: \citet{Markwardt_etal07}; 166: \citet{Manousakis_etal09}; 
167: \citet{Ferrigno_etal07}; 168: \citet{Orlandini_etal99}; 169: \citet{Barnstedt_etal08}; 170: \citet{JaisawalNaik_15}; 171: \citet{LaBarbera_etal01}; 
172: \citet{Rivers_etal10}; 173: \citet{Fuerst_etal11}; 174: \citet{Fuerst_etal12}; 175: \citet{Blay_etal05}; 176: \citet{Blay_etal06}; 177: \citet{Reig_etal16}; 
178: \citet{Wang_09}; 179: \citet{Wang_13}; 180 \citet{Corbet_etal07}; 181: \citet{Stoyanov_etal14}; 182: \citet{Reig_etal16}; 183: \citet{Naik_etal06};
184: \citet{TsygankovLutovinov_05};  185: \citet{Epili_etal16}; 186: \citet{DoroR_17}; 187: \citet{Staubert_etal17}; 188: \cite{Marcu-Cheatham_etal18}; 
189: \citet{Vybornov_etal18}; 190: \citet{Walton_etal18}; 191: \citet{Carpano_etal18}; 192: \citet{Cusumano_etal10}; 193: \citet{Corbet_etal10}; 
194: \citet{Corbet_etal10a}; 195: \citet{Pearlman_etal11}; 196: \citet{Bodaghee_etal06}; 197: \citet{Roy_etal17}; 198: \citet{Tsygankov_etal18};
199: \citet{Crampton_etal85}; 200: \citet{Farrell_etal06}; 201: \citet{Masetti_etal06}; 202: \citet{DenHartog_etal06}; 203: \citet{Bozzo_etal11}; 
204: \citet{Rodriguez_etal09}; 205: \citet{CorbetKrimm_09}; 206: \citet{Rodes-Roca_etal18};  207: \citet{Brumback_etal18}; 208: \cite{Hulleman_etal98}; 
209: \citet{Reig_04}; 210: \citet{Baykal_etal07}; 211: \citet{Baykal_etal00}; 212: \citet{Yan_etal16}; 213: \citet{Sugizaki_etal15}; 214: \citet{McClintock_etal77};
215: \citet{Maitra_etal18}; 216: \citet{Corbet_etal18}.; 217: \citet{Brightman_etal18};  218: \citet{LiuMirabel_05}; 219: \citet{Ibrahim_etal02}; 
220: \citet{Ibrahim_etal03}; 221: \citet{AlfordHalpern_16}; 222: \citet{Vurgun_etal18};  223: \citet{Fuerst_etal18}; 224: \citet{Shtykovsky_etal18}; 
225: \citet{HalpernGotthelf_07}; 226: \citet{Antoniou_etal18}, 227: \citet{Maravelias_etal18}.}
}

\newpage

\begin{table*}[]
 \caption[]{Candidate cyclotron line sources: cyclotron line(s) claimed, but doubtful and/or not confirmed.}
  \label{tab:candidates}
  \begin{center}
  \begin{tabular}{llclclllll}
  \hline\noalign{\smallskip}
System                     & Type$^{c}$  & P$_{spin}$ & P$_{orb}$  & Ecl. & E$_{cyc}$     & Instr. of       & Ref.$^{e}$        & Line         & other \\
                                 &                     &  (s)         &  (days)        &            & (keV)           & 1st Det.                 & 1st Det.   & conf.        & Ref.$^{e}$\\
                                 &                     &                &                   &            &                     &                              &                 &                &         \\
 \hline\noalign{\smallskip}                        
GX 1+4                       & LMXB$^{a}$ & 304?       & 138          & no        & 34?             & INTEGRAL          & 167        & no          & --  \\ 
     " " "                        &                      & 1161?     &                 &             &                    &                             &               &                &  \\ 
M51 ULX8$^{f}$         & ULX              &  --           & --              & no        & 4.5              & Chandra              & 217         & no         &  218 \\ 
LMC X-4                     & HMXB$^{b}$ & 1.4          & 13.5         & no        & 100?           & BeppoSAX           & 171        & no          & --  \\  
0052-723 (XTE J)       & Be XRB        & 4.78       & --               & ?          & 10.2            & NuSTAR              & 226         & no          & 227 \\  
0114+650 (3A,2S)      & B SG XRB     & 9520      & 11.6          & no       & 22, 44?        & INT$^{h}$            & 37          & no          & 38,199-202 \\ 
     " " "                        &                      &                &                 &             &                    &                             &                &               &  super-orbit: 30.75 d \\ 
054134.7-682550       & HMXB           & 80?         & 62            & no        & 10?,20?      & RXTE                   & 165        & no          & 166 \\ 
   " " " (XMMU)            &                      &                &                 &             &                    &                              &               &               & -- \\ 
1657-415 (OAO)         & HMXB           & 10.4        & 38            & yes      & 36?             & BeppoSAX            & 168        & no          & 169  \\ 
1700-37 (4U)              & HMXB            & 3.4         & ?              & yes       & 39?             & Suzaku                 & 170        & no          & --  \\  
1806-20 (SGR)$^{f}$  & SGR              & 7.47        & --              & no        & $\sim$5      & RXTE                   & 219        & no          & 220 \\ 
1810-197 (XTE J)$^{f}$ & AXP trans   & 5.54        & --              & no        & 1.1              & Chandra/XMM     & 221        & yes        & 222 \\ 
1843+009 (GS)           & Be XRB         & ?             & 29.5         & no        & 20?              & Ginga                   & 163       & no          & 164 \\ 
1901+03 (4U)             & Be trans         & 2.76        & 22.6         & no        & 10?             & RXTE                    & 140       & no           & --   \\ 
1908+075 (4U)           & HMXB            & 604         & 4.4           & ?          & 44?              & Suzaku                 & 21         & no          & 173,174,186 \\ 
2030+375 (EXO)        & Be trans         & 41.4        & 46            & no        & 36/63?         & RXTE/INT$^{d}$   & 89/90    & no          & --   \\ 
2103.5+4545 (SAX J) & Be XRB          & 359         & 12.7         & no        & 12?              & NuSTAR               & 207        & no          & 208-211  \\ 
 \noalign{\smallskip}\hline
  \end{tabular}
  \end{center}
$^{a}$ Low Mass X-ray Binary; $^{b}$ High Mass X-ray Binary; $^{c}$ a more specific definition of Type is given in Table~\ref{tab:companion}; 
$^{d}$ read: INTEGRAL; $^{e}$ References: see Table~\ref{tab:collection}; $^{f}$ candidates for a possible proton cyclotron line (see also Sect.~\ref{sec:proton_lines}).
\end{table*}

\newpage

\begin{table*}[]
  \caption[]{Additional information about the sources.}
  \label{tab:companion}
  \begin{center}
  \begin{tabular}{llllllllll}
  \hline\noalign{\smallskip}
System                     & Type$^{a}$  & RA$^{b}$     & DEC$^{b}$      & Optical         & Spectral    & Mass              & Mass               & Dist           & Ref.$^{c}$ \\
                                 &                     & (J 2000)       & (J 2000)          & compa         & Type         & compan          & NS              & (Gaia)$^{e}$  &       \\
                                 &                     & (h m sec)     & ($^\circ$ ' '' )    & nion             &                  & (M$_{\odot}$) & (M$_{\odot}$) & (kpc)         &        \\
 \hline\noalign{\smallskip}                        
Cen~X-3                  & 1263           & 11 21 15.78  & -60 37 22.7     & V779 Cen    & O6.5 II-III  & $22.1\pm1.4$ & $1.49\pm0.08$ & 8 (6.4)     & 50,104  \\ 
Cep~X-4                  & 1316           & 21 39 30.60  & +56 59 12.9    & V490 Cep    & B1-B2Ve   & --                    & --                    & 3.8 (10.2)   & 48 \\  
GX~301-2                & 1211           & 12 26 37.56  & -62 46 13.2     & WRAY 977  & B1.5 Ia      & $43\pm10$    & $1.9\pm0.6$   & 3 (3.5)        & 105 \\ 
GX~304-1                & 1316           & 13 01 17.10  & -61 36 06.6     & V850 Cen    & B2Vne       & --                  & --                    & 2.4 (2.0)    & 92,93 \\ 
Her~X-1                   & 1413           & 16 57 49.81  & +35 20 32.6    & HZ Her        & B0Ve-F5e  & $2.0\pm0.4$ & $1.1\pm0.4$    & 6.6 (5.0)     & 104,135 \\  
NGC~300                & 1340           & 00 55 04.85   & -37 41 43.5    & SN?             & --               & --                   & 1.4?                 & 1880          & 190,191 \\  
ULX1                       &                    &                       &                       &                     &                   &                      &                         &                   & \\
SMC~X-2                 & 1216           & 00 54 33.43   & -73 41 01.3    & star 5           & O9.5III-V    & --                  & --                    & 60              & 101 \\  
SXP~15.3                & 1310?         & 00 50 30.23   & -73 35 36.28  & Be star         & O9.5IIIe     & --                   & --                    & 60              & 215 \\ 
Vela~X-1                 & 1312           & 09 02 06.9     & -40 33 16.9    & HD 77581    & B0.5 Ib       & $24\pm0.4$  & $1.77\pm0.08$ & 2.0 (2.4)    & 104 \\ 
X~Per                      & 1310           & 03 55 23.0     & +31 02 59.0    & X Per           & B0Ve         & --                  & --                     & 0.7 (0.79)   & \\  
0115+63                  & 1386            & 01 18 31.9     & +63 44 24.0   & V635 Cas    & B0.2Ve       & --                 & --                     & 7 (7.2)        & \\     
0332+53                  & 1366            & 03 34 59.9     & +53 10 23.3   & BQ Cam      & O8.5Ve      & $\ge20$        & 1.44?               & 7.5 (5.1)     & 122 \\ 
0440.9+4431           & 1310            & 04 40 59.32   & +44 31 49.27 & LSV +44 17 & B0e            & --                  &                        & 3.3 (3.2)     & 102,103,212 \\ 
0520.5-6932           & 1316            & 05 20 30.90   & -69 31 55.0    & LMC8          & O8-9Ve      & --                  & --                     & 50               & 31 \\ 
0535+26                  & 1366           & 05 38 54.60   & +26 18 56.8    & V725 Tau    & O9.7IIIe     & $22\pm3$    & --                      & 2 (2.1)         & 124 \\  
0658-073                 & 1341           & 06 58 17.30   & -07 12 35.3     & M81-I33      & O9.7Ve      & --                 & --                      & 3.9 (5.1)      & 60 \\ 
1008-57                  & 1316            & 10 09 44.0     & -58 17 42.0    & --                  & B0 IIIVe     & 15                & --                      & $\sim$5      & 125 \\ 
1118-61                  & 1366            & 11 20 57.18   & -61 55 00.2    & Hen 3-640   & O9.5Ve      & --                 & --                      & 5 (2.9)        & 131 \\ 
1409-619                & 1310?          & 14 08 02.56   & -61 59 00.3    & 2MASS        & O/B?           & --                 & --                     & 14.5           & 79 \\ 
1538-52                  & 1213            & 15 42 23.3     & -52 23 10.0    & QV Nor        & B0 Iab        & $14.1\pm2.8$ & $1.0\pm0.1$  & 6.4             & 104 \\ 
1553-542                & 1316            & 15 57 48.3     & -54 24 53.1    & VVV star      & B1-2V         & --                 & --                      & $20\pm4$   & 82,146 \\ 
1626-67                  & 1410            & 16 32 16.79   & -67 27 39.3    & KZ TrA         & --                & 0.03-0.09     & --                      & 5-13           & 132,4 \\ 
1626.6-5156           & 1366            & 16 26 36.24   & -51 56 33.5    & 2MASSJ      & B0-2Ve       & --                 & --                      & --                & \\ 
16393-4643            & 1220            & 16 39 05.47   & -46 41 13.0    & B giant         & B giant        & $\ge7$?      & --                      & --                & 29 \\ 
16493-4348            & 1220            & 16 49 26.92   & -43 49 08.96  & 2MASS        & B0.5 Ib       & --                  & --                     & --                 & 30 \\ 
1744-28                  & 1495            & 17 44 33.09   & -28 44 27.0    & star a           & G4 III          & 0.2-0.7        & 1.4-2                & 7.5              & 106,108 \\
17544-2619            & ??                & 17 54 25.7     & -26 19 58       & GSC 6849    & O9Ib           & 25-28          & --                      & 2-4 (2.6)     & 129 \\ 
18027-2016            & 1360            & 18 02 39.9     & -20 17 13.5    & --                 & B1Ib            & $\ge11$       & --                      & 12.4            & 143,144 \\ 
18179-1621            & 1311            & 18 17 52.18   & -16 21 31.68  & 2MASS        & (IR)              & ??               &  ??                    & --                & 114 \\ 
1822-371                & 1417            & 18 25 46.81   & -37 06 18.6    & V691 Cra     & --                & $0.46\pm0.02$ & --                 & 2.5              & 116 \\ 
1829-098                & 1310?          & 18 29 43.98   & -09 51 23       & --                 & O/B             & --                  & --                      & $\sim$10    & 225 \\ 
1907+09                 & 1216            & 19 09 39.3     & +09 49 45.0   & ??                & O8-9 Ia       & $\sim$27      & $\sim$1.4         & 5 (4.4)         & 105 \\ 
19294+1816           & 1311            & 19 29 55.9     & +18 18 38.4   & 2MASS         & B1 Ve          & --                 & --                      & 11               & 203-206 \\  
1946+274               & ??                & 19 45 39.30   & +27 21 55.4   & --                   & B01-IVVe    & ??               & ??                     & ??               &  \\    
1947+300               & 1316            & 19 49 35.49   & +3012 31.8    & 2MASS         & B0Ve           & $>$3.4         & 1.4$^{d}$         & 9.5              & 183,184\\ 
2206+54                 & 1260            & 22 07 56.24   & +54 31 06.40 & BD+53 2790 & O9.5Vep     & 18                & --                      & 2.6 (3.3)      & 177,176 \\ 
  \noalign{\smallskip}\hline 
  \end{tabular}
  \end{center}
$^{a}$ The type of object is according to HEASARC coding: http://heasarc.gsfc.nasa.gov/W3Browse/class\_help.html:  \\
1000 - X-ray binary                    \\
1100 - HMXRB ---------------------- 10 - X-ray pulsar --------- 1 - flares  \\
1200 - HMXRB supergiant --------- 20 - burster ---------------- 2 - jets   \\
1300 - HMXRB ---------------------- 30 - black hole ------------ 3 - eclipsing   \\
1300 - HMXRB Be star ------------- 40 - QPO ------------------ 4 - ultra-soft trans.   \\ 
1400 - LMXRB ---------------------- 50 - QPO + black hole --- 5 - soft transient   \\
1400 - LMXRB ---------------------- 60 - QPO + pulsar -------- 6 - hard transient    \\
1500 - LMXRB Globular cluster -- 70 - QPO + bursts --------- 7 - eclipsing dipper \\ 
------------------------------------------ 80 - QPO,pulsar,bursts --- 8 - eclipsing ADC   \\
------------------------------------------ 90 - pulsar + bursts ------- 9 - dipper  \\                                                  
$^{b}$ Coordinates from SIMBAD;  $^{c}$ References: see Table~\ref{tab:collection}; $^{d}$ assumed; $^{e}$ distances from Gaia DR2 
in parenthesis \citep{BailerJones_etal18}.
\end{table*}


\newpage

\begin{table*}[]
  \caption[]{Changes of cyclotron line energy E$_{cyc}$ with pulse phase and luminosity, corresponding changes in photon index $\Gamma$
  with luminosity.}
  \vspace{-3mm}
  \label{tab:variation}
  \begin{center}
  \begin{tabular}{lllllllll}
  \hline\noalign{\smallskip}
System                    & E$_{cyc}$/phase    & Ref.   & E$_{cyc}$/L$_{x}$ & Change         & For factor   & Ref.$^{d}$  & $\Gamma$/L$_{x}$  & Ref.$^{d}$ \\
                                & (max-min)/             &               &  correl.              & in E$_{cyc}$    & in L$_{x}$   &                & correl.            &         \\
                                & mean                     &               &                          &                         &                     &               &                       &  \\
 \hline\noalign{\smallskip}                        
Cen~X-3                & $\sim$30\%            & 49,50     & --                     & --                       & --                 & --            & --                   & -- \\ 
Cep~X-4                & 15\% (at 28\,keV)   & 47          & pos$^{a}$        & +3\% for           & factor 3        & 46           & yes                & 46,48 \\  
      " " "                  & 20\% (at45\,keV)    & 47           & --                     & --                      & --                 & --             & --                   & -- \\  
      " " "                  & --                            & --            & pos, lin$^{b}$  & --                       & --                 & 189         & yes                & -- \\  
GX~301-2              & 25-30\%                 & 51,52     & pos?$^{c}$      & --                       & --                 & 53,52      & --                   & -- \\ 
GX~304-1              & $\sim$20\%            & 54          & pos$^{a}$        & +10\% for         & factor 10     & 20,23       & yes                & 23,59,133  \\ 
     " " "                    & --                            & --           & pos, flatten      & +20\% for          & factor 3       & 55            & yes                & 55  \\ 
Her~X-1                 & 25\%                       & 33,26    & pos, lin$^{a}$   & +5\% for           & factor 2        & 25,27       & yes                & 56  \\  
     " " "                    & --                            & --           & pos, lin$^{b}$   & +4\% for           & factor 2       & 56            & yes                 & 56  \\ 
NGC300 ULX1       & --                            & --           & --                      & --                      & --                 & --             & --                    & -- \\  
SMC~X-2               & 20\%                       & 91          & neg?$^{c}$      & not signif          & --                 & 91            & yes?$^{c}$    & 91 \\ 
SXP~15.3               & 50\%                       & 215        & --                     & --                      & --                 & --             & --                   & --  \\ 
Vela~X-1                & 25\%                       & 69          & pos, flat?          & +4\%                & factor 5       & 32            & yes                 & 32 \\ 
      " " "                   & 15\%-20\%             & 70,71     & --                      & --                      & --                & --              & --                    & --  \\ 
      " " "                   &                                &               & pos, flat            & +8\%                & factor 7       & 115          & --                    & -- \\ 
X~Per                     & 40\%                       & 16          & no                    & --                      & --                & --              & no                   & --  \\  
0115+63                    & 20\% 1st har 1990  & 66          & neg, lin$^{ac}$ & -40\% for         & factor 4       & 57,58       & yes                 & 59,133  \\    
        " " "                    & 35\%  2nd har 1990& 66          & neg, lin$^{bc}$ & -11\% for         & factor 2       & 56            & --                    & -- \\    
        " " "                    & 25\%  1st har 1991 & 66          & no                     & --                      & --                & 42           & --                    & -- \\   
        " " "                    & --                             & --           & no                     & --                      & --                & 41           & --                    & -- \\   
0332+53                 & small ($\sim$5\%)    & 65         & neg, lin$^{a}$   & -20\% for          & factor 10     & 62           & yes                 & 59,64,133 \\  
      " " "                    & 2-6\% (L$_{x}$ dep)& 67         & neg, lin$^{b}$  & $\sim$-2\% for  & factor 1.7    & 56           &                       & -- \\    
      " " "                    &                                 &              & pos, lin$^{b}$  & $\sim$1.2\% for & factor 3.5   & 189         &                        & 99 \\  
0440.9+4431           & --                             & --           &  --                    & --                       &                    & --            &  --                   & -- \\  
0520.5-6932           & 20\%                        & 31          & not signif         & $<$+0.6\%        & factor 1.09  & 31           & --                    & -- \\ 
0535+26                 & 15\%                        & 11,72     & no                    & --                      & --                & 73            & yes                 & 59,133  \\ 
      " " "                   &                                  &              & pos, lin             & 10\% for            & factor 5       & 74           & yes                 & 74,94  \\ 
0658-073                & no                            & 60,61     & no                    & --                       & --                 & --            & yes/no           & 59/60,133 \\ 
1008-57                  & $\sim$10\%              & 77          & no                    & --                       & --                & --             & yes                & 133,138 \\ 
1118-61                  & 20\%                        & 43          & no                    & --                       & --                & --             & yes                 & 133  \\ 
1409-619                & --                              & --           & --                      & --                       & --                & --             & --                   & --     \\ 
1538-52                  & 10\%                        & 81,87     & no                    & --                       & --                & 81,87       & yes                 & 139  \\ 
1553-542                & 20\%                        & 82          & no                    & --                       & --                & --             & yes                 & 82 \\  
1626-67                  & $\sim$20\%              & 83          & --                     & --                       & --                & --             & no                  &  83 \\ 
1626.6-5156           & 25\%                        & 15          & pos, lin             & $\sim$8\% for   & factor 6      & 15           & yes                 & 15 \\ 
16393-4643            & $\le3$\%                  & 29          & no                    & --                        & --               & --             & --                    & --  \\ 
16493-4348            & --                              & --            & --                     & --                        & --               & --             & --                    & --   \\ 
1744-28                  & --                              & --            & --                    & --                         & --               & --             & --                    & --   \\ 
17544-2619            & --                              & --            & --                    & --                         & --               & --             & --                    & --   \\ 
18027-2016            & 25\%                        & 142         & --                    & --                         & --               & --             & --                    & --  \\ 
18179-1621            & ??                            & ??          & ??                   & ??                       & ??              & ??            & ??                  & ??   \\ 
1822-371                & --                              & --            & --                    & --                         & --                & --             & --                    & --     \\ 
1829-098                & $\sim$9\%                & 224         & --                    & --                         & --               & --             & --                    & --  \\ 
1907+09                 & 25\%                        & 72           & pos?$^{c}$     & ??                       & ??              & 87            & --                   & --  \\ 
19294+1816           & --                              & --            & --                     & --                        & --                 & --             & --                   &  --    \\  
1946+274               & 30\%                        & 72           & no                    & ??                      & ??               & ??            & ??                  & ??    \\  
1947+300               & not signif                  & 88           & --                     & --                        & --                & --              & --                   & --    \\ 
2206+54                 &  --                             & --            & --                     & --                        & --                & --              & --                   & --    \\ 
\noalign{\smallskip}\hline 
  \end{tabular}\end{center}
$^{a}$ on long timescales; $^{b}$ on timescales of pulse period (``pulse amplitude selected''); $^{c}$ questionable / to be confirmed; 
$^{d}$ References: see Table~\ref{tab:collection}. \\

\end{table*}


\newpage

\begin{table*}[]
  \caption[]{Further details on cyclotron lines: centroid energy (from Table~\ref{tab:collection}), line width $\sigma$, central line depth D
  (the normalization constant of the line component, often called the ``strength''  of the line), ``optical depth'' $\tau$ of the line
   (with ``strength'' D = $\tau~\sigma_\text{cyc}\sqrt{2\pi}$), X-ray luminosity L$_{x}$, References and additional notes.}
  \vspace{-7mm}
  \label{tab:width}
  \begin{center}
  \begin{tabular}{lllllllll}
  \hline\noalign{\smallskip}
System                 & E$_{cyc}$      & Width            & ``strength''      & Optical           & L$_{x}$     & Ref.$^{c}$  & Notes  \\
                             &                       & $\sigma$      & (\texttt{gabs})& depth             & ($10^{36}$   &                 &  "tbc" = to be confirmed \\
                             & (keV)              & (keV)            & D                  & $\tau$            & ergs/s)          &                 &   \\
 \hline\noalign{\smallskip}                        
Cen~X-3               & 30/28             & 7/8                & --                  & 1.1/0.8           & --         & 128/186$^{a}$  &  \\  
       " " "                & 29                  & 4/6                 & 2.5               & 0.67,0.73      & 54               & 188/49,50  &  \\  
Cep~X-4               & 30                  & 5.8/4.9          & 20/16.6         & --                   & 1.4/5.5           & 48           & L$_{x}$  dependent  \\ 
       " " "                & 28/45             & 6-9/$\sim$11 & --                  & --                   & 1.4                 & 47           &   \\ 
GX~301-2            & 42                  & 8                     & --                  & 0.5                & --                & 128$^{a}$  & \\ 
     " " "                  & 35                  & 7.4/3.4           & --                  & 10.6/0.14       & 3.9/1.0           & 52/51      & \\ 
GX~304-1            & 56/60              & 10.1/9            & --                 & 0.9                 & --                    & 55           &   \\ 
     " " "                 & 43-58              & 4-12               & --                 & 0.6-0.9           & 17-380           & 55           & L$_{x}$ dependent\\ 
     " " "                 & 50-59              & 5-11               & 8.1               & 0.7                 & 2-16               & 24,54      &  \\ 
Her~X-1               & 40/42              & 6.4/6.5           & 9.5               & 0.66/0.74       & --         & 128/186$^{a}$  & L$_{x}$ dependent\\ 
     " " "                 & 37                   & $\sim$6         & --                  & 0.6-1.0           & 30                   & 136,137  & super-orbit: 34.85 d \\ 
NGC 300 ULX1   & 13                   & $\sim$3.5      & $\sim$5.5    & --                    & 100-3000       & 190         & tbc  \\  
SMC~X-2             & 27                   & 6.4-7.2          & --                 & 6-9                  & 180-550          & 91           & tbc, L$_{x}$ dependent \\ 
SXP~15.3            & 5-8                  & --                   & --                 & 0.4                   & 20-100           & 215          & tbc \\ 
Vela~X-1             & 24/25              & 0.9/4.7           & --                 & 0.16/0.26         & --           & 128/186$^{a}$ & L$_{x}$ dependent \\ 
      " " "                & 25                   & 4.05               & 3.8               & --                    & --                     & 188         &   \\ 
      " " "                & 26.5/53           & 3.5/7              & $\sim$15     & $\sim$0.4/--    & --                     & 32           &   \\ 
X~Persei              & 16.4/13.4        & 3.6/2.7            & --                 & 0.78/0.36       & --             & 128/186$^{a}$  & \\ 
0332+53              & 27/26              & 7.6/5.4            & --                 & 1.8/1.9           & 375/341            & 65/123    & L$_{x}$ dependent \\ 
      " " "                & 27.5-30.5        & 5-7                  & --                 & ~18                & 9-160               & 99            &  \\ 
      " " "                & 51/50              & 8.9/9.9            & --                 & 2.2/2.1           & 375/341           & 65/123     &  \\ 
      " " "                & 74/72              & 4.5/10.1          & --                 & 3.3/1.3           & 375/341           & 65/123     &  \\ 
0440.9+4431       & 32                   & 6(fix)               & --                 & --                    & --                     & 17            & tbc \\ 
0520.5-6932       & 31                   & 5.9                  & --                  & 0.6                 & 370-400          & 31             & tbc \\ 
0535+26             & 50                    & $\sim$10         & $\sim$0.1/1.6  & 0.5             & --/0.2               & 12,73/127 &  see also 75,94 \\ 
     " " "                 & 44-50              & 9-11                &                    & 3-13               & 6-49                & 74             & L$_{x}$ dependent \\ 
     " " "                 & 110                  & 10-15             &                    & 1/1.6               &                        & 73/126      & see also 74 \\ 
     " " "                 & 45.1                 & $9\pm1.3$     & 9                  &                        &                         & 188          & \\ 
0658-073            & 33                    & $\sim$12        & --                  & 0.4                 & --                     & 60             & same as MX 0656-072 \\ 
1008-57              & 78                    & $\sim$11         & --                 & --                    & 0.5-100           & 77, 78       &  \\ 
1118-616            & 55, 110?          & $\sim$14/8.8   & 60/15           & --                    & 9                     & 45/188      & \\ 
1409-619            & 44                    & 4 (fix)              & --                  & 16+14/-7         & 0.07                & 79            & tbc \\ 
1538-52              & 21                    & 2-3                  & --                  & $\sim$0.5        & 3.5-9       & 8,80,81,216    & super-orbit: 14.92 d \\ 
1553-542            & 23/27               & 10.8/6.44        & --/8.28          & --                     & 76                  & 82            & tbc \\ 
1626-67               & 39/38               & 6.6/4.3           & --                  & 2.1/1.4            & --          & 128/186$^{a}$   & \\ 
      " " "                & $\sim$38          & $\sim$5         & --                 & $\sim$20         & 1.5-100          & 83             & \\ 
1626.6-5156        & 10,18               & 0.6--1.3          & --                 & 0.1--0.3           & $\sim$30        & 15             & tbc, L$_{x}$ dependent \\ 
16393-4643        & 29                     & 4                    & --                  & 0.4                   & --                   & 29            & tbc \\ 
16493-4348         & 31                    & 10 fixed         & $\sim$0.6$^{b}$  & --               & --                   & 30,96       & tbc, super-orbit: 20.07\,d  \\ 
1744-28              & 4.3                   & 1.2                  & --                  & 0.12                 & 100               & 186           & tbc, Bursting Pulsar  \\ 
17544-2619         & 17/33              & 3.0/6.6            & --                  & 0.53/0.9           & 0.04              & 22             & tbc   \\ 
18027-2016         & 24                   & $\sim$5          & --                  & 0.2-0.5             & 3.1                & 142           & tbc  \\ 
18179-1621         & 21                   &  ??                 & ??                & ??                     & ??                 & ??            &  ??   \\ 
1822-371             & 0.7                   & 0.14               & --                  & 0.03                 & 180-250        & 85            &  tbc \\ 
      " " "                 & 33                    & $\sim$5         & --                  & --                     & --                   & 86            &  \\ 
1829-098              & 15                    & 2.3                & --                  & 0.52                 & $\sim$4.3      & 224          &  assumed D = 10\,kpc \\ 
1907+09               & 18                    & 1.6/1.9          & --                  & 0.26/0.26         & --        & 128/186$^{a}$    & L$_{x}$ dependent? \\ 
      " " "                 & 40                    & 3.06               & --                 & 2.3                   &                       & 186           & \\ 
19294+1816         & 43                    & 5.4                & --                  & 1.2                   & a few              & 198          & tbc \\ 
1946+274            & 35/39                & 4.8/9             & --                  & 0.25/1.1            & --        & 128/186$^{b}$   & \\ 
      " " "                 & 35                    & 2                    & 4                  & --                      & --                   & 188          & \\ 
1947+300             & 12                    & 2.5                 & 0.36-0.48     & (0.07)               & --                   & 88,98       & tbc \\ 
2206+54               & 32                    & --                   & --                  & --                      & --                   & --              & tbc  \\ 
 \noalign{\smallskip}\hline
  \end{tabular}
  \end{center}
$^{a}$ using the systematic studies by \textsl{RXTE}: \cite{Coburn_etal02} (Ref. 128) and \textsl{Beppo}/SAX: \cite{DoroR_17} (Ref. 186);
$^{b}$ using \texttt{cyclabs} for the description of the cyclotron line (see Sect.\ref{sec:spectral_fitting}); $^{c}$ References: see Table~\ref{tab:collection}.
\\
\end{table*}



\end{document}